\documentclass[lettersize,journal]{IEEEtran}
\usepackage{amsmath,amsfonts}
\usepackage{algorithmic}
\usepackage{algorithm}
\usepackage{array}
\usepackage{color}
\usepackage{authblk}
\usepackage{textcomp}
\usepackage{stfloats}
\usepackage{url}
\usepackage{verbatim}
\usepackage{graphicx}
\usepackage{cite}
\usepackage{amssymb}
\usepackage[justification=centering]{caption}
\usepackage{xcolor}
\usepackage{bm}
\usepackage{CJK}
\usepackage{subfigure}
\usepackage{amsthm}
\usepackage{booktabs} 

\newtheorem{myDef}{Definition}
\newtheorem{myTheo}{Theorem}
\newtheorem{myAssumption}{Assumption}

\begin{document}

\title{FlocOff: Data Heterogeneity Resilient Federated Learning with Communication-Efficient Edge Offloading\\
\thanks{Manuscript received November 30, 2023; revised March 23, 2024; accepted 21 May 2024. This work was partially supported by the National Nature Science Foundation of China (NSFC) under Grant U21B2002 and Grant 62202307, the Major Key Project of Peng Cheng Laboratory under Grant PCL2021A15, and Terminus Group under Grant R00035. Liekang Zeng and Yang Yang are the corresponding authors.}
\thanks{Mulei Ma, Chenyu Gong and Liekang Zeng are with the IoT Thrust and Research Center for Digital World with Intelligent Things (DOIT), The Hong Kong University of Science and Technology (Guangzhou), China. (e-mail: mma085@connect.hkust-gz.edu.cn; cgong040@connect.hkust-gz.edu.cn; liekangzeng@hkust-gz.edu.cn).}
\thanks{Yang Yang is with the IoT Thrust and Research Center for Digital World with Intelligent Things (DOIT), The Hong Kong University of Science and Technology (Guangzhou), China, also with Peng Cheng Laboratory, Shenzhen 518055, China, and also with Terminus Group, Beijing 100027, China. (e-mail: yyiot@hkust-gz.edu.cn).}
\thanks{Liantao Wu is with the Software Engineering Institute, East China Normal University, Shanghai 200062, China. (e-mail: ltwu@sei.ecnu.edu.cn).}
}


\author{
    \IEEEauthorblockN{Mulei Ma, \IEEEmembership{Graduate Student Member,~IEEE,}}
    \IEEEauthorblockN{Chenyu Gong, \IEEEmembership{Graduate Student Member,~IEEE,}}
    \IEEEauthorblockN{Liekang Zeng, \IEEEmembership{Member,~IEEE,}}
    \IEEEauthorblockN{Yang Yang, \IEEEmembership{Fellow,~IEEE,}}
    \IEEEauthorblockN{Liantao Wu, \IEEEmembership{Member,~IEEE}}
}




\ifodd 1
\newcommand{\rev}[1]{{\color{blue}#1}} 
\newcommand{\com}[1]{\textbf{\color{red} (COMMENT: #1)}} 
\newcommand{\comg}[1]{\textbf{\color{green} (COMMENT: #1)}}
\newcommand{\response}[1]{\textbf{\color{magenta} (RESPONSE: #1)}} 
\else
\newcommand{\rev}[1]{#1}
\newcommand{\com}[1]{}
\newcommand{\comg}[1]{}
\newcommand{\response}[1]{}
\fi

\maketitle

\begin{abstract}
Federated Learning (FL) has emerged as a fundamental learning paradigm to harness massive data scattered at geo-distributed edge devices in a privacy-preserving way. Given the heterogeneous deployment of edge devices, however, their data are usually Non-IID, introducing significant challenges to FL including degraded training accuracy, intensive communication costs, and high computing complexity. Towards that, traditional approaches typically utilize adaptive mechanisms, which may suffer from scalability issues, increased computational overhead, and limited adaptability to diverse edge environments. To address that, this paper instead leverages the observation that the computation offloading involves inherent functionalities such as node matching and service correlation to achieve data reshaping and proposes \underline{F}ederated \underline{l}earning based \underline{o}n \underline{c}omputing \underline{Off}loading (FlocOff) framework, to address data heterogeneity and resource-constrained challenges. Specifically, FlocOff formulates the FL process with Non-IID data in edge scenarios and derives rigorous analysis on the impact of imbalanced data distribution. Based on this, FlocOff decouples the optimization in two steps, namely : (1) Minimizes the Kullback-Leibler (KL) divergence via Computation Offloading scheduling (MKL-CO); (2) Minimizes the Communication Cost through Resource Allocation (MCC-RA). Extensive experimental results demonstrate that the proposed FlocOff effectively improves model convergence and accuracy by 14.3\%-32.7\% while reducing data heterogeneity under various data distributions.

\end{abstract}


\section{Introduction}

\begin{figure}[ht]
\centerline{\includegraphics[scale=0.3]{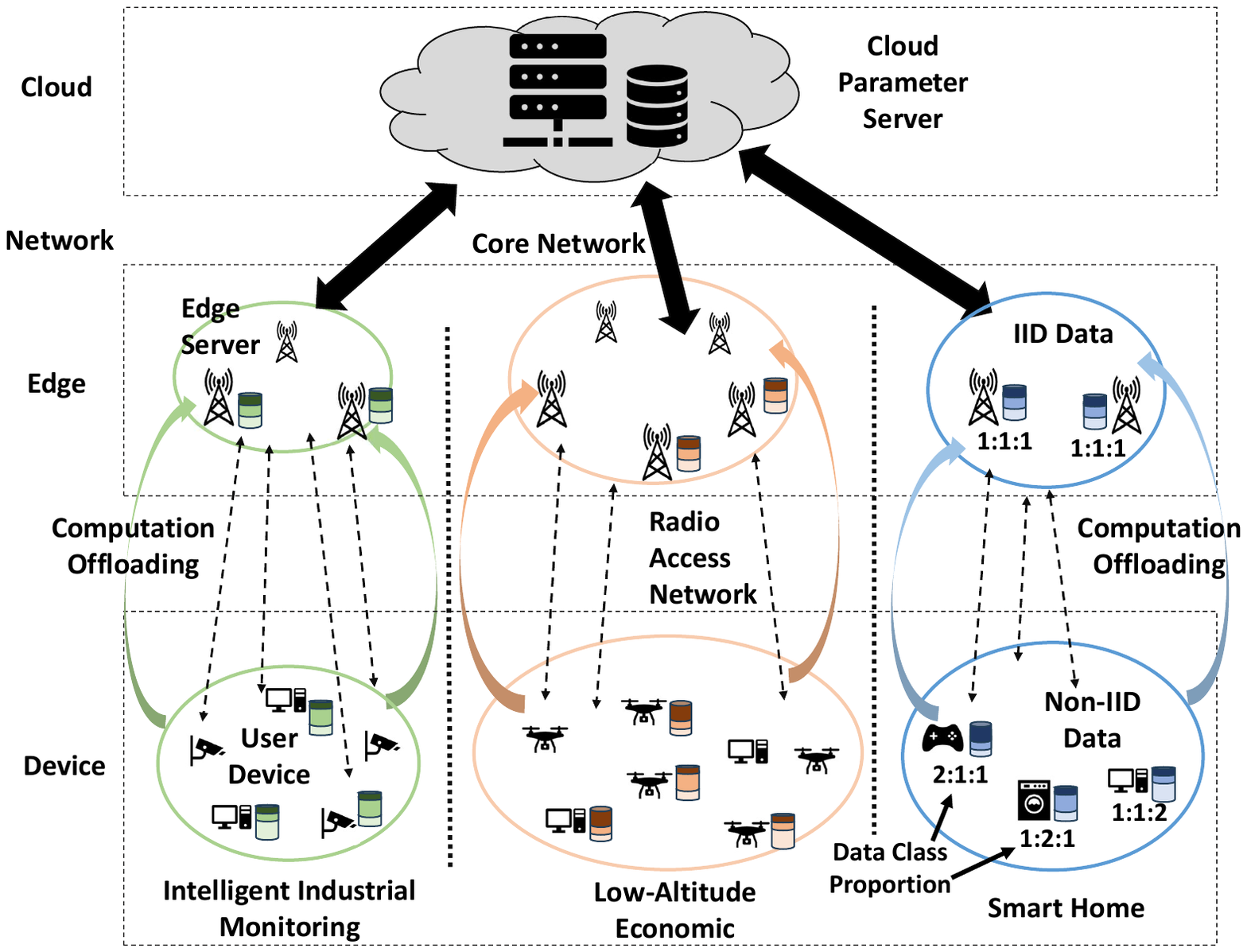}}
\caption{Illustration of the proposed Federated Learning framework based on edge computation offloading.}
\label{fig1}
\end{figure}

With the advent of the Internet of Things (IoT) and edge computing, an enormous volume of data is being generated on various clients, such as mobile phones, robotics, and wearable devices \cite{ref1}. Considering factors such as privacy and communication costs, data processing across numerous and dispersed devices has emerged as a future trend. With the proliferation of artificial intelligence applications, the demand for model training using distributed data has significantly increased \cite{ref2}. Among them, Federated Learning (FL) has become a feasible learning paradigm\cite{ref3}. Federated learning enables local model training on edge data sets. Subsequently, under the centralized control of the cloud parameter server, parameter aggregation and model delivery take place \cite{ref4}. This approach, which uploads parameters rather than original data, mitigates the risk of private data leakage and breaks data silos \cite{ref5}. Presently, FL has found application in several business scenarios, including smart healthcare, video detection, and virtual reality \cite{ref6, ref7, ref8}.

However, the training efficiency of FL can be severely affected by the fragmented nature of data, highlighting the substantial impact of data heterogeneity on the learning process. Fig. \ref{fig1} shows a three-layer FL service architecture that can be abstracted as "Cloud-Edge-Device". Heterogeneous data may accumulate at the device layer close to the user. In particular, from the perspective of data distribution, this can result in significant data imbalance across clients, leading to the issue of data heterogeneity. More preciously, data heterogeneity can be: (1) differences in the number of samples across clients, and (2) differences in the sample categories owned by each client \cite{ref9}. In Fig. \ref{fig1}, we represent the proportion of each category within the total sample using varying shades of color. A greater variation in color intensity indicates stronger data heterogeneity (in device layer), while uniform color intensity suggests higher data quality (in edge layer). For example, in intelligent industrial monitoring applications, numerous user devices storing Non-IID data (represented by cylinders of different colors) are distributed at the lowest level, as depicted in Fig. \ref{fig1}. If each participant in FL holds data samples from only a few classes, it can lead to the aggregated model not converging at all \cite{ref10}. Additionally, discrepancies between local and global data distributions result in significant differences between the aggregated local and global model parameters. This phenomenon becomes more pronounced with an increase in local iteration round \cite{ref11, ref12}. One example is the slightly inferior performance of the FedAvg algorithm compared to centralized training. Moreover, according to Zhu, H. et al.'s  \cite{ref13} investigation, FedAvg is highly sensitive to data distributions. In cases involving deep networks, convergence might not be achieved on severely Non-Independent and Identically Distributed (Non-IID) data.

To address data heterogeneity for FL, existing researchers have explored in several directions. Adaptive algorithms are one approach to addressing this issue. It autonomously adapts to heterogeneous data distributions, thereby enhancing the model's generalization capabilities. For instance, Zhang, J. et al. \cite{ref14} adjusts participant subsets and automatically controls hyperparameters to mitigate the negative impact of disparate data. However, these adaptive algorithms necessitate additional computational resources and time for the model's adaptive adjustments, increasing the training costs. Furthermore, due to the existence of heterogeneous data, the model's adaptive adjustments might lead to overfitting issues, reducing its generalization on new data. As an alternative option, weight divergence serves as another solution, aiming to enhance model stability by computing appropriate aggregation parameters based on data distribution characteristics. Wu, H. et al. \cite{ref15} dynamically adjusted contributions from each participant by assigning distinct weights to individual nodes. However, this weight divergence approach relies on accurately identifying and understanding data distribution characteristics. Inaccurate or incomplete comprehension of data distribution could potentially misguide parameter settings and result in erroneous aggregation outcomes.

Alternatively, we draw the following two observations from real-world FL deployment toward addressing data heterogeneity. First, in many FL use cases, edge devices within a dedicated domain can mutually share their data to eliminate data heterogeneity without violating privacy constraints. This can be relevant for many cross-silo FL applications, where a group of edge devices within a silo is usually managed by the same user or organization, e.g., a smart factory or a clinical hospital, and their data and computing resources can be effectively shared within a privacy-admitted scope \cite{ref16, ref17}. Second, computation offloading can be leveraged to enable heterogeneous data mixture among a group of trusted, cooperative edge devices, and thus can enhance the data resilience of FL. Besides, by offloading workload across facilities, the resource shortage of individual edge devices can be alleviated and thereupon accelerate the client-side training in FL.

Inspired by the above viewpoints, in this paper, we make a rigorous theoretical exploration of FL's data heterogeneity and propose the \underline{F}ederated \underline{l}earning based \underline{o}n \underline{c}omputing \underline{Off}loading (FlocOff) framework, which is an efficient FL system that integrates computation offloading for data heterogeneity resilience. Firstly, we investigate the impact of data distribution on the training efficiency of federated learning. It is proved that data heterogeneity is one of the factors affecting model convergence. Secondly, we formulate the data heterogeneity resilience optimization for FL and design the FlocOff framework, which proposes a two-fold approach: (1) Reshaping the Edge Dataset via Computation Offloading (\textbf{RED-CO}), and (2) Communication Cost Optimization in Edge Environments (\textbf{CCO-EE}). Specifically, for \textbf{RED-CO}, FlocOff employs computation offloading as a mapping algorithm to adjust the offloaded data and expand the data of minority classes in the local dataset. For \textbf{CCO-EE}, this paper obtains the optimal communication power allocation strategy through efficient numerical methods to efficiently utilize resources in the network. Finally, FlocOff is evaluated on two publicly available datasets, and the experimental results demonstrate that the reshaped dataset reduces data heterogeneity and significantly improves the accuracy of the federated learning model. The main contributions of this study are as follows:

\begin{itemize}

\item[1)]
We theoretically study the impact of data heterogeneity in FL and reveal the potential of leveraging computation offloading as a useful tool to empower FL with data heterogeneity resilience.

\end{itemize}

\begin{itemize}

\item[2)]
We formally formulate the offloading-assisted FL training efficiency optimization problem and show its mathematical properties for approximate subproblem decoupling.

\end{itemize}

\begin{itemize}

\item[3)]
We propose FlocOff, an efficient optimization framework that schedules offloading to eliminate data heterogeneity in FL. FlocOff introduces a data distribution aware mapping algorithm for associating edge devices with edge servers, and a communication-efficient resource allocation algorithm to minimize total system cost.

\end{itemize}

\begin{itemize}

\item[4)]
We conduct extensive experiments and show that FlocOff achieves up to 14.3\%-32.7\% accuracy improvement over baselines while reducing 48.1\% communication costs during FL training.

\end{itemize}

The rest of this article is as follows. Section \uppercase\expandafter{\romannumeral2} carries out our motivation and related work. Section \uppercase\expandafter{\romannumeral3} introduces the system model. Section \uppercase\expandafter{\romannumeral4} comes to the problem formulation. Section \uppercase\expandafter{\romannumeral5} explores the FlocOff algorithm and principle. Section \uppercase\expandafter{\romannumeral6} performs some experimental analysis and evaluations. Section \uppercase\expandafter{\romannumeral7} comes to the conclusion.

\section{Preliminary}
\subsection{Background and Motivation}
In this section, we conducted a preliminary experiment to validate the influence of data distribution on federated learning, thereby establishing the impetus for our proposed framework.

\begin{figure}[ht]
\centering
\subfigure[Data heterogeneity caused by dirichlet distribution.]
{
    \begin{minipage}[b]{.98\linewidth}
        \centering
        \includegraphics[scale=0.36]{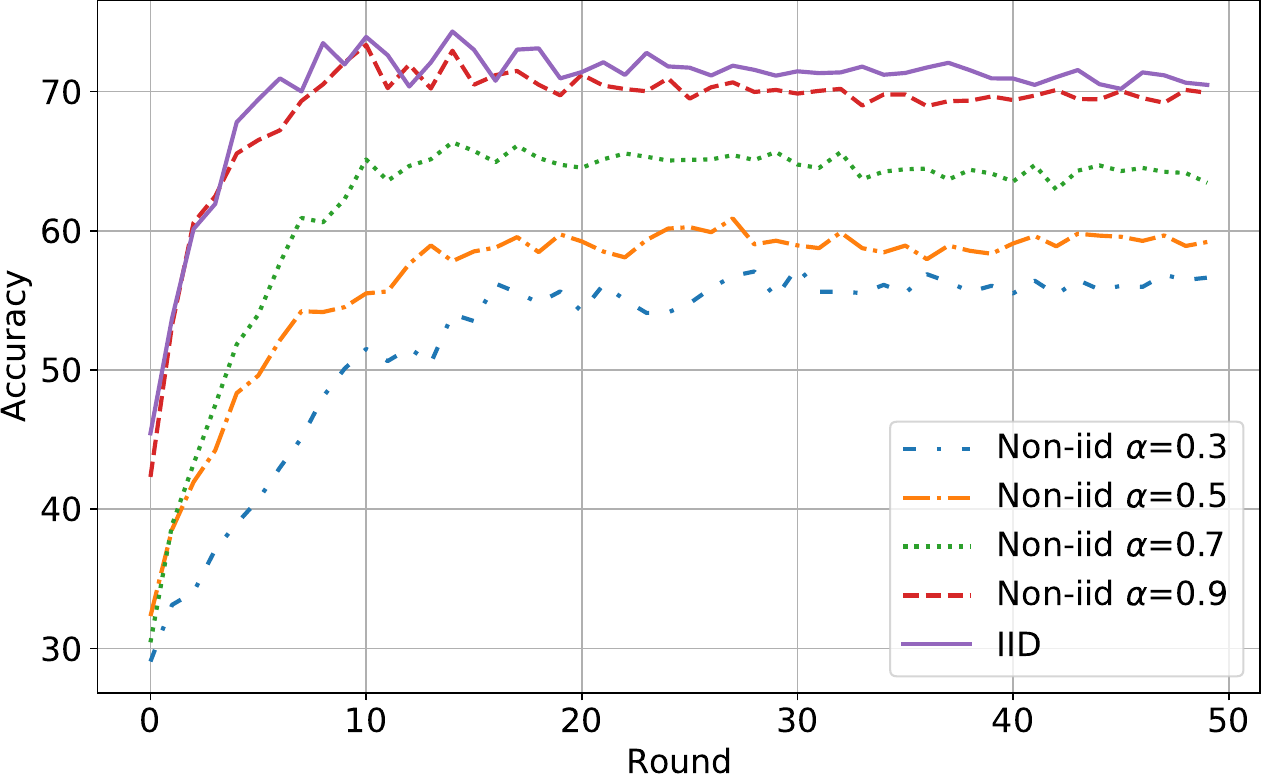}
    \end{minipage}
}
\subfigure[Data heterogeneity caused by imbalance between classes.]
{
 	\begin{minipage}[b]{.98\linewidth}
        \centering
        \includegraphics[scale=0.36]{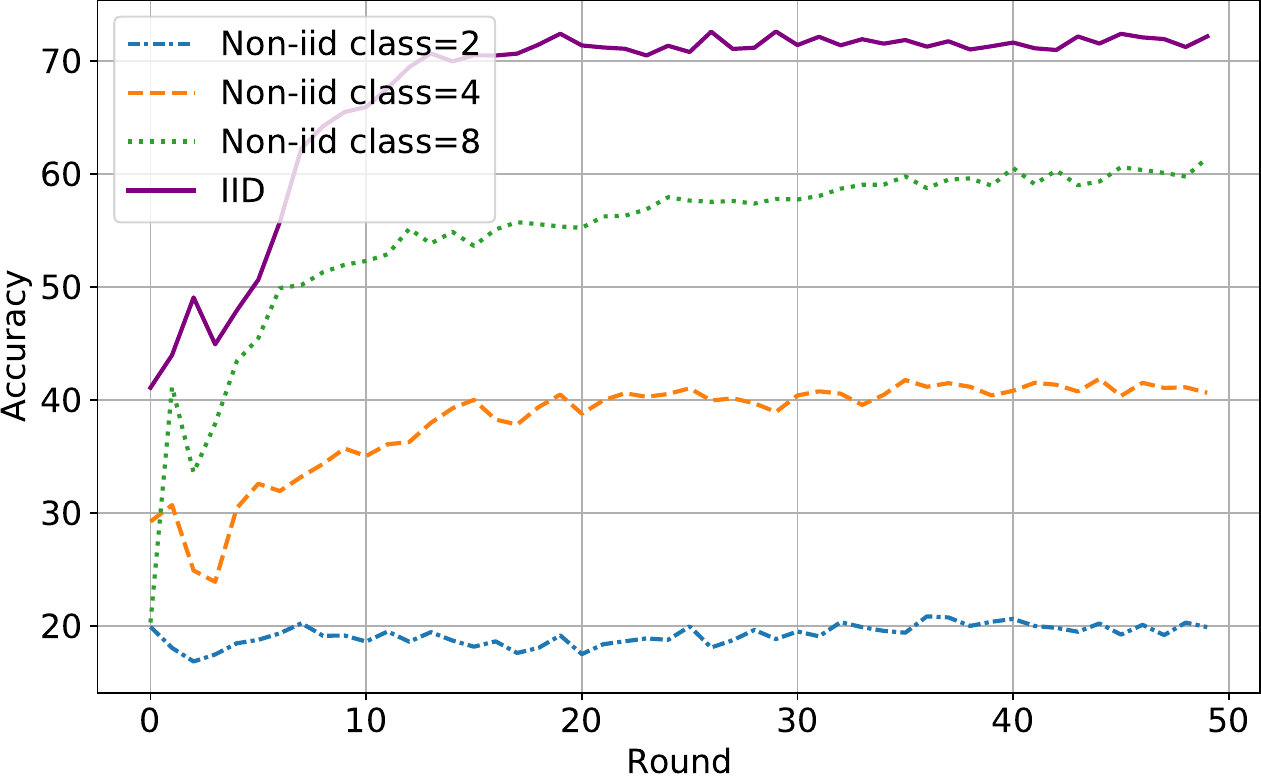}
    \end{minipage}
}
\caption{Impairment of model training by data heterogeneity.}
\label{fig2}
\end{figure}

The Dirichlet distribution plays a significant role in Bayesian models and is considered a suitable choice for simulating real data distributions\cite{ref18}. Specifically, the Dirichlet distribution is the conjugate distribution of the multinomial distribution. It can be defined as a set of continuous multivariate probability distributions, which can be represented as follows:
\begin{equation}
\Theta=\left\{\theta_1, \ldots, \theta_m\right\} \sim \operatorname{Dir}\left(\alpha_1, \ldots, \alpha_m\right),
\end{equation}

\begin{equation}
P\left(\theta_1, \ldots, \theta_m\right)=\frac{\Gamma\left(\sum_k \alpha_k\right)}{\prod_k \Gamma\left(\alpha_k\right)} \prod_{k=1}^m \theta_k^{\alpha_k-1},
\end{equation}

\noindent in which $\alpha$ represents as the distribution parameter (concentration or scaling parameter). The larger the value, the closer the distribution is to a uniform distribution, and the smaller the value, the more concentrated the distribution.

The training task depicted in Fig. \ref{fig2} (a) employs the CIFAR-10 dataset, which comprises RGB color images of diverse objects, classified into ten categories\cite{ref19}. The training model is based on the ResNet network architecture, and the depth of the network is set to 18 layers, with group normalization \cite{ref20}. To simulate the genuine data distribution, we leverage the Dirichlet distribution, which assumes the role of a continuous multivariate probability distribution parameterized by \emph{$\alpha$}. A higher value of \emph{$\alpha$} leads to a distribution closer to uniform, whereas a lower value increases data heterogeneity. The six curves portrayed in Fig. \ref{fig2} (a), corresponding to either IID or Non-IID data, display an upward trend, indicating the efficacy of the model training. Notably, the highest level is achieved for IID data, converging to 71\% after adequate training. The accuracy levels for \emph{$\alpha=0.9$}, \emph{$\alpha=0.7$}, \emph{$\alpha=0.5$} and  \emph{$\alpha=0.3$} diminish progressively, reaching 70\%, 64\%, 60\%, and 56\%, respectively. These findings demonstrate that Non-IID data significantly impairs training efficiency, and accuracy and convergence decline as the degree of data heterogeneity increases.

Fig. \ref{fig2} (b) illustrates the detrimental effect of class imbalance on training efficacy. The figure uses different colors to denote the number of classes on each server, where blue, yellow, green, and purple represent two, four, eight, and ten classes (IID) on each server respectively. The results indicate that severe data heterogeneity, particularly when the number of classes is equal to 2 or 4, leads to a significant drop in accuracy. However, when the number of classes is equal to 8 or 10 (IID), improvements in data distribution can enhance model accuracy.

Upon conducting preliminary experiments, we have uncovered a set of noteworthy characteristics pertaining to the training of federated learning models: (1) The quality of model training is intimately linked to the distribution of data. Specifically, Non-IID data distributions such as those marked by an uneven allocation of samples across classes can significantly compromise the accuracy of federated learning algorithms. (2) We have noted that the rate of convergence in the model training process is inversely proportional to the degree of Non-IID in the dataset. In the federated learning framework, Non-IID data will increase the number of training rounds and parameter aggregation times, increasing system communication costs. Leveraging these insights, in this paper we conduct a retrospective examination of the impact of data heterogeneity on federated learning. We subsequently propose the adoption of computation offloading to revamp private samples and enhance dataset quality. Finally, we revisit the issue of communication efficiency in the federated learning framework with the aim of streamlining system costs whilst concurrently optimizing model accuracy.

\subsection{Related Work}

\begin{figure}[ht]
\centerline{\includegraphics[scale=0.32]{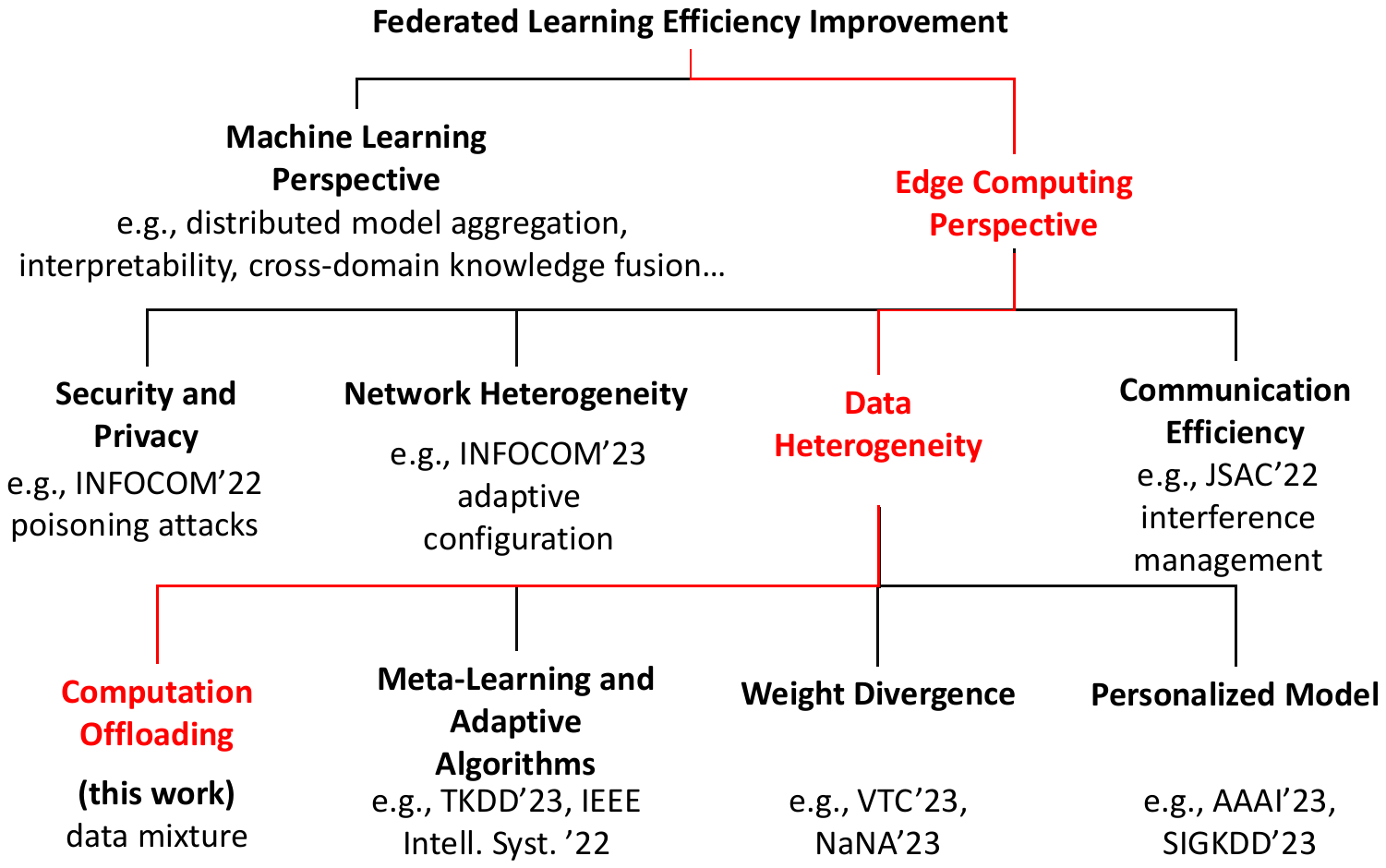}}
\caption{Classification of federated learning efficiency improvement and various approaches to address data heterogeneity. Our solution employs computing offloading to address issues of data heterogeneity.}
\label{fig3}
\end{figure}

In this section, as shown in Fig. \ref{fig3} we provide a related work overview where we present the two underlying perspectives to improve federated learning efficiency: machine learning perspective and edge computing perspective \cite{ref21}.

Different from the perspective of edge computing and communication, machine learning focuses on optimizing models and knowledge itself. Methods to enhance the efficiency of federated learning involve various aspects. Firstly, distributed model aggregation concentrates on merging local model updates from different participants to achieve a higher quality global model. Interpretability stands as another critical area that aims to make model decisions more interpretable and trustworthy, thus enhancing the transparency and controllability of federated learning systems. Additionally, cross-domain knowledge fusion is an important direction in leveraging information from different domains to enhance the generalization ability and learning efficiency of federated learning models. These methods focus on optimizing the way model updates occur and aim to elevate the performance and efficiency of federated learning through more efficient information and model integration.

From the edge computing perspective, enhancing the efficiency of federated learning involves several key domains. Firstly, Security and Privacy demands safeguarding participants' privacy during data transmission to prevent data leakage or malicious attacks. For instance, Ma, Z. et al. \cite{ref22} introduced the ShieldFL framework, employing dual Trapdoor homomorphic encryption to address poisoning in PPFL's encrypted models. Secondly, Network Heterogeneity acknowledges that federated learning participants may use varied network connections, necessitating effective model update transmission across diverse network environments. This may require designing new communication protocols and optimization algorithms adaptable to diverse network conditions. Liao, Y. et al. \cite{ref23} utilized the FedHP framework to optimize local updating frequencies and network topology in decentralized federated learning, addressing heterogeneity for quicker convergence and improved accuracy. Another critical area is Data Heterogeneity, where different participants in federated learning may possess disparate types and qualities of data. Addressing how to effectively train and update models in the presence of significant data differences is essential. Finally, Communication Efficiency focuses on reducing communication latency, improving data transmission speed, and efficacy. This might involve compressing model updates, reducing data transmission volume, or optimizing network scheduling strategies. Wang, Z. et al. \cite{ref24} optimized the AirComp and multi-cell FL framework to mitigate inter-cell interference, enhancing learning performance in wireless networks.

Data heterogeneity stands as a significant factor impacting the efficiency of federated learning. It denotes the differences among various data sources on devices, encompassing uneven data distribution, inconsistent feature distributions, and more. To mitigate the adverse effects of data heterogeneity on federated learning efficiency, researchers have proposed several methodologies. One such approach involves meta-learning and adaptive algorithms. These techniques employ meta-learning to automatically adapt to data heterogeneity, thereby enhancing the model's generalization capability and performance. A recent investigation \cite{ref25} employs meta-learning and consistency constraints, allowing servers to exchange parallel model information to eliminate data heterogeneity. Yang, L. et al. \cite{ref26} introduced a framework named G-FML, utilizing adaptive grouping of clients and leveraging meta-learning within groups to achieve personalized models, effectively addressing highly heterogeneous environments.

Another solution involves weight divergence, which aims to influence the aggregation of parameters based on data distribution characteristics, thereby enhancing the convergence and stability of the model. Naas, S.-A. et al. \cite{ref27} proposed a merging algorithm utilizing weight transfer values quantifying device contributions and implemented a client selection mechanism based on a Bayesian model. Liu, T. et al. \cite{ref28} implemented an optimization method based on class-difficulty based weights, addressing the stability and convergence issues of Non-IID data in federated learning. Additionally, personalized models serve as another approach to tackling data heterogeneity. This method tailors individualized models for each device or data source, considering data heterogeneity extensively, and enabling the model to adapt better to the characteristics of each data source. Tan, Y. et al. \cite{ref29} proposed the FedStar framework using structural embedding and independent structure encoders for federated graph learning, aiming to extract shared structural knowledge in non-uniformly distributed cross-dataset and cross-domain graph data. Qin, Z. et al. \cite{ref30} combined mutual and ensemble learning, addressing non-IID data without prior knowledge, achieving high accuracy on diverse data distributions in cross-silo federated learning.

Our methodology introduces an approach to addressing data heterogeneity in federated learning environments, focusing on optimizing data distribution via computing offloading, rather than merely adjusting algorithms post-distribution. This strategy aims to reshape edge datasets by strategically deploying computational tasks across devices based on their capabilities and data characteristics. By directly tackling the challenges of data heterogeneity, this method enhances the learning process's efficacy and efficiency across devices, diverging from traditional methods like client selection or simple data sharing. Notably, through the implementation of algorithms such as the Federated Client Cluster and Latency-Prediction Selection (FCCPS) by Xin, F. et al. \cite{ref31}, and the utility function modeling based on clients' data and resources heterogeneity by Zhang, R. et al. \cite{ref32}, our approach seeks to reduce the skewness caused by data heterogeneity, optimizing the learning process by ensuring data processing is most representative where it occurs.

In contrast to existing approaches that mainly focus on algorithmic adjustments to tackle data heterogeneity, such as Morafah, M. et al. \cite{ref33} incorporating a proximal term in local optimization, modifying the model aggregation scheme at the server side, or Miao, J. et al. \cite{ref34} employing federated clustering methods like FedDec for improved model accuracy through hyperparameters similarity, our work prioritizes a data-centric strategy. By integrating computational offloading as a key technique, FlocOff aims to enhance the training efficiency of edge federated learning by proactively optimizing data distribution and processing, thus addressing the root cause of data heterogeneity. Our research, unlike prior works that primarily focus on algorithmic adjustments or selective participation to combat heterogeneity, delves into the fundamental reorganization of data processing.

In the context of communication resource allocation methods, directly allocating resources specifically to address data heterogeneity is not commonly observed. Instead, several studies have addressed this challenge by integrating communication strategies with other methods to manage data heterogeneity indirectly. For instance, Huang, X. et al. \cite{ref35} tackled the reduction of data volume transmitted in federated learning via communication compression, and introduced a method employing stochastic control to manage the heterogeneity of data processing. Others \cite{ref36} have minimized the impact of Non-iid data by optimizing user associations and the uplink channel signal-to-noise ratio (SNR), leveraging wireless resource allocation to reduce learning latency. Contrary to strategies aimed specifically at data heterogeneity, communication resource allocation often addresses scenarios involving device heterogeneity. For example, Cui, Y. et al. \cite{ref37} developed a mixed-integer linear programming strategy optimized for communication resource allocation to mitigate training inefficiencies caused by device heterogeneity and limited communication resources. Furthermore, Ruby, R. et al. \cite{ref38} explored energy-efficient computing and communication resource allocation in a two-tier federated learning network involving complex networks of IoT nodes and aerial platforms.

However, these methods failed to incorporate the characteristics of data distribution into the mechanism design. Fragmented information exacerbates the dynamic nature of edge environments, compromising the quality of federated learning datasets. In this paper, we introduce the concept of computation offloading into the federated learning landscape. Computation offloading, as a core technology in edge computing, enables mapping strategies from data to service nodes \cite{ref39}. Leveraging computation offloading techniques, we reshape the federated learning dataset to address data heterogeneity.

As shown in Fig. \ref{fig3}, our study presents a pioneering approach by introducing computation offloading to address the issue of data heterogeneity in federated learning. Our research focuses on data distribution reshaping by offloading decision-making to mitigate the impact of Non-IID. Specifically, our proposed framework, named FlocOff, involves the offloading and mapping of fragmented data at the edge to service nodes. This strategy enables the flexible adjustment of the data set, which enhances the efficiency of federated learning training. Furthermore, to minimize communication costs, we employ an efficient resource allocation algorithm. Subsequently, we conduct a series of experiments to verify the effectiveness and superiority of the FlocOff framework. Our approach demonstrates the remarkable potential for improving the quality of federated learning datasets, enabling the realization of efficient training and the achievement of accurate models.

\section{System Model}

In this section, we first introduce the general form of federated learning. Then, we briefly introduce the network communication model of this paper. It serves as a foreshadowing for the problem formulation and solution below. The symbols in this article are summarized in Table 1.

\begin{table}  
\caption{Summary of Notations}  
\begin{tabular*}{9cm}{ll}  
\hline  
Notation & Description \\  
\hline  
\emph{S}  & Number of ES \\ 
\emph{U}  & Number of UD \\ 
\emph{$D_u$}  & Local dataset of UD \emph{$u$} \\ 
\emph{$D_s$}  & Local dataset of ES \emph{$s$} \\ 
\emph{$D$}  & Global virtual data set of CPS \\ 
\emph{$\bar{D}_s$}  & Complement set of \emph{$D_s$} with respect to \emph{$D$}. \\ 
\emph{$w_s$}  & Weight of the model in ES \emph{$s$} \\ 
\emph{$w$}  & Weight of the model in CPS \\ 
\emph{$F_s(w_s)$}  & Loss function of ES \emph{$s$} \\ 
\emph{$F(w)$}  & Global loss function of CPS \\ 
\emph{$\phi$}  &  Learning rate \\ 
\emph{$h_{us}$}  & Communication gain between \emph{$s$} and \emph{$u$} \\ 
\emph{$p_{us}$}  & Communication power of task \emph{$u$} uploads to ES \emph{$s$} \\ 
\emph{$P_{\text{max}}$}  & Maximum power available to UD \emph{$u$} \\ 
\emph{$B$}  & Total system bandwidth \\ 
\emph{$K$}  & The number of subcarriers \\ 
\emph{$B_k$}  & Bandwidth of subcarrier \emph{$k$} \\ 
\emph{$I^k$}  & Set of users occupying subcarrier \emph{$k$} \\ 
\emph{$\sigma_{\mathrm{u}}$}  & Interference of \emph{$u$} \\ 
\emph{$\sigma$}  & Background noise \\ 
\emph{$R_u$}  & Task upload rate \\ 
\emph{$d_u$}  & Size of private dataset in \emph{$u$} (bit) \\ 
\emph{$T_u$}  & Time to transfer a task from \emph{$u$} to \emph{$s$} \\ 
\emph{$v_s$, $v$}  & Weight of the model in ES and CPS when IID \\ 
\emph{$\gamma_s$}  & Gradient distance to \emph{$F_s(w)$} and \emph{$F(w)$} \\ 
\emph{$L_s$,$L$}  & Lipschitz smooth gradient of \emph{$F_s(w)$} and \emph{$F(w)$} \\ 
\emph{$A$}  & Binary offloading policy matrix \\ 
\emph{$P$}  & Power allocation policy \\ 
\emph{$J$}  & Total communication cost of the system \\ 
\emph{$C$}  & Data category \\ 
\emph{$P_u(c)$}  & Proportion of category \emph{$c$} in the \emph{$D_u$} \\ 
\emph{$\Gamma$}  & Data Volume Threshold \\ 
\emph{$U_s$}  & Serviceable UD set for \emph{$s$} \\ 
\hline  
\end{tabular*}  
\end{table}

\subsection{Federated Learning Model}

Our objective is to establish a federated learning framework in edge computing scenarios, consisting of a Cloud Parameter Server (CPS) and a set \emph{$\mathcal{S}$} of S Edge Server (ES). Each ES and CPS possess local and global models, respectively, which enable them to execute model training and model aggregation tasks. The federated learning training procedure involves three stages\cite{ref40}: (1) Each ES trains their local models using their respective data; (2) Upon completing the local training, the ES uploads the trained model to the CPS, which aggregates all the uploaded models and performs parameter aggregation; (3) Subsequently, CPS disseminates the synthesized model to all service nodes. These steps ensure that all nodes collaborate and contribute to the federated learning process while preserving data privacy and security.

In the subsequent analysis, we concentrate on the local training phase of federated learning. For a given ES \emph{$s \in \mathcal{S}$}, we assume that its local dataset is represented by \emph{$D_s$}, and its model is represented by \emph{$w_s$}. During the local training process, the loss function can be denoted by \emph{$F_s(w_s)$}. The augmented update equation for the local training of ES \emph{$s$}, incorporating multiple local updates, is defined as follows:
\begin{equation}
w_s=w_s^{\prime}-\phi \sum_{i=1}^M \nabla F_s\left(w_s^{(\prime, i)}\right), \label{eq3}
\end{equation}

in this formulation, \emph{$w_s^{\prime}$} signifies the local parameters from the previous round, \emph{$M$} denotes the number of local update steps, and \emph{$\phi$} stands for the learning rate. This process entails the accumulation of gradients across multiple local updates, which are then adjusted by the learning rate \emph{$\phi$}, to update the local parameters before their transmission to the CPS.

Subsequent to a certain number of training rounds incorporating multiple local updates, the ES uploads its refined local model parameters to the CPS. The CPS undertakes the collection and aggregation of all local models utilizing the following model update formula:

\begin{equation}
w=\frac{\sum_{s \in S}\left|D_s\right| w_s}{|D|}, \label{eq4}
\end{equation}

\noindent in which, \emph{$w$} represents the weight of the model after CPS aggregation. $|D_s|$ represents the number of samples in the local dataset \emph{$D_s$}. \emph{$D$} represents the global virtual data set of CPS, and \emph{$|D|$} represents the global data sample size. Therefore, the loss function of the aggregation model can be expressed as follows:
\begin{equation}
F(w)=\frac{\sum_{s \in S}\left|D_s\right| F_s\left(w_s\right)}{|D|}, \label{eq5}
\end{equation}

\noindent among them, \emph{$F(w)$} represents the global loss function, and \emph{$F_s(w_s)$} represents the local loss function.

\subsection{Communication Model}

Our proposed federated learning framework is depicted in Fig. \ref{fig1}. This can be abstracted into a multi-layer network architecture with Cloud-Edge-Device, where a plethora of User Devices (UD) at the end layer, such as smartphones, cameras, or drone devices, are connected to the ES on the upper layer through the wireless access network \cite{ref41}. Each UD in the end layer possesses private data to be offloaded, and each node in the ES layer stores the AI model to be trained. In Fig. \ref{fig1}, the data stored by each UD is represented by cylindrical shapes, with sample category information indicated by different colors. The data samples within UD exhibit non-uniform distribution across sample categories, falling under the category of Non-IID data. The underlying device-edge computing scenario can be abstracted as Multi-Task Multi-Helper (MTMH), as discussed in \cite{ref42}. The ES will select the UD for service based on the service scope and data distribution. The UD will transfer the data to the specified ES according to the service matching result and wait for the response to be sent back. Once the ES training is completed, the model is uploaded to the CPS via the core network. CPS aggregates the models for global broadcasting. Also, it is worth noting that the framework illustrated in Fig. \ref{fig1} still embodies privacy protection features in a broader sense. Computational offloading and local model training occur within departmental or branch-level edge business scenarios. Global model aggregation takes place among departments or companies. Therefore, this framework prevents data exposure to the public, breaks down data silos, and still adheres to the original intent of FL design.

The edge layer, comprising multiple servers, is crucial for handling data offloading and AI model training without causing bottlenecks or service disruptions \cite{ref43, ref44}. To mitigate risks associated with server failures and ensure continuous learning processes, our framework utilizes a network of edge servers. This setup not only prevents any single server from becoming overloaded but also enhances system robustness by facilitating load balancing and intelligent service scheduling. Such a design ensures that the federated learning process remains efficient, reliable, and aligned with high-availability systems typically seen in modern server cluster architectures of user service-providing companies  \cite{ref45, ref46}.

Assuming that any UD \emph{$u$} needs to offload tasks to ES \emph{$s$}, the communication gain between the two can be represented by \emph{$h_{us}$}. The communication power of task upload is denoted by \emph{$p_{us}$}, and the strategy for power allocation in the uplink is represented by \emph{$P=\{p_{us}|0\leq p_{us}\leq P_{\text{max}}\}$}. Here, \emph{$P_{\text{max}}$} represents the maximum power available to UD \emph{$u$}. The total system bandwidth is denoted by \emph{$B$}, which is divided into \emph{$K$} subcarriers, each with a bandwidth of \emph{$B_k=B/K, k \in K$} \cite{ref47,ref48}. OFDMA is selected as the communication access solution for edge scenarios \cite{ref49}. In this scheme, each subcarrier is orthogonal in the frequency domain, so the transmission link is free from interference within the cell \cite{ref50}. However, interference may occur when multiple tasks occupy the same subcarrier frequency band between cells. Let \emph{$I^k$} represent the set of users occupying subcarrier \emph{$k$}. The Signal-to-Interference-Noise Ratio (\emph{SINR}) between UD \emph{$u$} and ES \emph{$s$} can be expressed as:
\begin{equation}
\operatorname{SINR}=\frac{h_{u s} p_{us}}{\sigma_u^2}, \label{eq6}
\end{equation}
\begin{equation}
\sigma_{\mathrm{u}}^2=\sigma^2+\sum_{v \in I^k \backslash\{u\}} p_v h_{v s},
\end{equation}

\noindent where \emph{$\sigma_{\mathrm{u}}$} represents the interference of \emph{$u$}. \emph{$\sigma$} represents the background noise. \emph{$\sum_{v \in I^k \backslash\{u\}} p_v h_{v s}$} represents the uplink interference between cells occupying the same subcarrier. The task upload rate from user \emph{$u$} to \emph{$s$} can be expressed as follows:
\begin{equation}
R_u=B_k \log _2\left(1+\operatorname{SINR}_u\right),
\end{equation}

\noindent where \emph{$R_u$} represents the task upload rate. \emph{$B_k$} represents the subcarrier bandwidth occupied by the task. Assuming UD \emph{$u$}’s private dataset size is \emph{$d_u$}(bit), then the time to transfer a task from \emph{$u$} to \emph{$s$} \emph{$T_u$} can be expressed as:
\begin{equation}
T_u=\frac{d_{u}}{R_u}.
\end{equation}

\section{Problem Formulation}  

Given the highly fragmented nature of edge data, the deployment of federated learning based on heterogeneous data may result in multiple rounds of parameter aggregation and reduced accuracy, thereby engendering unnecessary communication and resource wastage in resource-constrained edge scenarios\cite{ref51}. To address this challenge, we propose a two-pronged solution: (1) Analyzing the correlation between data heterogeneity and model training efficiency to identify potential means of reducing the loss function of the model from a data distribution standpoint, and (2) Modeling the communication cost to optimize communication efficiency in resource-constrained environments.

\subsection{Problem Definition}

Our objective is to enhance the training efficiency of federated learning models and minimize communication expenses in edge environments. We introduce the binary offloading policy matrix \emph{$A$}, where the entry \emph{$a_{us}$} is a binary indicator that determines whether participant \emph{$u$} transfers their data to the ES \emph{$s$}. Specifically, \emph{$a_{us}=1$} indicates that the user offloads their data to the ES \emph{$s$}, while \emph{$a_{us}=0$} implies that no action is taken. We denote the cost function of the system by \emph{$J$}, which can be mathematically formulated as follows:
\begin{equation}
J(A, P)=\sum_{s \in S} \sum_{u \in U} a_{u s} p_{u s} T_u=\sum_{s \in S} \sum_{u \in U} a_{u s} p_{u s} \frac{d_u}{R_u}.
\end{equation}

The cost function J involves computation offloading strategy \emph{$A$} and power allocation strategy \emph{$P$}. To facilitate discussion, we define \textbf{P0} as minimizing the cost function, denoted as $\min J(A, P)$. This can be expressed in the following form:
\begin{align}
\text { \textbf{P0}: } \min_{a_{us}, p_u} & J(A, P), \label{eq11}\\
\text { s.t. } & a_{u s} \in\{0,1\}, u \in U, s \in S, \label{eq12}\\
& \sum_{s \in S} a_{u s}=1, u \in U, s \in S, \label{eq13}\\
& p_u>0, \quad u \in U, \label{eq14}\\
& p_u<P_{\text{max}}, \quad u \in U. \label{eq15}
\end{align}

Wherein, \emph{$P_{\text{max}}$} represents the maximum transmission power of device \emph{$u$}.

The first item of the restriction expresses the value rule of \emph{$a_{u s}$}. The latter two items indicate that the transmission power allocated by user u to s should be within the range of the maximum power. Solving \textbf{P0} is challenging. Firstly, the objective function of \textbf{P0} constitutes a non-convex function. Within this function, there exists a fractional denominator \emph{$\log_2\left(1+\frac{h_{u s} p_u}{\sigma_u^2}\right)$}, which includes \emph{$p_u$}. While the internal components of this denominator the \emph{$\log$} function and the fraction \emph{$\frac{h_{u s} p_u}{\sigma_u^2}$} are both convex functions, the entire denominator remains non-convex. This is attributed to the nature of the \emph{$\log$} function, altering the convex nature of the entire denominator as the fraction within it varies, resulting in an overall non-convex objective function.

Secondly, the constraints further compound the complexity of the problem. Particularly, the constraint involving the variable \emph{$a_{u s}$} being a binary variable, that is, \emph{$a_{u s} \in \{0,1\}$}. Such binary constraints often make the problem more challenging to solve as they confine the search space to discrete values. Hence, \textbf{P0} is a continuous-discrete mixed problem. Additionally, the offloading strategy directly determines the data distribution on ESs, indirectly impacting model training accuracy. In large-scale edge networks, the prevalence of imbalanced and Non-IID data poses a common challenge. Designing offloading strategies effectively to enhance data quality becomes a significant hurdle. Due to \textbf{P0}'s optimization problem being a mix of continuous-discrete elements and non-convex in nature, directly solving the overall problem presents considerable difficulty.

\subsection{Theoretical Analysis}

In the previous section of Problem Definition, our aim is to enhance the efficiency of federated learning model training while reducing communication costs in edge environments. One of the key focal points in the analysis of the proposed solutions is the impact of Non-IID data on the convergence of federated learning models. It is worth noting that the optimization variable \emph{$a_{us}$} in \textbf{P0} determines the offloading decision, which are intrinsically linked to the efficiency of FL training. Effective offloading decisions can enhance data distribution and consequently improve the convergence rates of FL training processes. Before attempting to find a solution for \textbf{P0}, it is imperative to delve into understanding how Non-IID data influences the convergence of models. Non-IID data refers to a scenario in federated learning where participants possess locally distributed and possibly non-uniform data, potentially resulting in inconsistent training effects across participants, thus affecting the overall convergence speed and accuracy of the model. Hence, our subsequent analysis will concentrate on delineating the impact of Non-IID data on the convergence of federated learning models. As a foreshadowing, we first define some symbols and principles.

To facilitate our analysis, we consider a scenario where each federated learning participant has the same amount of data and the data distribution is identical across all participants, i.e., the ideal data distribution is an IID scenario. Under this assumption, we use \emph{$v_s$} and \emph{$v$} to denote the model parameters of ES and CPS, respectively. The corresponding parameter update formula can be expressed as follows:
\begin{equation}
v_s=v_s^{\prime}-\phi \nabla F_s\left(v_s^{\prime}\right),
\end{equation}

\noindent for the convenience of comparison, we assume that the model \emph{$w_s$} based on Non-IID data is trained synchronously with the model \emph{$v_s$} based on IID data. We will trace the difference when the two models are updated, but before referring to the work in \cite{ref52}, we give the following assumptions:

\begin{myAssumption}

For the model \emph{$w_s$} on any ES, its loss \emph{$F_s\left(w_s\right)$} is convex. And \emph{$F_s\left(w_s\right)$} is \emph{$L_s$} Lipschitz smooth, that is, the following formula is established:
\begin{equation}
\left\|\nabla F_s\left(w_s\right)-\nabla F_s\left(w_s^{\prime}\right)\right\| \leqslant L_s\left\|w_s-w_s^{\prime}\right\|,
\end{equation}

\end{myAssumption}

\noindent among them, the local model parameters of the current and next round are denoted as \emph{$w_s$} and \emph{$w_s^{'}$}, respectively. Traditional SVMs or neural networks with linear activation functions and square loss have convex properties. In fact, our proposed algorithm has demonstrated its effectiveness even on non-convex neural networks, as evidenced in the experimental results presented in Section \uppercase\expandafter{\romannumeral6}. Furthermore, we extend this property to the aggregated model, whereby the following formula holds:

\begin{myAssumption}

In any weight aggregation cycle, for the global model \emph{$w$} on CPS, its loss \emph{$F\left(w_s\right)$} is convex. And \emph{$F\left(w\right)$} is \emph{$L$} Lipschitz smooth, that is, the following formula is established:
\begin{equation}
\left\|\nabla F\left(w\right)-\nabla F\left(w^{\prime}\right)\right\| \leqslant L\left\|w-w^{\prime}\right\|.
\end{equation}

\end{myAssumption}

Assumption 2 can be derived from Assumption 1 and the triangle inequality, as demonstrated in Appendix A. It is worth noting that the model gradient \emph{$\nabla F_s\left(w_s\right)$} in Assumption 1 is sensitive to the distribution of training samples. To facilitate analysis, we define the following terms:

\begin{myDef}

We use \emph{$\gamma$} to describe the difference in model update gradients. For example, \emph{$\gamma_s$} represents the gradient distance to \emph{$F_s(w)$} and \emph{$F(w)$}.
\begin{equation}
\left\|\nabla F_s(w)- \nabla F(w)\right\|=\gamma_s.
\end{equation}

\end{myDef}

To investigate the impact of data heterogeneity on federated learning, we first defined the potential impact of data distribution through Definition 1. For instance, the variation in the ratio of each category in the dataset will lead to differences in the direction of model update. Then, we will explore global aggregation model updates, specifically focusing on Non-IID and IID data. To probe the convergence efficiency of the two models, we performed an analysis of \emph{$||w-v||$}\cite{ref52}. Utilizing the definition of \emph{$\gamma_s$}, Assumption 1 and 2 were derived, leading to the following theorem:

\begin{myTheo}

\emph{$w$} and \emph{$v$} are model parameters trained based on Non-IID and IID data, respectively. Assume that in the weight aggregation period t (\emph{$t \textgreater 1$}), the update frequencies of \emph{$w$} and \emph{$v$} are synchronized. The following formula is established:
\begin{equation}
\left\|w-v\right\|\leqslant\left\|w^{\prime}-v^{\prime}\right\|+\frac{\phi \sum_{s \in S} D_s \gamma_s\left(\phi L_s+1\right)^t}{D},
\end{equation}

\end{myTheo}

\noindent among them, \emph{$w^{'}$} and \emph{$v^{'}$} represent the model parameters of the previous cycle respectively. \emph{$\phi$} and \emph{$L_s$} represent the learning rate and Lipschitz smooth gradient, respectively. Theorem 1 addresses the evolution of the discrepancies between model parameters trained under Non-IID conditions (\emph{$w$}) and IID conditions (\emph{$v$}) over successive aggregation periods. The theorem quantifies how the distance between these parameters develops as a function of the aggregation cycles (\emph{$t$}). Refer to Appendix B for the proof of Theorem 1.

It is readily apparent that the expression \emph{$\frac{\phi \sum_{s \in S} D_s \gamma_s\left(\phi L_s+1\right)^t}{D} > 0$} is strictly positive. As such, it follows that as the aggregation period \emph{$t$} increases, the distance between the local model parameters \emph{$w$} and the aggregated model parameters \emph{$v$} will become increasingly pronounced. It is worth noting the position of the data heterogeneity coefficient \emph{$\gamma_s$} in Theorem 1, which is a crucial determinant of the impact of heterogeneous data on \emph{$w$}. Specifically, the damage incurred by \emph{$w$} is found to be a contributing factor in the degree of heterogeneity in the data distribution, as manifested by the value of \emph{$\gamma_s$}. When \emph{$w_s^*$} is trained on i.i.d. data, \emph{$\left\|\nabla F_s(w_s^*)- \nabla F_s(v_s)\right\|=\gamma_s^*=0$}, implying that \emph{$\frac{\phi \sum_{s \in S} D_s \gamma_s^*\left(\phi L_s+1\right)^t}{D} = 0$}. Conversely, when the heterogeneity of the data distribution increases, \emph{$\gamma_s$} increases and hinders the convergence rate of the federated learning algorithm.

We have thoroughly analyzed the impact of Non-IID data on the convergence rate and accuracy of the overall model. Mathematically, we have demonstrated the significant influence of data heterogeneity on the model's training. As the heterogeneity in data distribution increases, it leads to an increase in \emph{$\gamma_s$}, thereby impairing the convergence rate of federated learning.

\subsection{Algorithmic Insights}

\begin{figure}[ht]
\centerline{\includegraphics[scale=0.3]{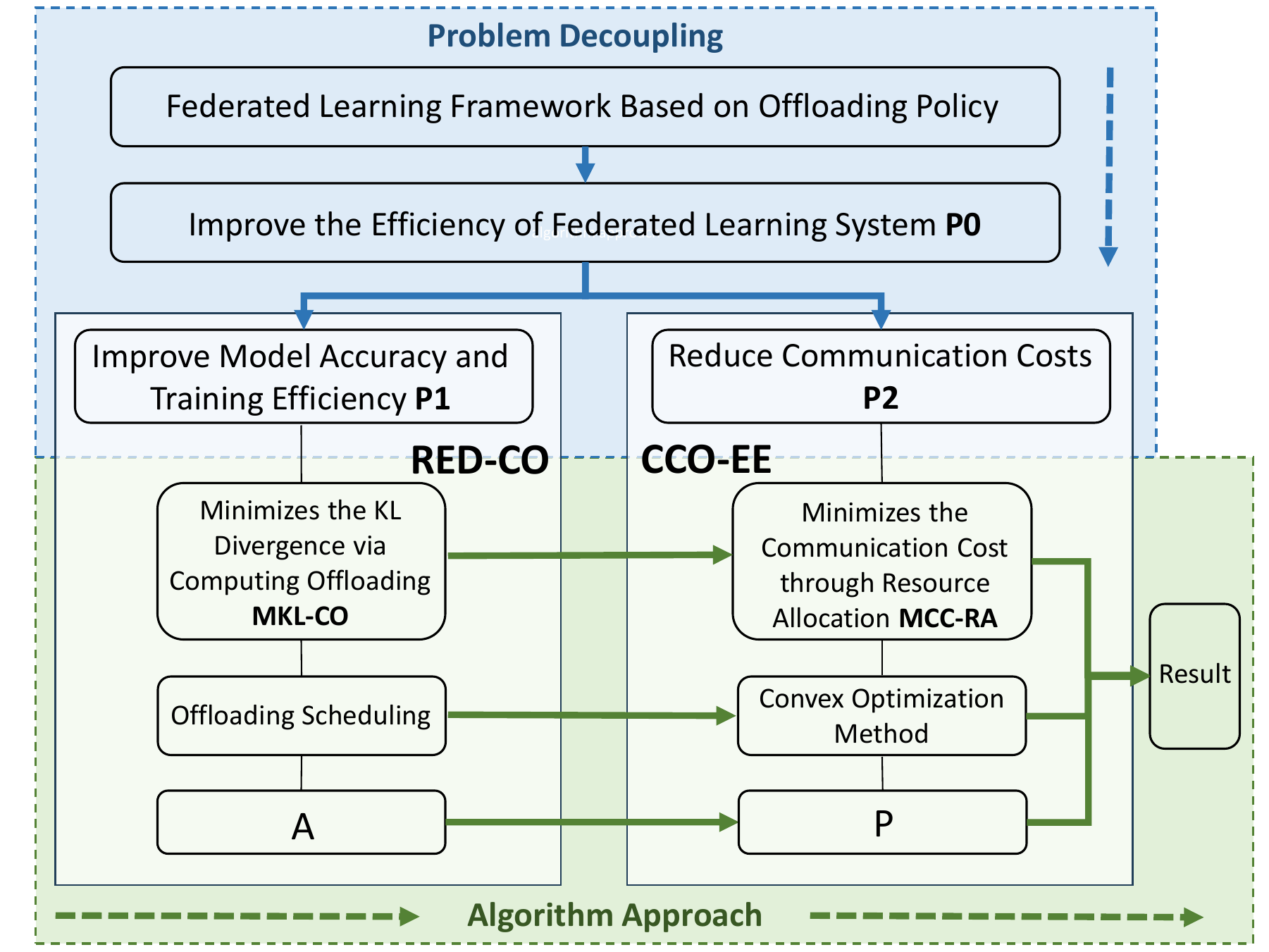}}
\caption{Process of FlocOff algorithm based on computation offloading scheduling. }
\label{fig4}
\end{figure}

The primary aim of this paper is to leverage computation offloading to mitigate data heterogeneity, thereby enhancing the convergence and accuracy of federated learning. Transitioning from problem analysis to the primary objective of this paper, it is evident that managing data heterogeneity is a pivotal factor in improving the performance of federated learning. This understanding forms the foundation for the development of an effective algorithmic framework. From the problem analysis, we can derive several key observations that aid in the design of the algorithmic framework:

\begin{itemize}

\item[1)]

We recognize that data heterogeneity results in the increase of certain parameters, consequently affecting the efficacy of federated learning. Therefore, when devising the algorithmic framework, it is essential to purposefully address how to mitigate the adverse effects of data heterogeneity.
\end{itemize}

\begin{itemize}

\item[2)]

We acknowledge that computation offloading techniques can ameliorate data heterogeneity, allowing the algorithmic framework to focus on optimizing the utilization of computation offloading in federated learning applications.
\end{itemize}

The significance of this issue lies in comprehending the challenge that data heterogeneity poses to federated learning. Consequently, we have uncovered a crucial insight: by employing computation offloading techniques to reduce data heterogeneity, we anticipate an enhancement in the convergence rate and accuracy of federated learning. In essence, the algorithmic framework we propose aims to mitigate the challenges posed by data heterogeneity through computation offloading and subsequently optimize the effectiveness of federated learning.

So, it is necessary for us to introduce computation offloading into the framework and simplify the overall problem. Nevertheless, due to the complexity of \textbf{P0}, which is a challenging non-convex problem, solving it as a whole poses significant difficulties. However, upon careful observation of the variables and their respective implications within the equations, we note that \textbf{P0} encompasses two pivotal aspects of federated learning: the computing facet and communication optimization. These facets hold distinct optimization objectives and constraints. Furthermore, it becomes apparent that decoupling \textbf{P0} into two independent sub-problems enables efficient resolution.

Firstly, in the objective function of \textbf{P0}, each term is associated with the computation offloading strategy \emph{$a_{us}$} and the power allocation strategy \emph{$p_u$}. Yet, these strategies exist as products within the objective function without explicit coupling. Additionally, the constraints of \textbf{P0} can be divided into two parts: one pertaining to the constraints of \emph{$a_{us}$} (Eq. (\ref{eq12}), Eq. (\ref{eq13})), and the other concerning the constraints of \emph{$p_u$} (Eq. (\ref{eq14}), Eq. (\ref{eq15})). These constraints are separable across the two sub-problems, allowing each sub-problem to address its specific set of constraints independently.

Drawing from these insights, as depicted in Fig. \ref{fig4}, we decouple the original problem \textbf{P0} into two distinct sub-problems: (1) Reshaping the Edge Dataset via Computation Offloading (\textbf{RED-CO}), denoted as \textbf{P1}; and (2) Communication Cost Optimization in Edge Environments (\textbf{CCO-EE}), referred to as \textbf{P2}. 

As shown in Fig. \ref{fig4}, regarding \textbf{P1}, we propose the Minimizes the KL Divergence via Computation Offloading scheduling (\textbf{MKL-CO}). \textbf{MKL-CO} focuses on reshaping edge datasets through computation offloading, aiming to enhance training efficiency and model accuracy. This is closely related to the term associated with \emph{$a_{us}$} in the objective function. The variable \emph{$a_{us}$} represents the strategy for computation offloading, determining which computing tasks should be executed on edge devices. Concerning \textbf{P2}, we introduce the Minimizes the Communication Cost through Resource Allocation algorithm (\textbf{MCC-RA}). \textbf{MCC-RA} aims to optimize communication power allocation strategies to reduce communication costs and enhance transmission efficiency. This is closely related to the term associated with \emph{$p_u$} in the objective function. The variable \emph{$p_u$} represents the allocation strategy of communication power, utilized for optimizing communication efficiency while considering channel attenuation and the impact of noise.

Moreover, after decoupling the original problem \textbf{P0} into two sub-problems \textbf{MKL-CO} and \textbf{MCC-RA}, the solutions obtained from these sub-problems yield either the optimal or sub-optimal solutions to the original problem. A detailed proof of this can be found in Appendix C.

\section{FlocOff Algorithm Design}

\subsection{Framework Overview}

The workflow within the FlocOff framework involves several critical network elements, including user devices, edge servers, and the collaborative utilization of computation and communication resources. Below, we will provide a detailed explanation of the workflow for each network element and offer some practical insights:

\begin{itemize}
 
  \item \textbf{User Device}: \newline
  \emph{Data Distribution Analysis}: Initially, user devices analyze their own data distribution, encompassing their local dataset $D_u$. UDs employ methods like KL divergence to quantify the similarity or dissimilarity of their data distribution with that of the ES. This aids UD in assessing the extent of dissimilarity between their data and the ES. \newline
  \emph{Offloading Decision}: Based on the results of data distribution analysis, UD decides whether to offload a portion of the data to ES for federated learning. The offloading decision is driven by the aim of maximizing federated learning efficiency and minimizing communication overhead in resource-constrained edge environments. \newline
  \emph{Communication}: Once the offloading decision is made, UD is responsible for transferring the relevant data to ES for use in federated learning. This may involve the selection of data transfer protocols, as well as data compression and encryption.
  
  \item \textbf{Edge Server}: \newline
  \emph{Data Reception and Storage}: ES receives offloaded data from UD and stores it locally. This data is used in conjunction with ES's local data for training the federated learning model. \newline
  \emph{User Device Selection}: ES employs an offloading algorithm to select suitable UD, minimizing KL divergence and improving data distribution similarity to the greatest extent possible. This aids in choosing UD that are particularly useful for model training. \newline
  \emph{Communication Power Allocation}: ES is responsible for communication power allocation in the collaborative process to ensure the effective transmission of offloaded data. This may involve addressing interference and limited communication resources to minimize communication costs.
  \item \textbf{Resource Sharing and Collaboration}: \newline
  \emph{Data Aggregation and Model Updates}: After receiving data from different UD, ES aggregates the data into a global model for model updates and training. This enables all UD to collectively contribute data, enhancing model performance. \newline
  \emph{Resource Management and Performance Optimization}: Efficient sharing of computing and communication resources is essential among network elements. This may entail optimizing computation offloading decisions and communication power allocation in resource-constrained edge environments to maximize system efficiency.

\end{itemize}

The operational mechanism of the FlocOff framework holds significant potential for real-world applications. For instance, in edge computing environments, mobile devices can act as UD, uploading local data to edge servers to enhance model training. This has promising applications in devices like smartphones, sensor networks, and IoT devices, such as improving personalized recommendations for mobile apps or optimizing performance in smart home systems. Furthermore, by optimizing the collaborative utilization of communication and computing resources, the FlocOff framework is suitable for a variety of edge scenarios, including smart cities and intelligent transportation systems, aimed at improving the efficiency of data analysis and model training.

In this section, we first solve the computation offloading decision \emph{$A$}, that is, service mapping between ES and UD. Then, we optimize the communication power allocation strategy.

\subsection{Computation Offloading Scheduling Algorithm}

KL divergence is a crucial metric in information theory for measuring the difference or information loss between two probability distributions. It is commonly employed to assess the disparity in uncertainty of one probability distribution relative to another. In academic terms, KL divergence \emph{$D_{KL}(P|Q)$} is utilized to compare the dissimilarity between probability distributions P and Q, where P represents the reference distribution, and Q represents the target distribution. Here, we employ KL divergence to quantitatively compare the differences in data set distributions.

In this paper, we define the private data distribution of UD \emph{$u$} to be denoted by \emph{$D_u$}. For ease of subsequent description, we first define \emph{$c \in C$} to represent data categories. Secondly, we define \emph{$P_u(c)$} to denote the proportion of samples of the category \emph{c} in the data set \emph{$D_u$}. Specifically, the computation of \emph{$P_u(c)$} is as follows:
\begin{equation}
P_u(c)=\frac{\sum_{u \in U} a_{u s} n_c}{\left|D_u\right|},
\end{equation}

\noindent in which, \emph{$n_c$} represents the number of samples in class \emph{c}. To quantify the dissimilarity between different data distributions, this paper introduces KL divergence. Our goal is to make the offloaded data distribution as close as possible to the global data distribution. We represent the distribution dissimilarity between User Device \emph{$u$} and Edge Server \emph{$s$} as \emph{$KL\left(D_u \mid D_s\right)$}. Therefore, the final problem can be formulated as stated in \cite{ref53}:
\begin{align}
\textbf{P1}: \min_{a_{us}} & KL\left(D_u \mid D_s\right) = \sum_{c \in C} P_u(c) \log \frac{P_u(c)}{P_s(c)}. \label{eq22} \\
\text{s.t.} & \quad a_{us} \in \{0,1\}, u \in U, s \in S, \notag \\
 & \sum_{s \in S} a_{us} = 1, u \in U. \notag
\end{align}

Therefore, the smaller the KL divergence in Eq. (\ref{eq22}), the closer the data distribution between \emph{$u$} and \emph{$s$} is. Reducing KL divergence is equivalent to minimizing $\gamma_s$ as in Eq.(19). Please refer to Appendix D for detailed proof.

Next, we present an offloading algorithm designed to facilitate computation offloading and service association between UD and ES. The algorithm, denoted as Algorithm 1, is characterized by several key steps. To begin with, we initialize the global data distribution \emph{$D$}. Ideally, the global data is more comprehensive and includes samples from every category. Thus, in the initialization phase of the global data distribution \emph{$D$}, we ensure that the total number of samples is evenly divided among each category. Also, we initialize the Data Volume Threshold \emph{$\Gamma$}, a variable that serves as an upper limit for the number of samples during ES training.

Next, we determine the serviceable UD set \emph{$U_s$} for each \emph{$s \in S$} and carry out the following operations: (1) Compute the complement set \emph{$\bar{D}_s$} of the current remaining data \emph{$D_s$} with respect to the global data \emph{$D$}. (2) Traverse the set of UD \emph{$u \in U_s$} to compute the KL divergence between \emph{$D_u$} and \emph{$\bar{D}_s$}, which is denoted as \emph{$K L\left(D_u \mid \bar{D}_s\right)$}. (3) Identify the UD with the smallest KL divergence, which is marked as \emph{$u^{*}$}. (4) Solve the communication power allocation strategy \emph{$P$}. The specific method for solving \emph{$P$} is detailed in the subsequent subsections. (5) Transfer all data belonging to \emph{$u^{*}$} to \emph{$s$}. (6) Verify if the current data size of \emph{$D_s$} exceeds \emph{$\Gamma$}. If this is the case, terminate the unloading and initiate local training. Otherwise, loop back to steps (1)-(5) and continue the algorithm.

\begin{algorithm}[ht] 
\caption{The workflow of MKL-CO algorithm.} 
\label{alg:Framwork} 
\begin{algorithmic}[1] 
\STATE Initialize global data distribution \emph{$D$}.
\STATE Monitoring the data distribution of UD and ES as \emph{$D_u$} and \emph{$D_s$}, \emph{$u \in U, s \in S$}.
\STATE Initialize Data Volume Threshold \emph{$\Gamma$}.
\FOR{\emph{$s \in S$}}
    \FOR{episode number t in {1,2,...T}}
        \STATE Compute the complement of the distribution \emph{$D_s$}: \emph{$\bar{D}_s=D-D_s, s \in S$}.
        \STATE Determine the service scope of \emph{$s$}, and the UD set within the scope is expressed as \emph{$U_s$}.
        
        \FOR{\emph{$u \in U_s$}}
            \STATE calculate \emph{$K L\left(D_u \mid \bar{D}_s\right)$}.
            \STATE Find the largest index expressed as \emph{$u^*$}.
            \STATE Solve the communication power allocation strategy \emph{$P$}.
            \STATE Offload all data of \emph{$D_{u^*}$} to \emph{$D_s$}.
            \STATE Record the current data size of \emph{$D_s$}, expressed as \emph{$\Gamma_s$}.
            \IF{\emph{$\Gamma_s \leqslant \Gamma$}}
                \STATE \textbf{Continue}.
            \ELSE
                \STATE ES \emph{$s$} offload complete. 
                \STATE \textbf{Break}.
            \ENDIF
        \ENDFOR
        
    \ENDFOR
    \STATE Train a local model \emph{$w_s$} based on \emph{$D_s$} on \emph{$s$}.
    \STATE Upload \emph{$w_s$} to CPS for model aggregation.
\ENDFOR
\end{algorithmic}
\end{algorithm}

In Algorithm 1, we encounter two nested loops. The outer loop traverses all service nodes in \emph{$S$}, while the inner loop iterates through all time points processed by each service node, denoted as \emph{$T$}. The overall complexity of these loops will be \emph{$O(ST)$}. Within the inner loop, we perform computations involving calculating the complement of specific distributions, determining the scope of service nodes, and computing the KL divergence for each UD. The total complexity of these steps is \emph{$O(US)$}. Lastly, identifying the UD with the maximum KL divergence requires \emph{$O(U)$} time. Therefore, the complete algorithm exhibits a time complexity of \emph{$O(U^2D)$}, where \emph{$U$} represents the number of user devices, and D stands for the dataset's size. This implies that the time complexity will increase quadratically with the number of user devices and linearly with the dataset size.

\subsection{Efficient Numerical Power Allocation Strategy}

Next, we introduce the communication power allocation strategy. Since Algorithm 1 provides an offload map \emph{$A$}, the original \textbf{P0} problem is transformed as follows:

\begin{align}
\textbf{P2} : \Lambda(P) = \min_{p_u} & \sum_{s \in S} \sum_{u \in U} a_{u s} p_u \frac{d_u}{B_u \log _2\left(1+\frac{h_{u s} p_u}{\sigma_u^2}\right)}, \label{eq23} \\
\text{s.t.} \quad p_u &> 0, \quad u \in U, \notag \\
p_u &< P_{\text{max}}, \quad u \in U. \notag
\end{align}

Eq. (\ref{eq23}) is non-convex with respect to \emph{$p_{us}$}, given that \emph{$\sigma_u$} encompasses interference resulting from transmissions that occupy the same subcarrier \emph{$k$} as \emph{$u$}. To simplify \emph{$\sigma_u$}, we adopt the parameter substitution approach proposed in \cite{ref54}. Specifically, we substitute the rated power \emph{$P_v$} for \emph{$p_v$}, and replace the original interference term with \emph{$\hat{\sigma}_u=\sigma^2+\sum{v \in I^k \backslash{u}} P_v h_{v s}$}. We then introduce \emph{$\hat{\Lambda}$} to denote the objective function in Eq. (\ref{eq23}). Subsequently, we demonstrate that \emph{$\hat{\Lambda}$} is quasi-convex concerning \emph{$p_{us}$}. The proof is detailed in Appendix E. Finally, we employ the bisection method \cite{ref55} to solve for the optimal \emph{$p_{us}$}.

It is easy to know that \emph{$q_1(p_{u s})=a_{u s} d_u p_{u s}$} is a convex function, \emph{$B_k \log _2\left(1+\frac{h_{u s} p_{u s}}{\hat {\sigma_u}^2}\right)$} is a concave function. If \emph{$\hat{\sigma_u}^2$} is quasi-convex, then there exists a series of \emph{$\phi_t(x)$} such that the following conditions are equivalent: (1)\emph{$\bar{\Lambda}(p_{us} ) \leqslant t$}, (2)\emph{$\phi_t(p_{us}) \leqslant 0$}.

Let \emph{$\phi_t(p_{us})=a_{u s} d_u p_{u s}-B_k \log _2\left(1+\frac{h_{u s} p_{u s}}{\hat{\sigma_u}^ 2}\right)$}. On the domain of \emph{$p_{us}$}, \emph{$\phi_t(p_{us})$} is a convex function. Then \emph{$\bar{\Lambda}(p_{us}) \leqslant t$} if and only if \emph{$\phi_t(p_{us}) \leqslant 0$}. Therefore for fixed \emph{$t$}, Eq. (\ref{eq23}) translates into the following convex feasibility problem:
\begin{equation}
\begin{aligned}
& \phi_t(p_{us})  \leq 0,  \quad u \in U,  s \in S, \\
& 0<p_{u s}<P, \quad u \in U, s \in S.
\end{aligned}
\end{equation}

\begin{algorithm}[ht] 
\caption{The workflow of MCC-RA algorithm.} 
\label{alg:Framwork} 
\begin{algorithmic}[1] 
\STATE Initialize tolerance \emph{$\tau > 0$}.
\STATE Initialize \emph{$l<p^*$}, \emph{$r=\bar{\Lambda}(p_{us})$}.
\WHILE{\emph{$r-l>\tau$}}
    \STATE Set \emph{$m=(l+r)/2$}
    \STATE Solve convex feasibility problem \emph{$\phi_t(p_{us})$}, \emph{$0<p_{u s}<P$}
        \IF{Feasible (have \emph{$p_{us}$} satisfies the constraints)}
            \STATE Set \emph{$r = m$}
            \STATE \emph{$p_{us}$} can be any solution of the feasible problem
        \ELSE
            \STATE Set \emph{$l = m$}
        \ENDIF
\ENDWHILE
\end{algorithmic}
\end{algorithm}

In Algorithm 2, \emph{$p^{\star}=\inf \left\{\bar{\Lambda}(p_{us}) \mid 0<p_{u s}<P, \quad u \in U, s \in S\right\}$}. We use \emph{$r$} and \emph{$l$} to denote the upper and lower bounds of the algorithm. Iterates using the bisection method and solves the convex feasibility problem.

The time complexity of this algorithm consists mainly of two components: the number of iterations in binary search and the complexity of solving the convex feasibility problem within each iteration. Firstly, as this involves a binary search, the number of iterations depends on the initial search range (\emph{$r-l$}) and the allowed error range \emph{$\tau$}. In the worst-case scenario, it requires \emph{$O(\log\left(\frac{r-l}{\tau}\right))$} iterations. Next, for the constrained convex feasibility problem within each iteration, where \emph{$p_{us}$} represents the variable to be solved, with \emph{$u \in U$} and \emph{$s \in S$}, the dimensionality of this problem is \emph{$S \times U$}. This is because for each pair \emph{$(u, s)$}, we need to solve for a \emph{$p_{us}$}. For such problems, interior-point methods are commonly employed for resolution. In the worst-case scenario, the time complexity of interior-point methods can reach \emph{$O((S \times U)^{3.5})$}. Therefore, the overall time complexity of the algorithm becomes \emph{$O((S \times U)^{3.5} \log\left(\frac{r-l}{\tau}\right))$}.

\section{Performance Evaluation}

In this section, we present a comprehensive evaluation of the proposed framework, aiming to assess its effectiveness on data distribution and federated learning performance. To this end, we conduct a series of experiments using the PyTorch platform, with an Intel Core i5-1135G7 2.40GHz CPU and 16 GB memory. The experimental setup includes several benchmarks and datasets. The results are statistically analyzed and presented in the following sections.

\subsection{Experimental Settings}

In this section, we perform a series of experiments to assess the impact of the proposed framework on data distribution and federated learning performance. The communication environment is modeled as a multi-cell system where users' geographic locations are generated randomly, and servers are located at the center of gravity of hexagonal cells. The communication power of each server is set to 0.5-1.0W. Unless specified otherwise, we consider 10 servers and 1000 users. Each server serves 100 users, and users are sliced based on their location, ensuring that they are only sent to servers in their service area. For wireless access, we create stochastic networks through channel gain, and the background noise power is set to -100dBm. We set the number of OFDMA subbands to 128 and the bandwidth to 5MHz.

\emph{Data Set Selection} In terms of selecting data sets, our research focused on visual classification tasks, and we chose two datasets of moderate size: the MNIST handwritten digit recognition and the CIFAR-10 universal object recognition. The \textbf{MNIST dataset} comprises a total of 70,000 images, consisting of 60,000 training images and 10,000 test images. Each grayscale image in the dataset is a 28 by 28-pixel matrix of handwritten digits ranging from 0 to 9\cite{ref56}. The dataset provides labels for 10 classes, which makes it suitable for supervised learning classification tasks. On the other hand, the \textbf{CIFAR-10 dataset} is a smaller dataset, consisting of 10 classes of RGB color images. Each sample in the dataset has three channels and a size of 32 by 32. The images contain common objects such as birds, cats, airplanes, cars, and deer, among others. The CIFAR-10 dataset contains 50,000 training images and 10,000 test images. Compared to the MNIST dataset, the CIFAR-10 dataset contains real-world objects that exhibit greater diversity and noise, making it more challenging for recognition.

In terms of model implementation, we initially implemented a deep neural network model featuring two hidden layers, each with 200 neurons activated by the Rectified Linear Unit (ReLU) activation function. We implemented the ResNet-18 model on the CIFAR-10 dataset \cite{ref57}. The ResNet-18 model is a deep convolutional neural network comprising 18 layers, which incorporates residual blocks each containing two 3x3 convolutional layers and employs a "skip connection" mechanism. To simulate Non-IID distribution, we had to allocate data distribution on the client-side. In this regard, we referred to the dataset generation method in literature \cite{ref58} to slice and group experimental samples. Firstly, in real-world scenarios, classes often appear in pairs. To replicate this phenomenon, we grouped samples from two classes into one group for both the MNIST and CIFAR-10 datasets, each of which consists of 10 classes divided into 5 groups. Secondly, in certain scenarios, samples from some classes may appear more frequently than others. To account for this, we split the samples and assigned one random group as the high probability occurrence group. The remaining four groups were considered as low probability occurrence groups with samples generated with varying distributions. This approach offers flexibility in generating sample distributions that can be tailored to realistic scenarios. The extent of Non-IID data is adjusted by altering the parameters of the generated distribution. By default, unless set otherwise by subsequent experiments, each user's private data is characterized by the following distribution: two classes were randomly designated as high probability classes, following a Gaussian distribution of X$\sim$N(50,20), while the remaining eight classes were categorized as low-probability classes and followed a Gaussian distribution of X$\sim$N(10,2).

From an integrated perspective, enhancing the overall efficiency and Quality of Service (QoS) in federated learning is a significant task \cite{ref59}. The QoS is a comprehensive indicator incorporating multiple key factors. In federated learning, we typically decompose QoS into two components: training accuracy and communication costs. Training accuracy directly determines the quality of the model, serving as a crucial criterion for measuring QoS. Meanwhile, communication costs are pivotal in influencing the overall efficiency of federated learning. In a distributed federated learning system, nodes need to exchange and share data through the network. If communication costs are excessively high, they can significantly impede the overall learning efficiency.

\subsection{Impact of server threshold on computation offloading scheduling.}

\begin{figure*}[ht]
\centering
\subfigure[Performance when threshold \emph{$\Gamma=500$}.]
{
        \centering
        \includegraphics[scale=0.263]{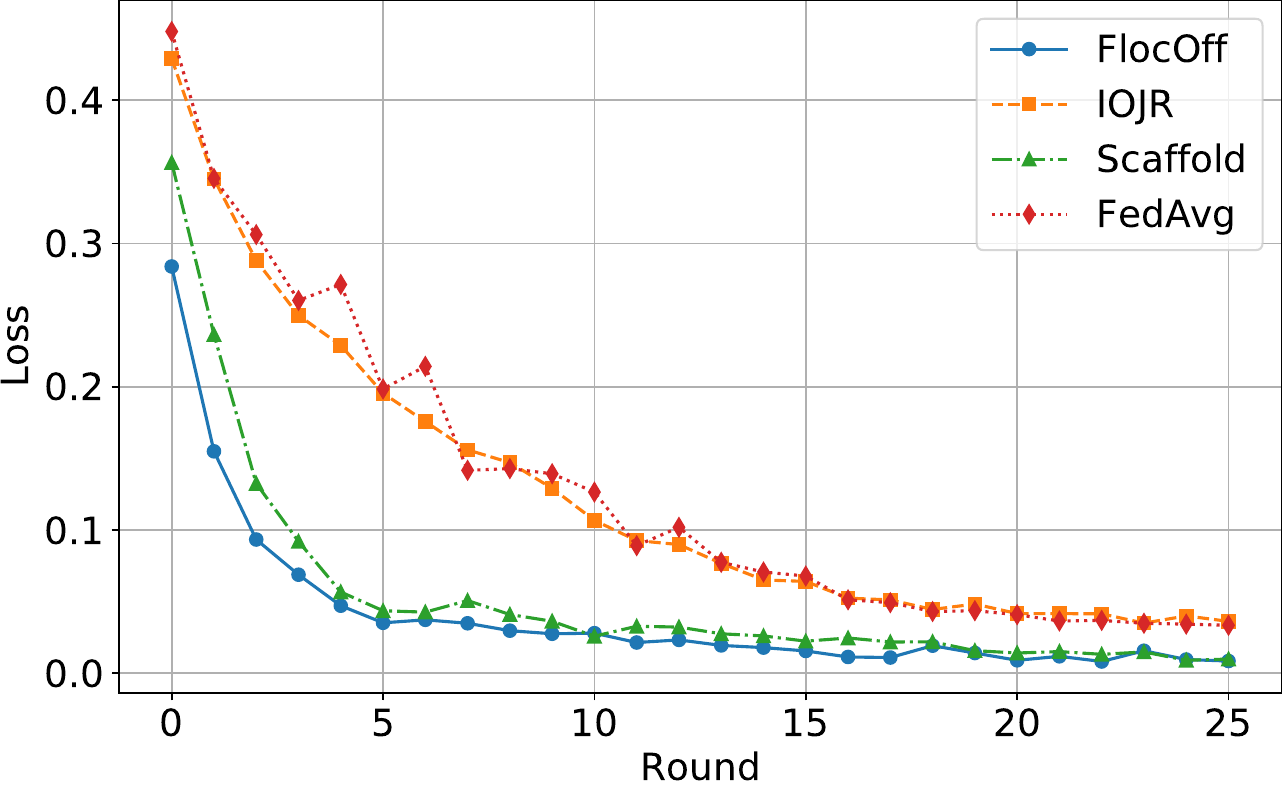}
}
\subfigure[Performance when threshold \emph{$\Gamma=1000$}.]
{
        \centering
        \includegraphics[scale=0.263]{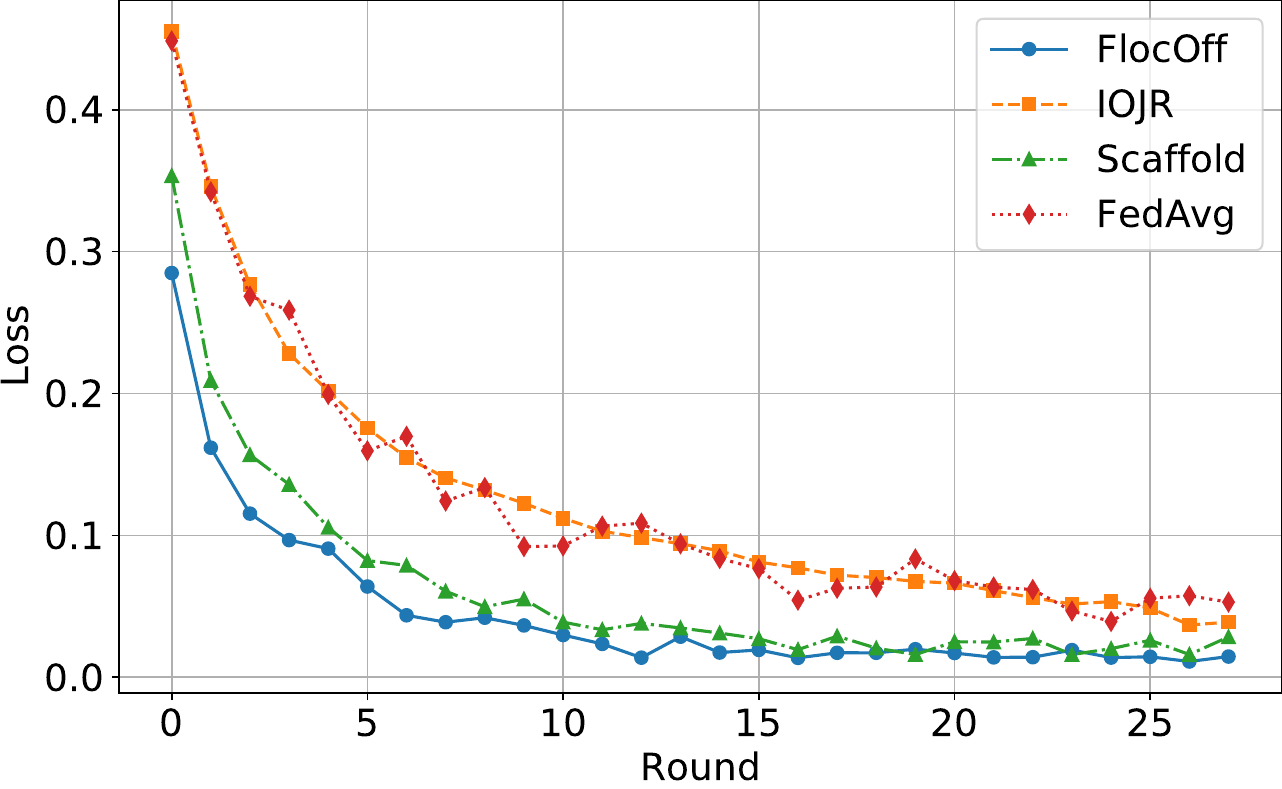}
}
\subfigure[Performance when threshold \emph{$\Gamma=1500$}.]
{
        \centering
        \includegraphics[scale=0.263]{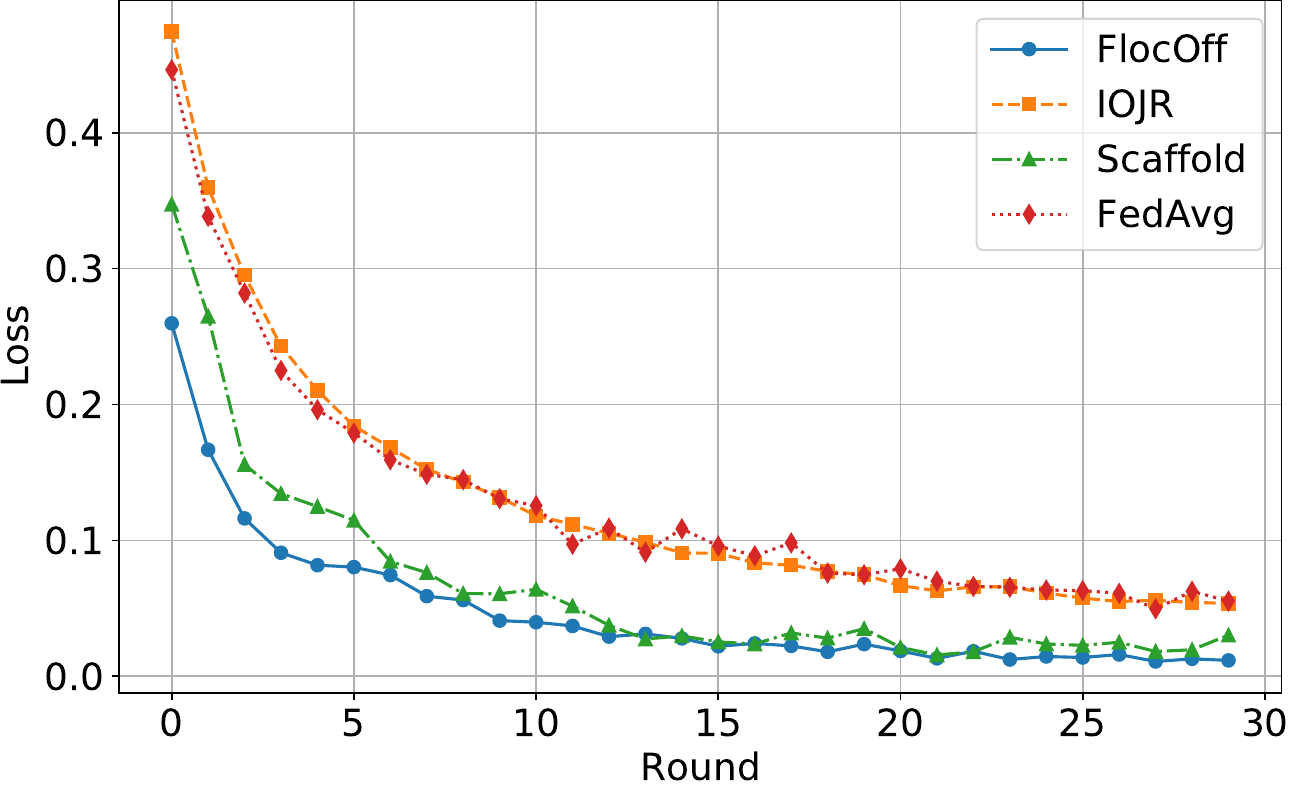}
}
\caption{Impact of Server threshold.}
\label{fig10}
\end{figure*}

Our initial investigation focused on the impact of server data collection thresholds in federated learning. Due to the rich applications of edge scenarios, the data volume is close to the business scenarios. We employed varying offloading thresholds to adjust the training cycle, tailored to both task sparsity and task-intensive scenarios. Specifically, we tested three thresholds, namely \emph{$\Gamma=500,1000,1500$}. We set the baselines: 

\begin{itemize}

  \item \emph{Federated Learning based on Computing Offloading} (\textbf{FlocOff}): Our method, which dynamically adjusting offloading decisions based on the distribution of samples to optimize Non-IID at the data level. Concurrently, it utilizes numerical algorithms to optimize power allocation strategies, thereby reducing system costs.

  \item \emph{Federated Averaging Algorithm} (\textbf{FedAvg}): Classic federated averaging algorithm, designed to handle Non-IID data effectively in a decentralized learning setting. \cite{ref60}

  \item \emph{FL-based Independent Offloading and Joint Resource Allocation} (\textbf{IOJR}): Each task is greedily assigned to the nearest ESs and allocates communication resources to it, regardless of data distribution. 

  \item \emph{Stochastic Controlled Averaging Federated Learning} (\textbf{Scaffold}): SCAFFOLD estimate update biases between each client and server model, thereby mitigating client-drift inherent in local updates, which enables effective handling of highly Non-IID data distributions and accelerates the overall convergence rate. \cite{ref61}
\end{itemize}

The graphical representation of the impact of the offloading algorithm is depicted in Fig. \ref{fig10}, wherein the horizontal axis represents the number of rounds performed by the offloading algorithm, and the vertical axis measures the change in KL divergence of the server dataset with respect to the global data. In Fig. \ref{fig10}, the lines in blue, orange, green, and red represent the FlocOff, IOJR, Scaffold, and FedAvg algorithms, respectively. As the offloading strategies are implemented, the loss trends of all four curves exhibit a decline. Notably, FlocOff achieves the fastest convergence and lowest loss across various threshold settings, significantly outperforming the other three baseline algorithms. Scaffold demonstrates strong adaptability to Non-IID data, closely approximating the performance of the blue curve. However, IOJR and FedAvg, lacking mechanisms to counteract data heterogeneity, exhibit the slowest convergence rates and higher losses upon model stabilization. Moreover, it is important to note that the similar trends across the curves suggest that different threshold settings have a limited impact on the performance of FlocOff.

It is noteworthy that a common feature among the three images in Fig. \ref{fig10} is the position of the blue line at the bottom, indicating that FlocOff achieved optimal convergence. Additionally, the FlocOff curve demonstrates the most significant reduction in loss within the initial 10 rounds, suggesting that the algorithm rapidly reshaped the server dataset. As the algorithm progresses, the loss approaches zero, markedly reducing the discrepancy between the offloaded data and the global dataset. Specifically, when the threshold \emph{$\Gamma=500$} is applied, the KL divergence is reduced to below 0.05. Under these conditions, FedAvg required 18 rounds to converge, whereas FlocOff completed the task in just 5 rounds, enhancing efficiency by 72.2\%.

\subsection{Impact of ES Initial Data Distribution on Computation Offloading}

\begin{figure*}[ht]
\centering
\subfigure[Initial distribution of ES is IID.]
{
        \centering
        \includegraphics[scale=0.26]{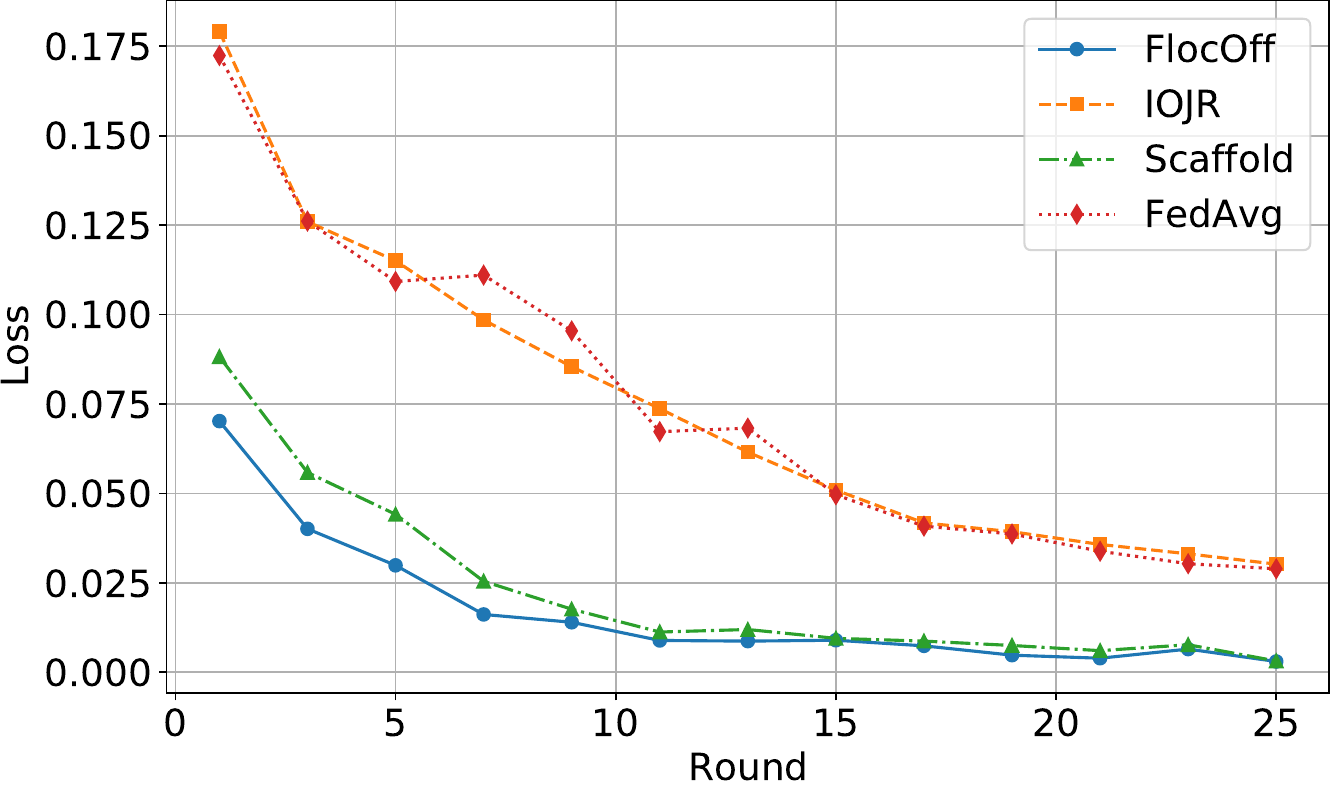}
}
\subfigure[Initial distribution of ES is Blank.]
{
        \centering
        \includegraphics[scale=0.26]{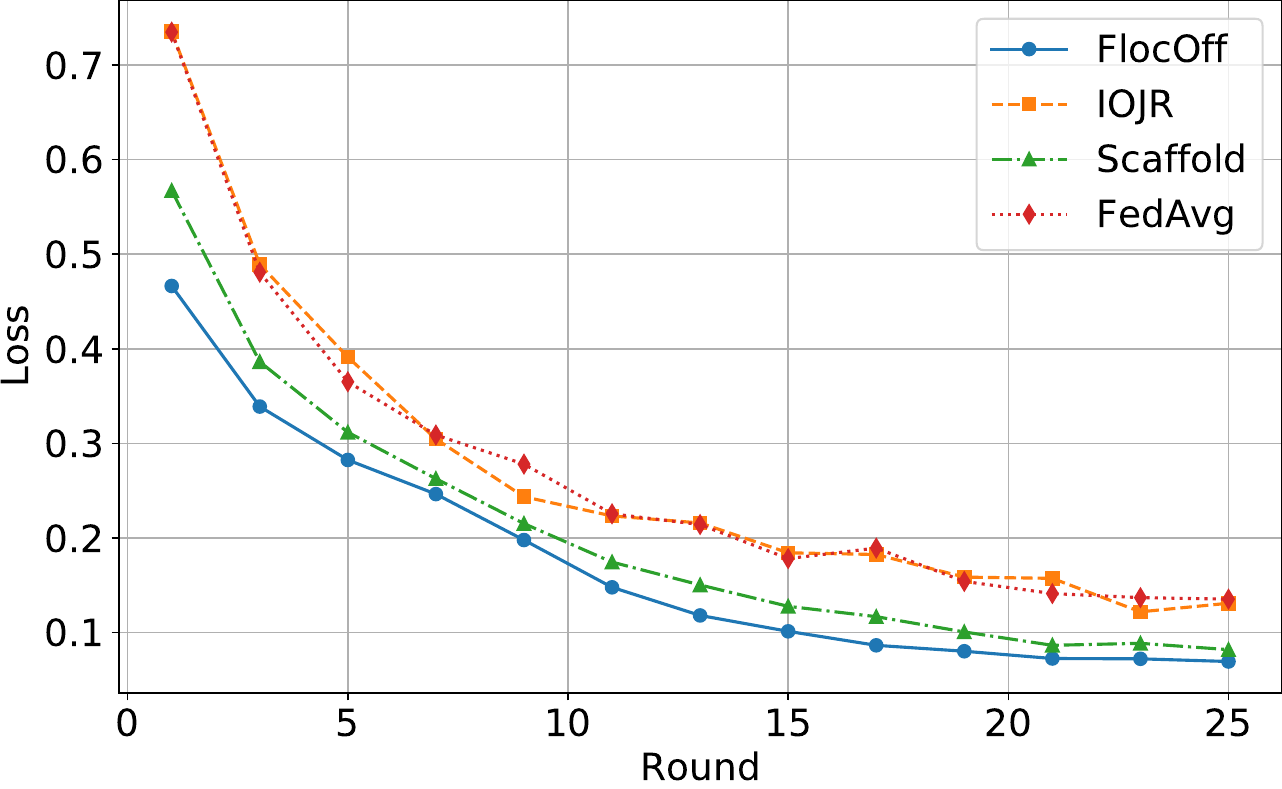}
}
\subfigure[Initial distribution of ES is Non-IID.]
{
        \centering
        \includegraphics[scale=0.26]{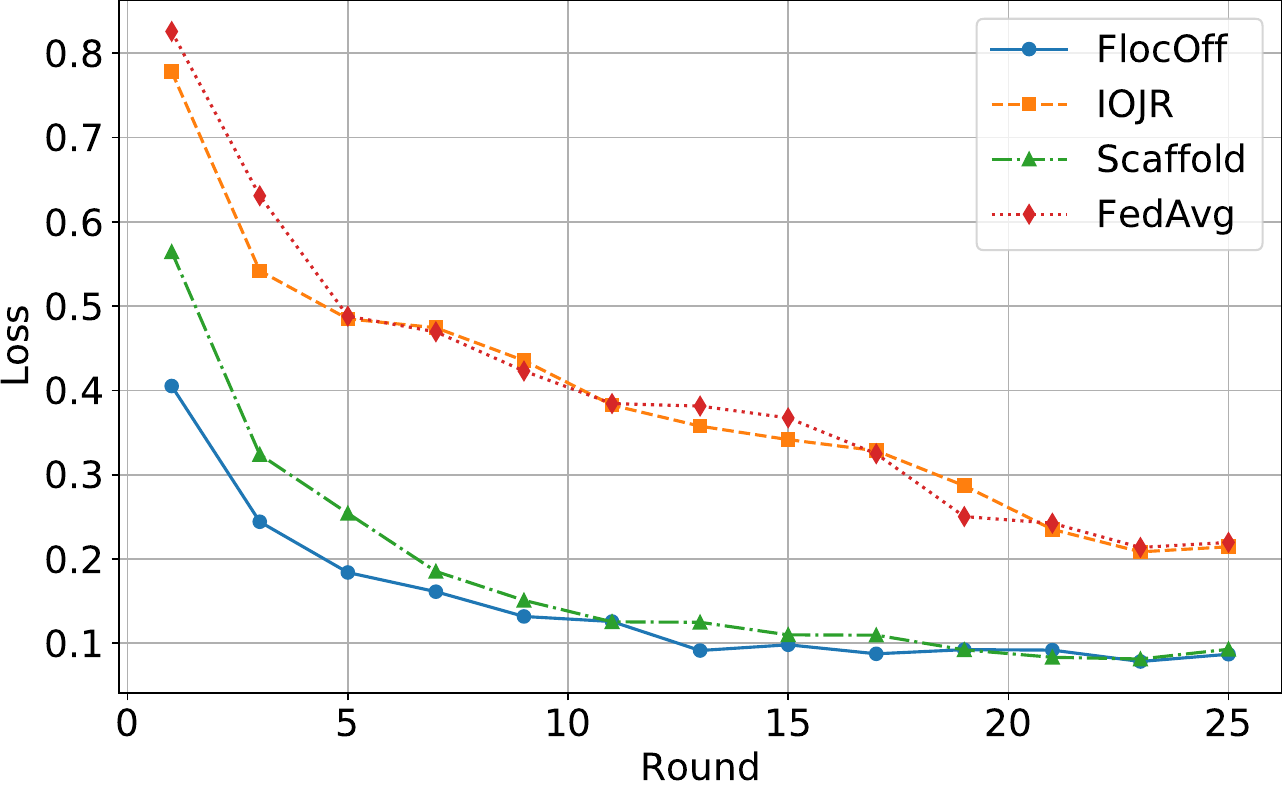}
}
\caption{Impact of ES initial data distribution.}
\label{fig11}
\end{figure*}

Secondly, we experiment with the initial data distribution of the server. As the ESs serve as a model training platform, they may have accumulated data that could affect offloading results. Thus, it is essential to investigate the impact of initial data on the performance of our model. We delineated three distinct distributions of server initialization data, namely: (1) Blank, indicating a server with no data initially and is reliant on offloaded data for model training; (2) IID, where each class of data X follows a Gaussian distribution with X$\sim$N(30,10); (3) Non-IID, wherein two classes are randomly designated as high probability classes, following a Gaussian distribution of X$\sim$N(50,20), while the remaining eight classes are low-probability, obeying Gaussian distributions from X$\sim$N(10,2).

Fig. \ref{fig11} illustrates the impact of various initial data distributions on loss at a threshold of \emph{$\Gamma=500$}. The x-axis represents the number of rounds, while the y-axis charts the changes in loss. A prominent feature across all three graphs is the consistent positioning of the blue curve below the orange, red, and green curves. This indicates that the FlocOff algorithm outperforms the three baseline algorithms across different initial data distributions. The performances of IOJR and FedAvg are closely matched, whereas Scaffold achieves a lower loss. By comparing the results, we can identify the differences caused by different initial data distributions. Notably, the initial data based on IID exhibits the smallest KL divergence, hovering around 0.15, which is significantly lower than that for the blank and Non-IID settings. Furthermore, the consistent downward trend of the blue curve across all three graphs suggests that varying initial data distributions have a minimal impact on the performance of FlocOff.

\subsection{Algorithm Stability in Non-IID Data}

\begin{figure}[ht]
\centering
\subfigure[Algorithm performance when \emph{$\Gamma=500$}.]
{
    \begin{minipage}[b]{.98\linewidth}
        \centering
        \includegraphics[scale=0.36]{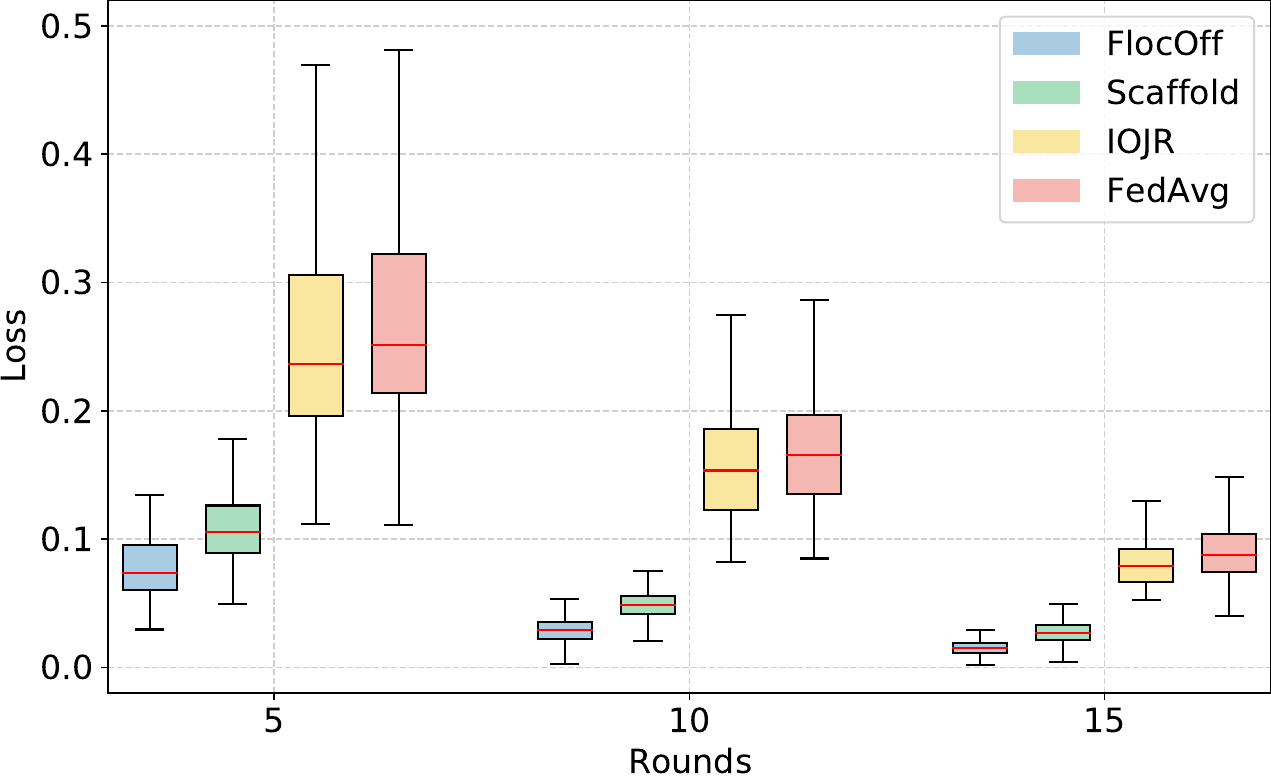}
    \end{minipage}
}
\subfigure[Algorithm performance when \emph{$\Gamma=1000$}.]
{
 	\begin{minipage}[b]{.98\linewidth}
        \centering
        \includegraphics[scale=0.36]{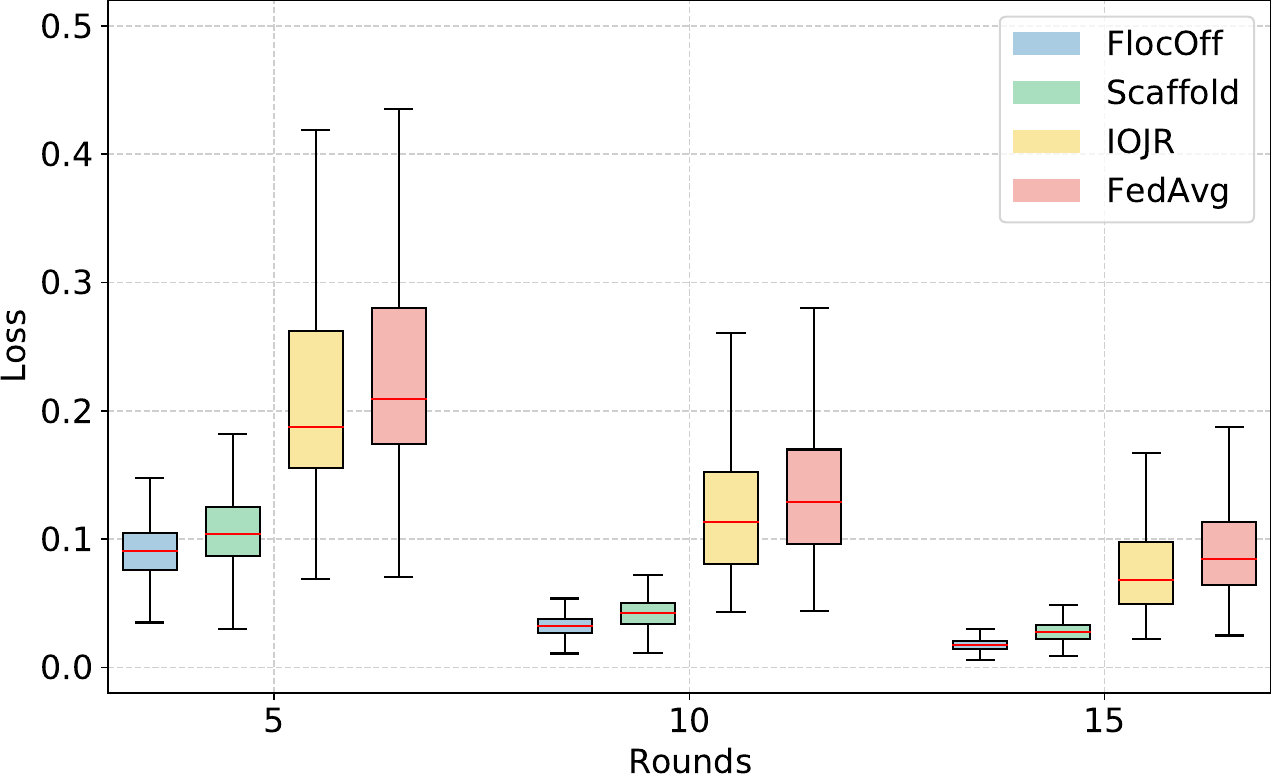}
    \end{minipage}
}
\subfigure[Algorithm performance when \emph{$\Gamma=1500$}.]
{
 	\begin{minipage}[b]{.98\linewidth}
        \centering
        \includegraphics[scale=0.36]{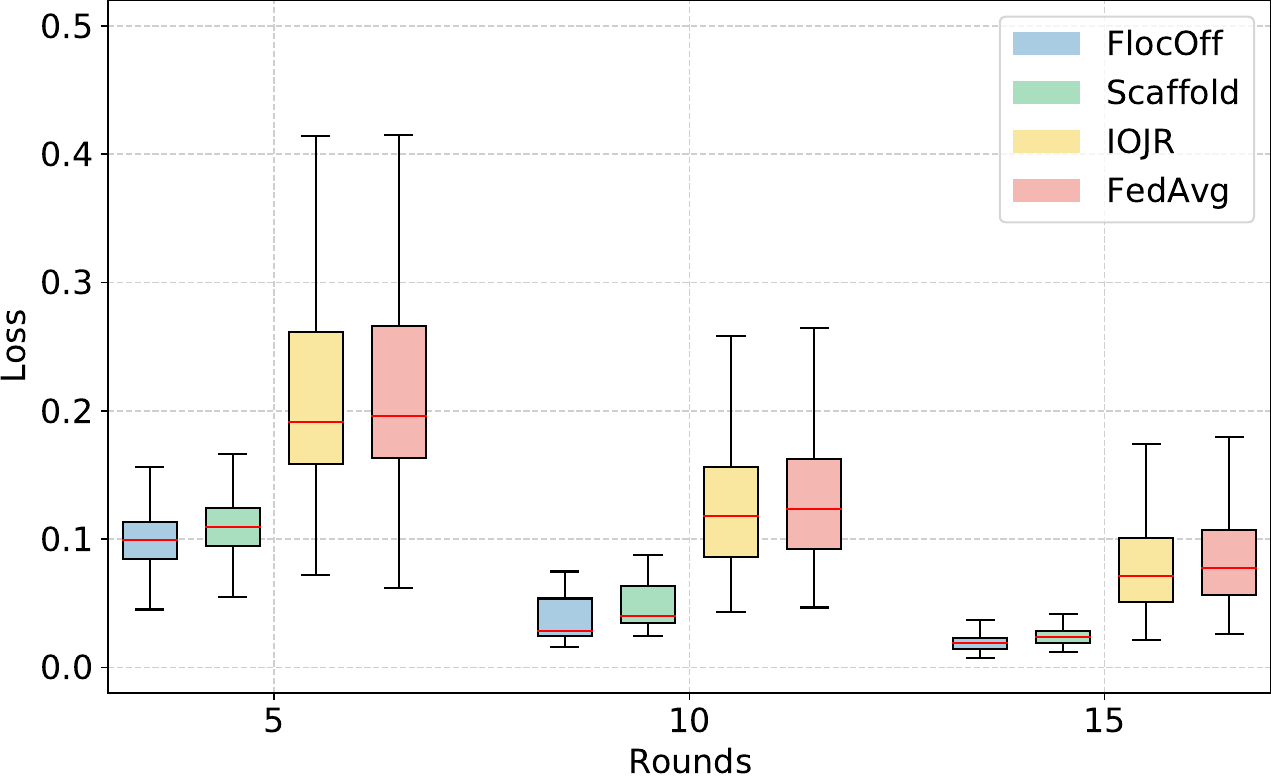}
        
    \end{minipage}
}
\caption{Algorithm stability in Non-IID data.}
\label{fig12}
\end{figure}

We proceeded to examine the stability of the computation offloading strategy under FlocOff. To illustrate the algorithm's fluctuations under various \emph{$\Gamma$}, we conducted 30 iterations for each algorithm and plotted the box plots. Fig. \ref{fig12}(a) depicts the fluctuations of the FlocOff and the three baseline when \emph{$\Gamma=500$}. The red solid line in the figure represents the median. The FlocOff in Fig. \ref{fig12}(a) exhibits the highest resilience in round 5, corresponding to 0.06, 0.07, and 0.10 for the lower, median, and upper edges of the box, respectively. On the other hand, the corresponding lower, median, and upper edges of the Scaffold in Fig. \ref{fig12}(a) are 0.09, 0.11, and 0.14, respectively, which are slightly higher. However, the IOJR and FedAvg clearly exhibit higher loss values, with medians reaching 0.23 and 0.24, respectively, and also display greater variance. Furthermore, the median of this two algorithms for rounds 10 and 15 are 0.15, 0.16 and 0.08, 0.09, respectively, whereas the corresponding values of FlocOff are 0.003 and 0.002, resulting in a KL divergence reduction of 81.25\% and 77.8\%, respectively. This demonstrates that FlocOff is more effective at reshaping the dataset throughout the unloading cycle, thereby achieving smaller loss.

In addition, it is noteworthy that the IOJR and FedAvg exhibit box heights of 0.06 and 0.03 at rounds 10 and 15, respectively, whereas the FlocOff algorithm shows significantly lower box heights of 0.014 and 0.008, representing a 76.7\% and 73.3\% decrease, respectively. This observation demonstrates that FlocOff is associated with lower fluctuations and greater stability compared to the IOJR and FedAvg, leading to superior convergence performance.

Fig. \ref{fig12}(b) illustrates the loss of FlocOff under \emph{$\Gamma=1000$}. Notably, the box plot is accompanied by a line segment at the top and bottom, indicating the maximum and minimum values, respectively. For the FlocOff algorithm, loss decreases rapidly in the initial 15 rounds, with median values of 0.09, 0.03, and 0.01, respectively. In contrast, the Scaffold exhibits a somewhat poorer performance, with median values for the three groups being 0.10, 0.04, and 0.02, respectively, and it demonstrates a notably higher variance compared to FlocOff. With regards to stability, the box heights of FlocOff in the initial 15 rounds remain consistently low at 0.028, 0.016 and 0.009. In contrast, Fig. \ref{fig12}(b), the FedAvg's three groups have box heights of 0.101, 0.065, and 0.042, which are significantly higher than those observed for FlocOff. This higher susceptibility of FedAvg to heterogeneous data highlights the greater stability of the FlocOff algorithm when dealing with Non-iid data.

Likewise, Fig. \ref{fig12}(c) exhibit similar trends, further highlighting the superiority of FlocOff when the threshold \emph{$\Gamma$} is increased to 1500. Specifically, the median value of the FlocOff for the 5th round is 0.100, demonstrating a noteworthy 48.7\% reduction compared to the FedAvg's median value of 0.195. Additionally, the corresponding box height of FlocOff is also 72.3\% lower. At the 15th cycle, when the two algorithms begin to converge, the median value of the FedAvg is 0.073, while the median value of FlocOff is 0.016, with loss that is 78.1\% lower. The aforementioned results unequivocally underscore the efficacy of FlocOff in reshaping superior quality datasets and surpassing the other baselines in terms of stability.

\subsection{Effect of federated Learning and Training Efficiency}

\begin{figure*}[ht]
\centering
\subfigure[Performance in LN-IID when \emph{$\Gamma=500$}.]
{
        \centering
        \includegraphics[width=5.7cm]{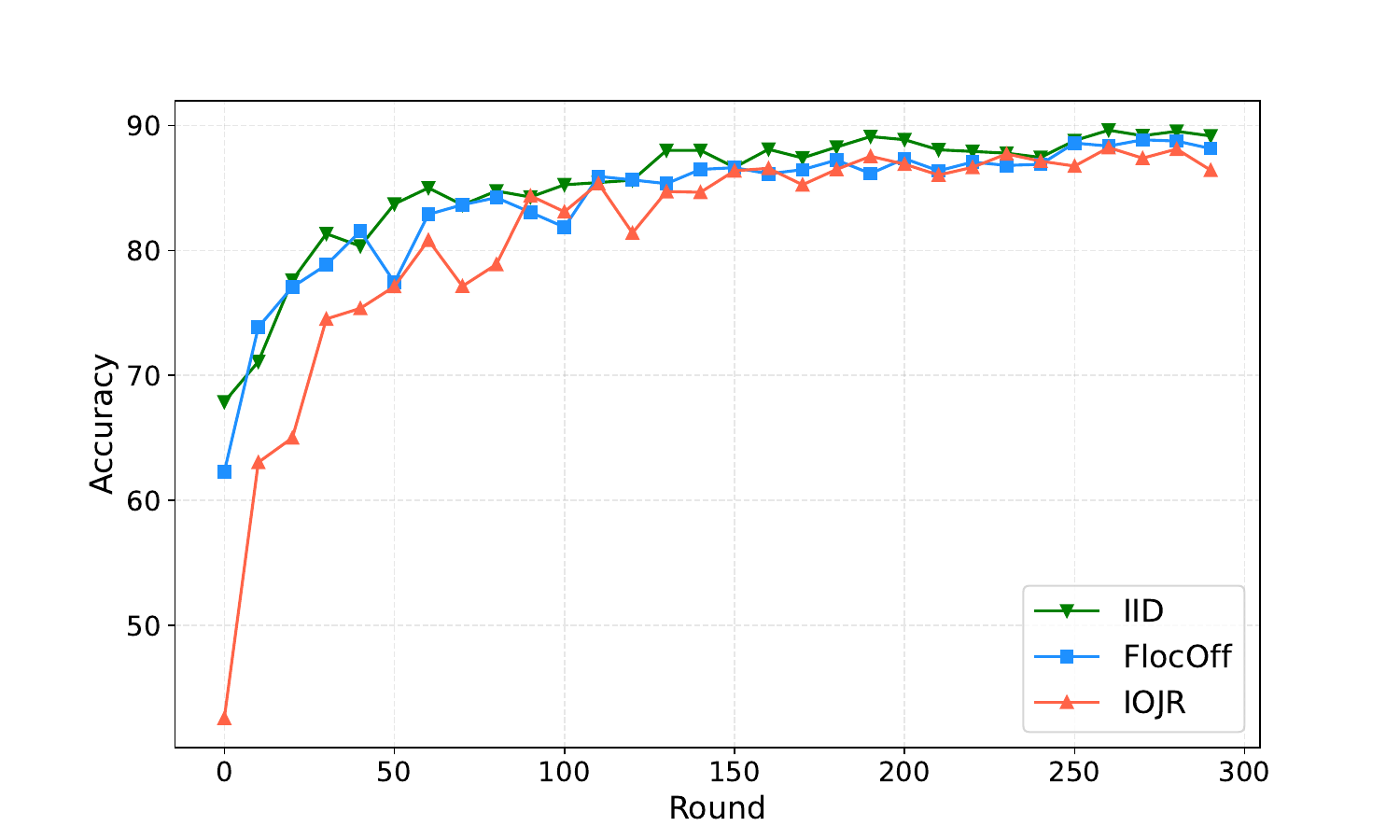}
}
\subfigure[Performance in LN-IID when \emph{$\Gamma=1000$}.]
{
        \centering
        \includegraphics[width=5.7cm]{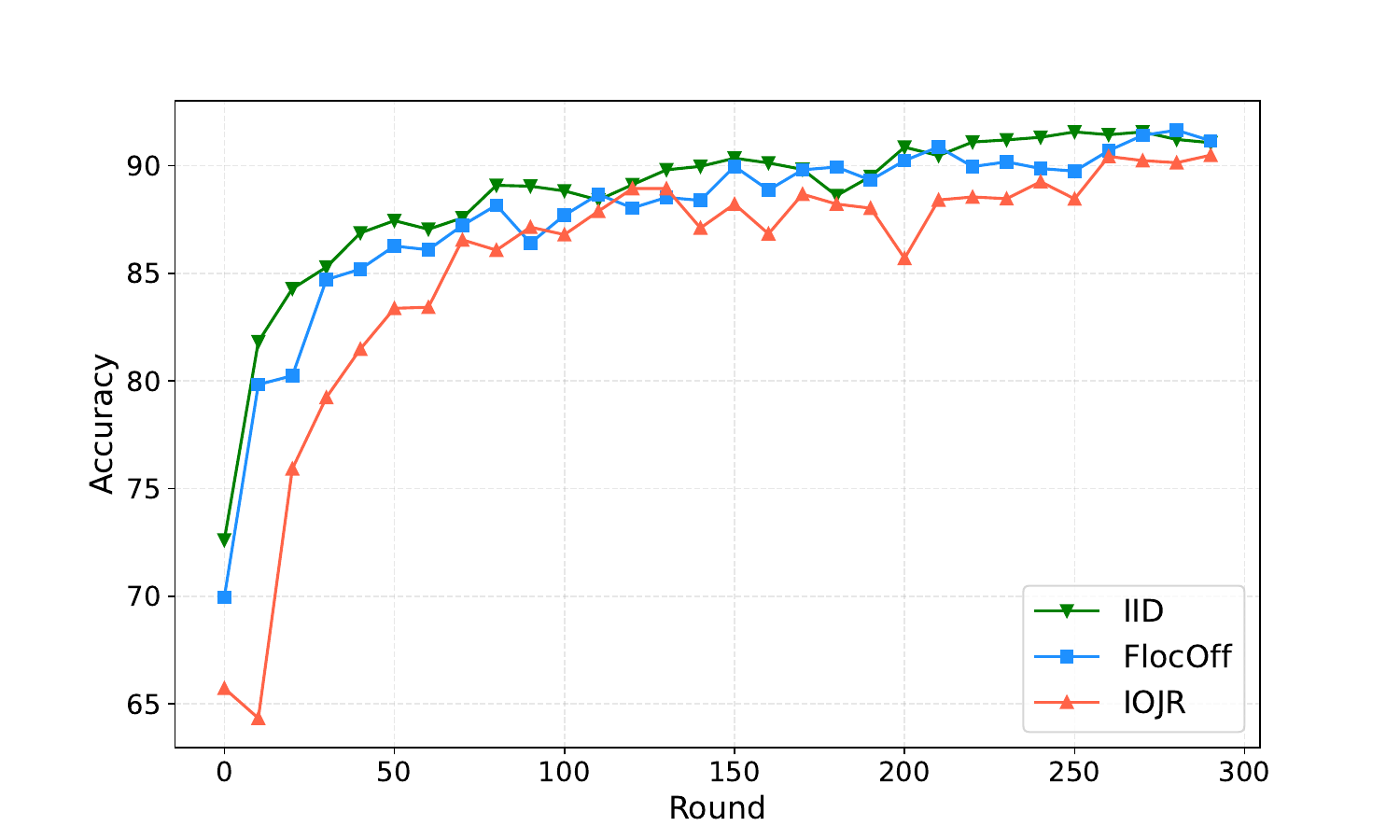}
}
\subfigure[Performance in LN-IID when \emph{$\Gamma=1500$}.]
{
        \centering
        \includegraphics[width=5.7cm]{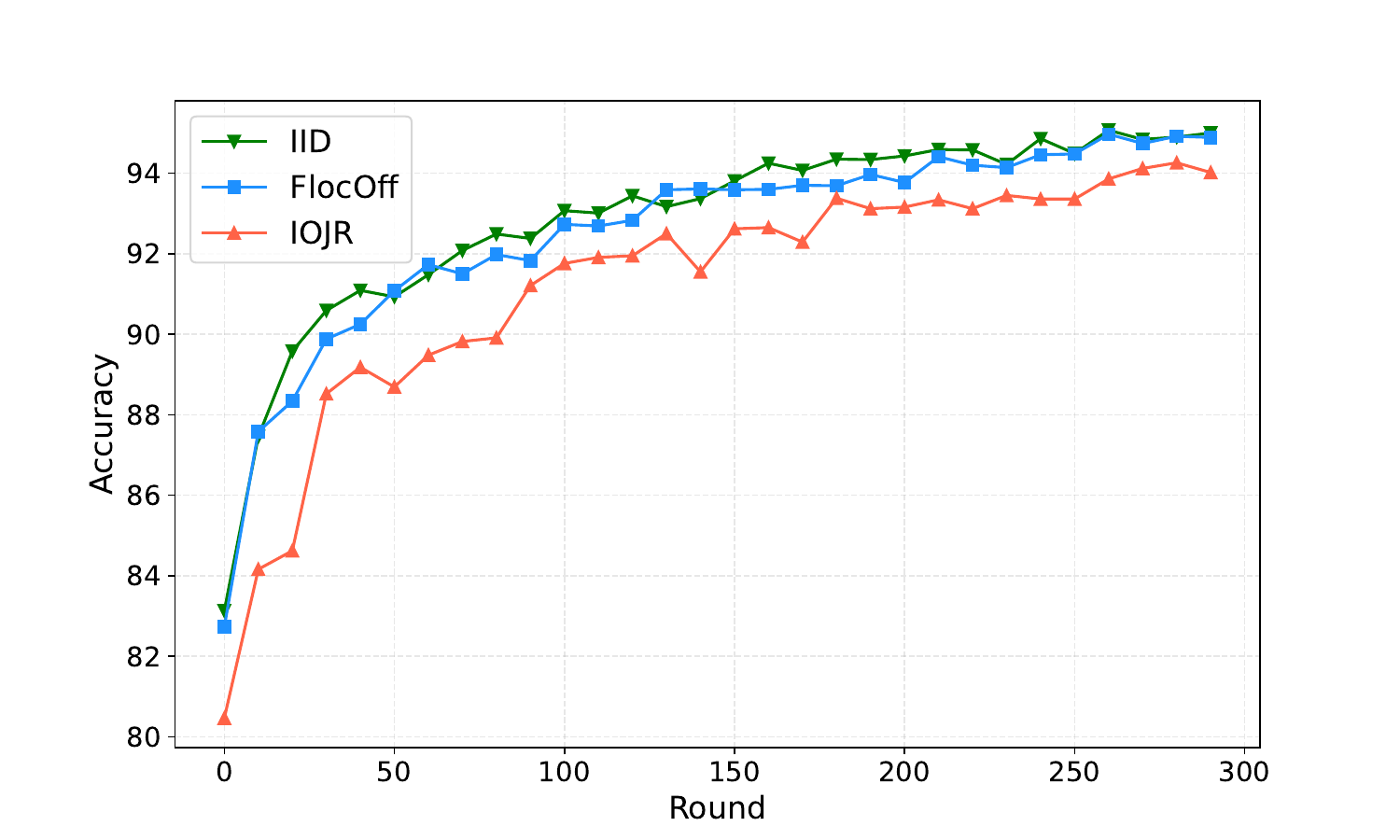}
}
\\ 
\centering
\subfigure[Performance in HN-IID when \emph{$\Gamma=500$}.]
{
        \centering
        \includegraphics[width=5.7cm]{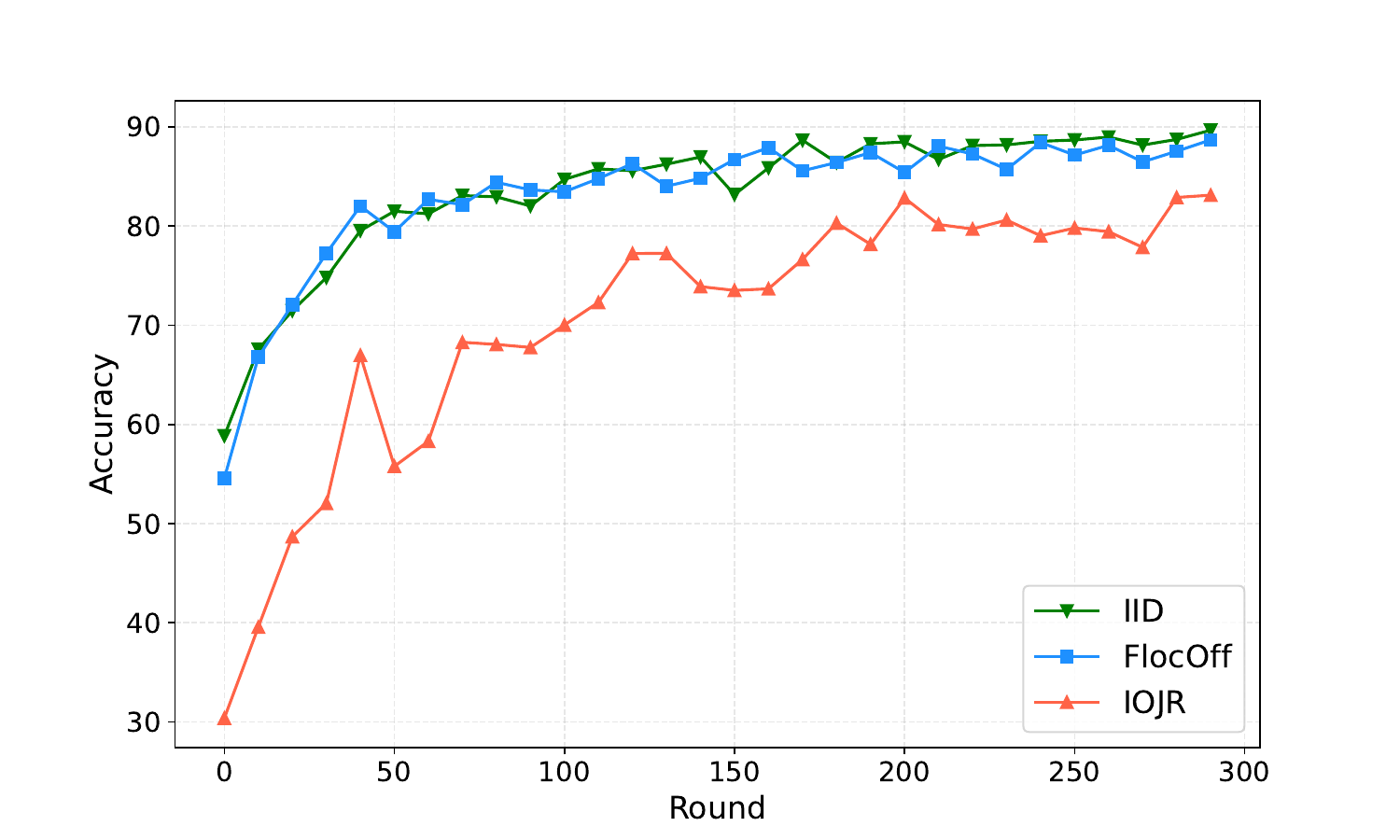}
}
\subfigure[Performance in HN-IID when \emph{$\Gamma=1000$}.]
{
        \centering
        \includegraphics[width=5.7cm]{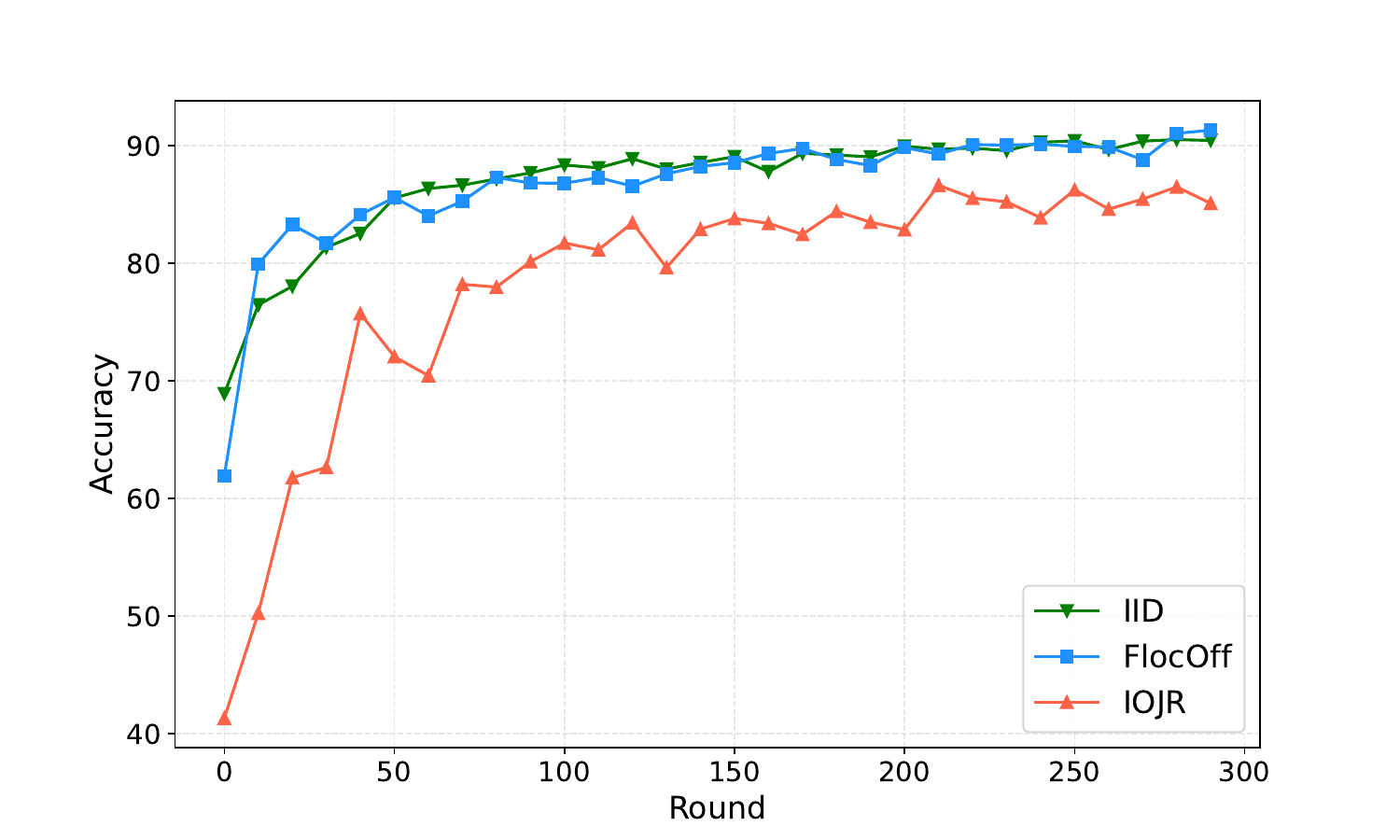}
}
\subfigure[Performance in HN-IID when \emph{$\Gamma=1500$}.]
{
        \centering
        \includegraphics[width=5.7cm]{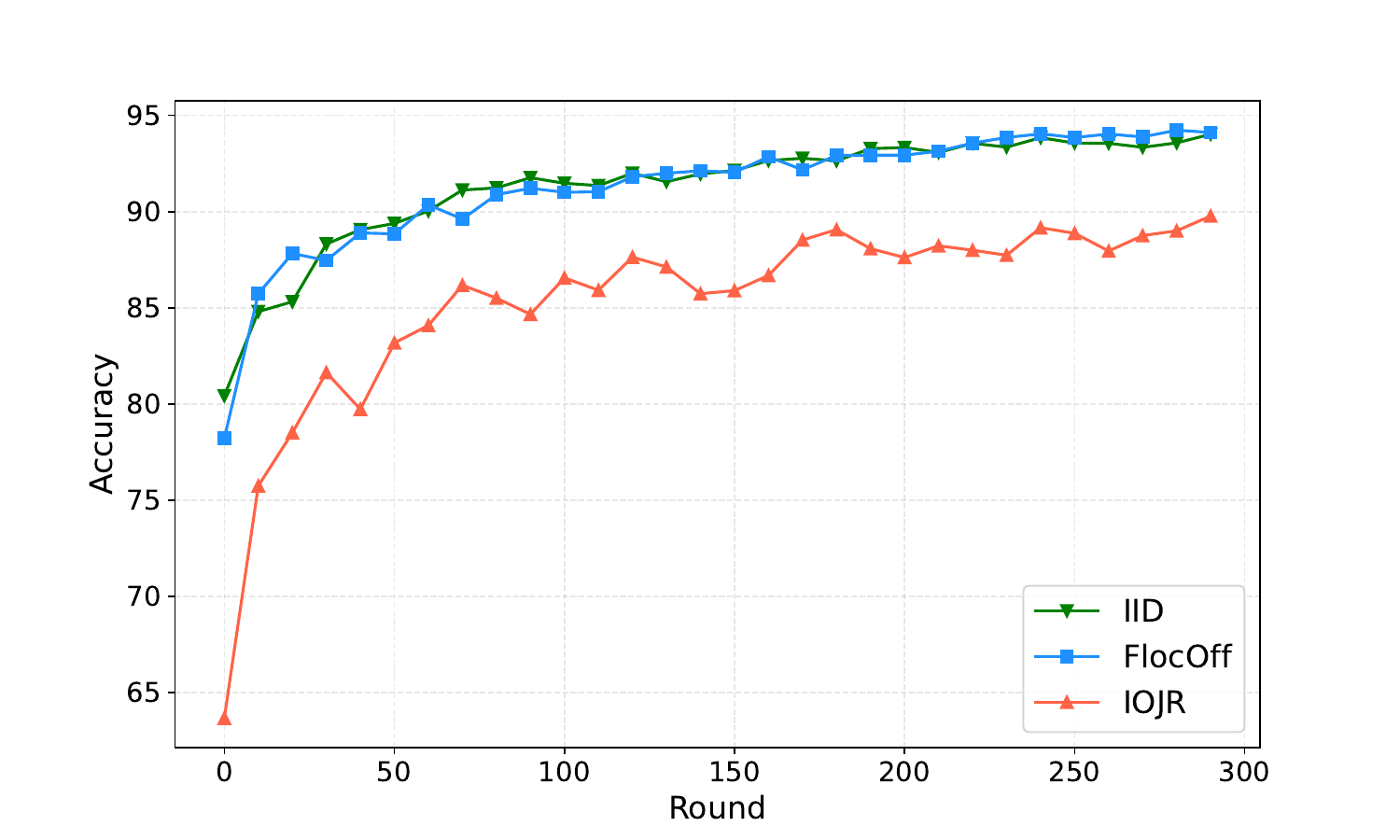}
}
\caption{Performance of federated learning in MNIST.}
\label{fig13}
\end{figure*}

\begin{figure*}[ht]
\centering
\subfigure[Performance in LN-IID when \emph{$\Gamma=500$}.]
{
        \centering
        \includegraphics[width=5.5cm]{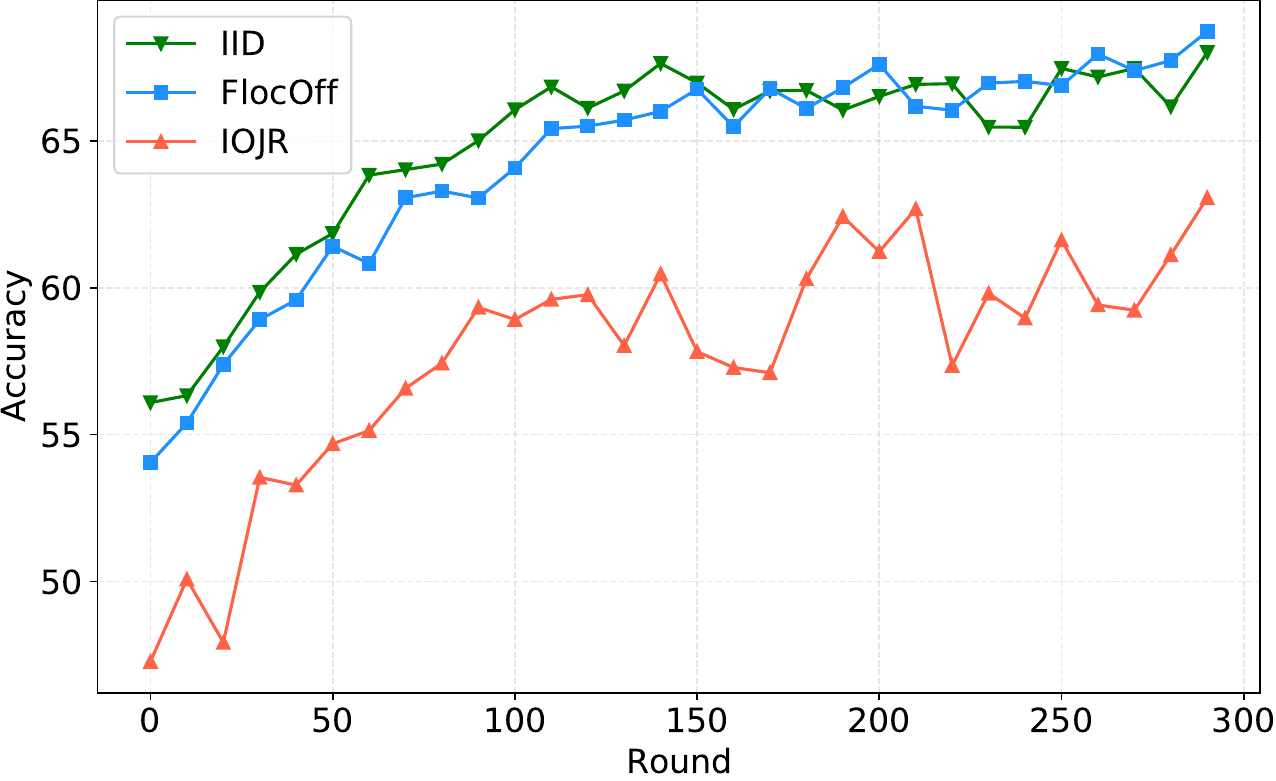}
}
\subfigure[Performance in LN-IID when \emph{$\Gamma=1000$}.]
{
        \centering
        \includegraphics[width=5.5cm]{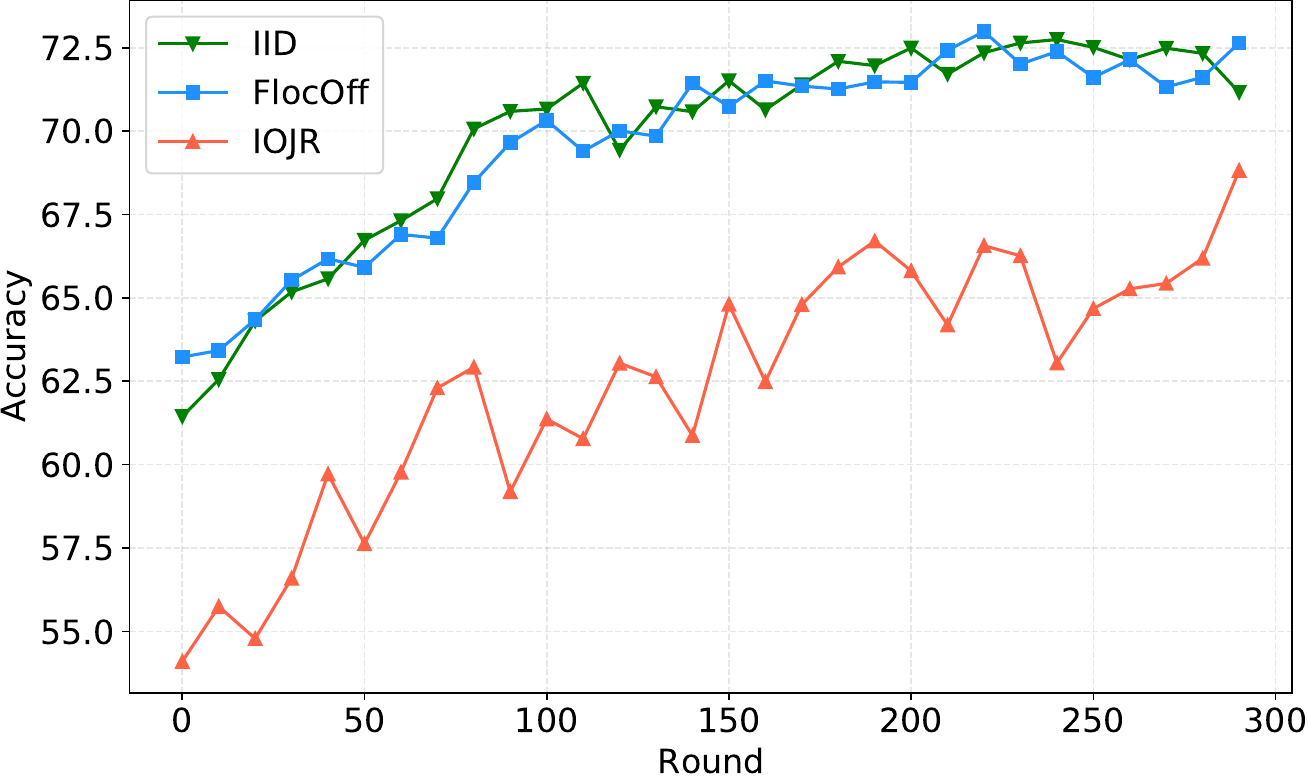}
}
\subfigure[Performance in LN-IID when \emph{$\Gamma=1500$}.]
{
        \centering
        \includegraphics[width=5.5cm]{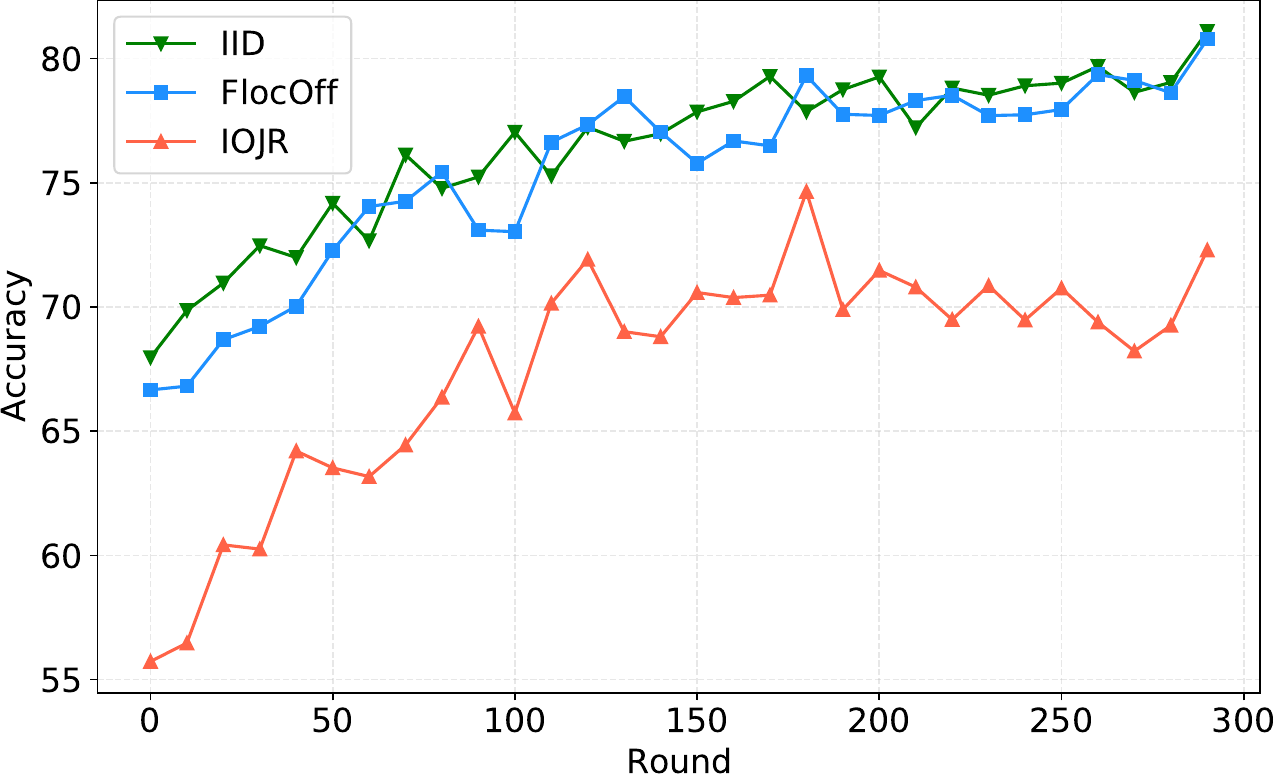}
}
\\ 
\centering
\subfigure[Performance in HN-IID when \emph{$\Gamma=500$}.]
{
        \centering
        \includegraphics[width=5.5cm]{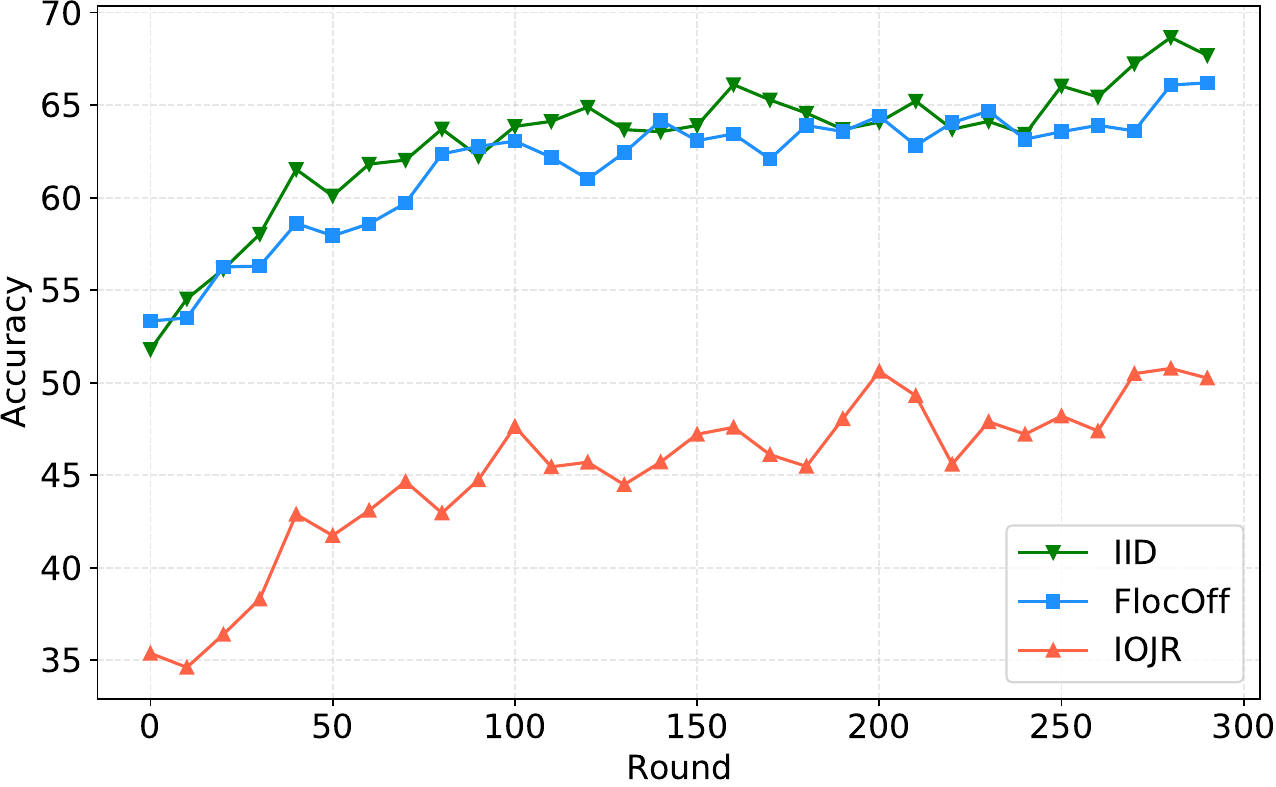}
}
\subfigure[Performance in HN-IID when \emph{$\Gamma=1000$}.]
{
        \centering
        \includegraphics[width=5.5cm]{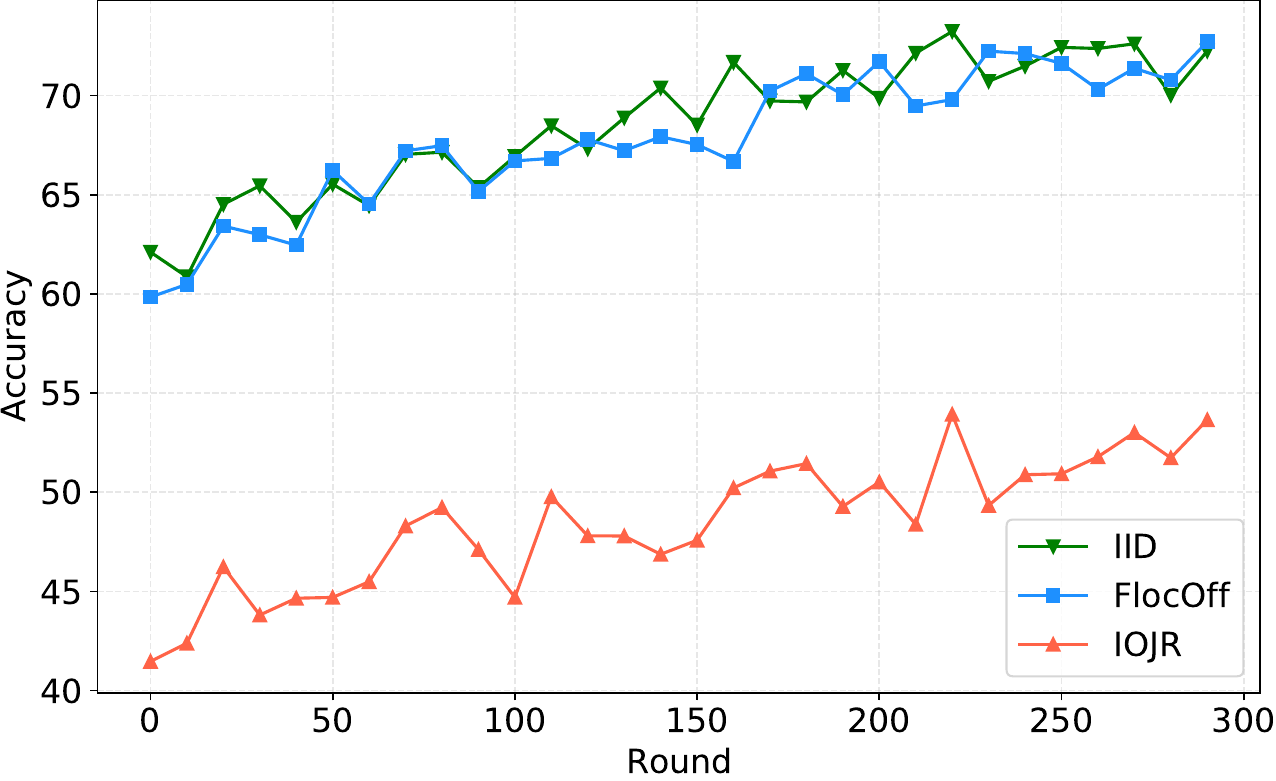}
}
\subfigure[Performance in HN-IID when \emph{$\Gamma=1500$}.]
{
        \centering
        \includegraphics[width=5.5cm]{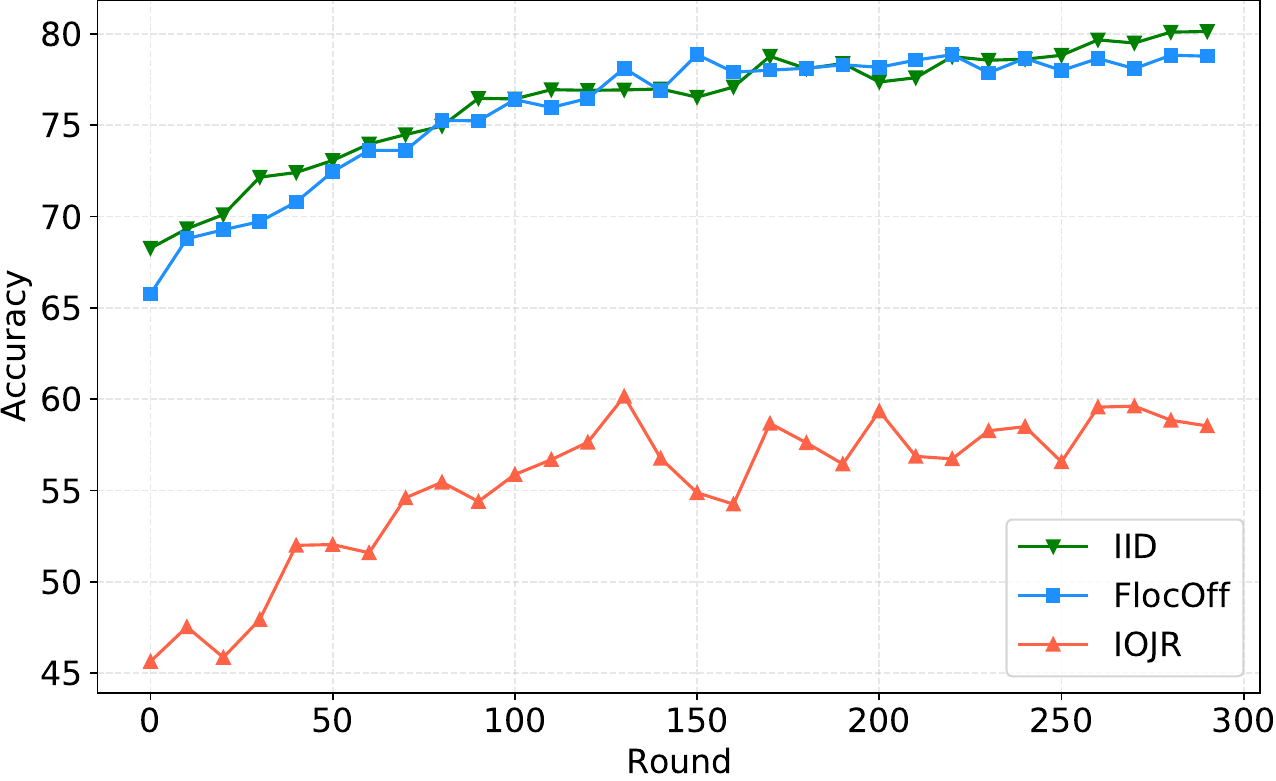}
}
\caption{Performance of federated learning in CIFAR-10.}
\label{fig14}
\end{figure*}

Here, we sought to evaluate the training efficiency of FlocOff by examining its performance on two distinct types of Non-IID datasets. To ensure a realistic simulation of edge-end heterogeneous datasets, we employed a sample generation method outlined in subsection A. Specifically, we designed two types of Non-IID data: Scenario (1) Light Non-IID data (LN-IID), which randomly set two high-frequency occurrence classes following Gaussian distributions of X$\sim$N(50,20), while the remaining eight classes were low-probability classes following a Gaussian distribution of X$\sim$N(10,2); Scenario (2) Heavy Non-IID data (HN-IID), which also randomly set two high-frequency occurrence classes following Gaussian distributions of X$\sim$N(50,20), but with a larger number gap between categories compared to the Light Non-IID data. The remaining eight classes in the Heavy Non-IID data were low-probability classes following Gaussian distributions of X$\sim$N(10,2).

Subsequently, we applied the federated learning approach to assess its performance on two Non-IID datasets. Additionally, we established an IID environment as a baseline to compare with FlocOff performance in Non-IID conditions. In the IID scenario, each edge dataset is independently and identically distributed. This represents an ideal condition, eliminating the need for offloading algorithms to adjust data distribution. The homogeneity of data in IID environments facilitates optimal model convergence and accuracy. This scenario serves as a perfect baseline to underscore the effectiveness of FlocOff. Through this approach, we aim to demonstrate the FlocOff capability to counteract the adverse impacts of data heterogeneity on model training in Non-IID conditions.

Fig. \ref{fig13}(a)-(c) demonstrate the training process of federated learning based on offloaded data with varying threshold values \emph{$\Gamma$} ranging from 500 to 1500. The horizontal axis represents the training period, and the vertical axis denotes the accuracy rate. The figures exhibit three curves in different colors, where the green, blue, and red curves correspond to IID, FlocOff, and IOJR algorithms, respectively. Notably, the green and blue curves in all three images are proximate to each other and consistently outperform the red curve, indicating that FlocOff can effectively enhance the dataset's quality after executing computation offloading and achieve similar accuracy as IID data, thereby significantly improving the accuracy compared to the IOJR algorithm. Specifically, FlocOff and IID jointly converge around 40 rounds, while the IOJR algorithm reaches its inflection point at approximately 80 rounds. Furthermore, the green and blue curves exhibit higher accuracy at the end of the training, with FlocOff and IID achieving accuracy rates of 89\%, 92\%, and 94\% when \emph{$\Gamma$} equals 500, 1000, and 1500, respectively. In contrast, the IOJR algorithm achieves corresponding accuracy rates of 87\%, 90\%, and 93\%, respectively. Hence, these findings clearly demonstrate that FlocOff can efficiently integrate offloaded data, improving the model's convergence and accuracy.

In contrast to the experimentation with light Non-IID data, the results of the heavy Non-IID data experiments displayed in Fig. \ref{fig13}(d)-(f) show a similar trend. The green and blue curves remain proximate to each other and both surpass the red curve. This is a testament to FlocOff's ability to transform the dataset to be akin to the IID data. The numerical analysis revealed that the accuracy of the converged FlocOff and IID models were equal to 89\%, 91\%, and 94\% for \emph{$\Gamma$} values of 500, 1000, and 1500, respectively. On the other hand, the IOJR algorithm corresponded to accuracies of 82\%, 85\%, and 88\%, respectively. In comparison to the light Non-IID data experiments, the heavy Non-IID data significantly lowered the accuracy of the model. Nonetheless, the effect of the FlocOff on heavy Non-IID data remained consistent with the IID data. This signifies that FlocOff can effectively counteract different degrees of heterogeneous data and increase the accuracy of federated learning training.

Fig. \ref{fig13} is based on the MNIST handwritten font, whereas Fig. \ref{fig14} replicates the experiment using the CIFAR-10 dataset, which contains colored images of common objects and thus presents a greater challenge for recognition due to its increased number of channels, pixels, and features compared to the black and white MNIST dataset. Fig. \ref{fig14}(a)-(c) demonstrate the variation of accuracy curves based on the CIFAR-10 light Non-IID data. It is evident that the green and blue curves are remarkably similar and higher than the red curve, indicating that FlocOff can produce comparable results to IID on the CIFAR-10 dataset. In terms of accuracy, the converged accuracy of the green and blue curves in the three images are 68\%, 72\%, and 80\%, respectively, while the red curve corresponds to an accuracy of 62\%, 67\%, and 70\%, reflecting a decrease of 9.7\%, 7.5\%, and 14.3\%, respectively.

Fig. \ref{fig14}(d)-(f) display the accuracy curves based on the CIFAR-10 heavy Non-IID data and provide a similar conclusion. The dataset reshaped by FlocOff exhibits the same training performance as the IID data. Notably, the FlocOff attains an accuracy of 65\%, 72\%, and 79\% for \emph{$\Gamma$} values equal to 500, 1000, and 1500, respectively. In contrast, the IOJR's accuracy is 49\%, 52\%, and 57\%, which represents a substantial decrease of 32.7\%, 38.5\%, and 38.6\%, respectively. It is noteworthy that CIFAR-10's severe Non-IID data leads to more significant training damage, and accuracy decreases remarkably when compared to Fig. \ref{fig14}(a)-(c). Furthermore, the FlocOff demonstrates robustness against the impact of data heterogeneity on both MNIST and CIFAR-10 datasets. Our proposed framework effectively achieves near-IID performance on datasets with varying degrees of Non-IID.

\subsection{Performance of Model Accuracy and Data Distribution}

\begin{table}[htbp]
	\centering
	\caption{Different Degrees of Data Heterogeneity. While MOHP and MOLP means Mean of High Prob and Mean of Low Prob respectively.}
	\begin{tabular}{ccccccccc}
		\toprule  
		Group&  1&  2&  3&  4&  5&  6&  7& 8\\
		\midrule  
		MOHP&  40.0&  36.0&  32.0&  28.0&  24.0&  20.0&  16.0& 12.0\\
        MOLP&  4.0&  5.0&  6.0&  7.0&  8.0&  9.0&  10.0& 11.0\\
		\bottomrule  %
	\end{tabular}
\end{table}

\begin{figure}[ht]
\centering
\subfigure[Relationship between data distribution and accuracy of MNIST.]
{
    \begin{minipage}[b]{.98\linewidth}
        \centering
        \includegraphics[scale=0.53]{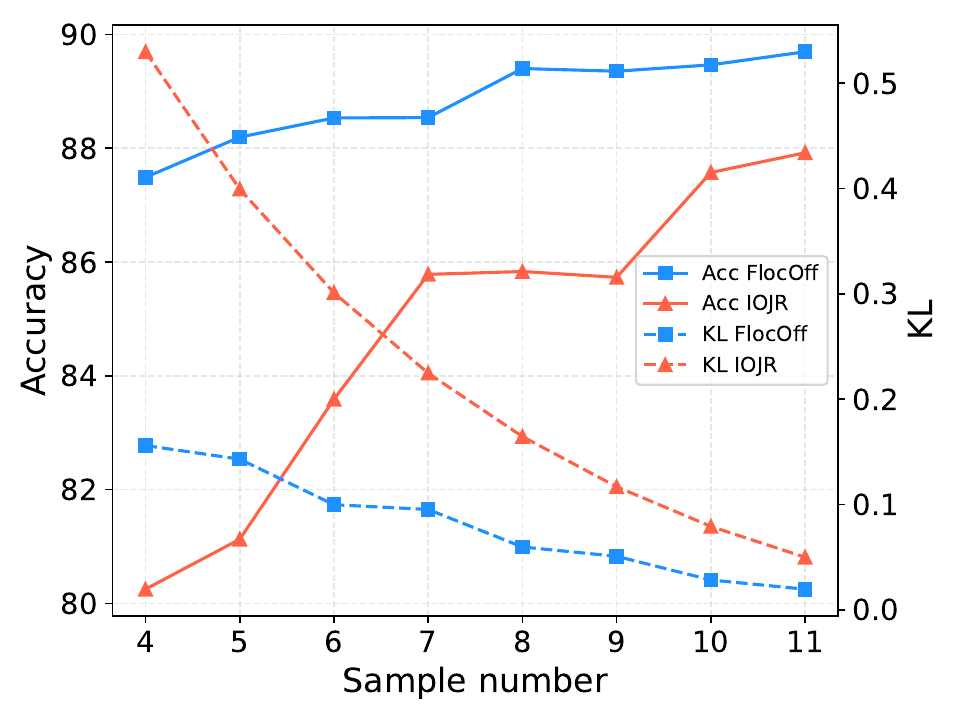}
    \end{minipage}
}
\subfigure[Relationship between data distribution and accuracy of CIFAR-10.]
{
 	\begin{minipage}[b]{.98\linewidth}
        \centering
        \includegraphics[scale=0.53]{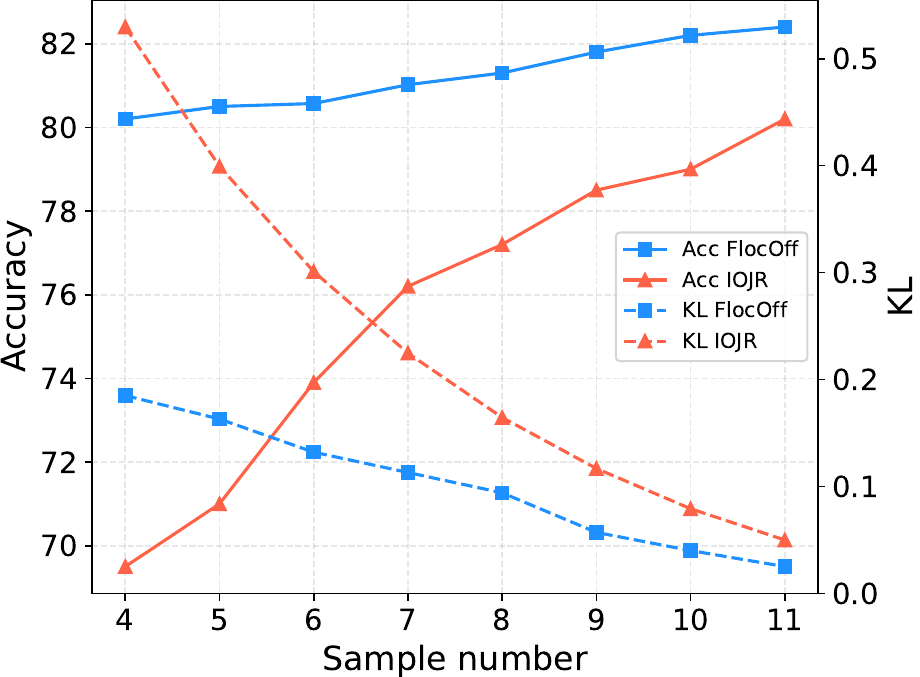}
    \end{minipage}
}
\caption{Performance of model accuracy and data distribution.}
\label{fig15}
\end{figure}

To investigate the relationship between model accuracy and data distribution under the FlocOff framework, we conducted a series of experiments using eight different data distribution methods. In the first set of experiments, we intentionally created severe data heterogeneity, with two high probability classes following a Gaussian distribution of X$\sim$N(40,5), and the other low-probability class following a Gaussian distribution of X$\sim$N(4,2). We then gradually reduced the severity of the data heterogeneity by varying the meanwhile keeping the variance constant. In the second set of experiments, the data distribution is modified to have the two high probability classes follow a Gaussian distribution of X$\sim$N(36,5), while the other low probability classes follow a Gaussian distribution of X$\sim$N(5,2). This pattern is continued in the subsequent experiments, with the data distribution adjusted until the mean values of the last set of high and low probability classes were 12 and 11, respectively, which closely resembled IID data. Table \uppercase\expandafter{\romannumeral2} presents the mean values of the data generation distributions, providing a detailed overview of the experimental setup.

The blue and red solid lines in Fig. \ref{fig15} (a) illustrate the accuracy variations of FlocOff and IOJR on the MNIST dataset, respectively. As we traverse from group 1 to group 8, we observe an upward trend in both curves, indicating improved data quality and reduced KL divergence. It is noteworthy that the accuracy of FlocOff ranges from 87.4\% to 89.7\%, surpassing the IOJR's accuracy range of 80.2\%-88.0\%. In Fig. \ref{fig15} (b), the same trend is evident on the CIFAR-10 dataset, with the KL divergence of FlocOff dropping from 0.19 to 0.01 and model accuracy increasing from 80.1\% to 82.3\% across groups 1 to 8. Similarly, the IOJR experienced a decrease in KL divergence from 0.52 to 0.05 and a corresponding increase in model accuracy from 69.5\% to 80.2\%. This demonstrates that our proposed algorithm can effectively enhance model accuracy.

\subsection{System Communication Cost}

\begin{figure}[ht]
\centering
\subfigure[Communication cost performance of CIFAR-10.]
{
    \begin{minipage}[b]{.98\linewidth}
        \centering
        \includegraphics[scale=0.5]{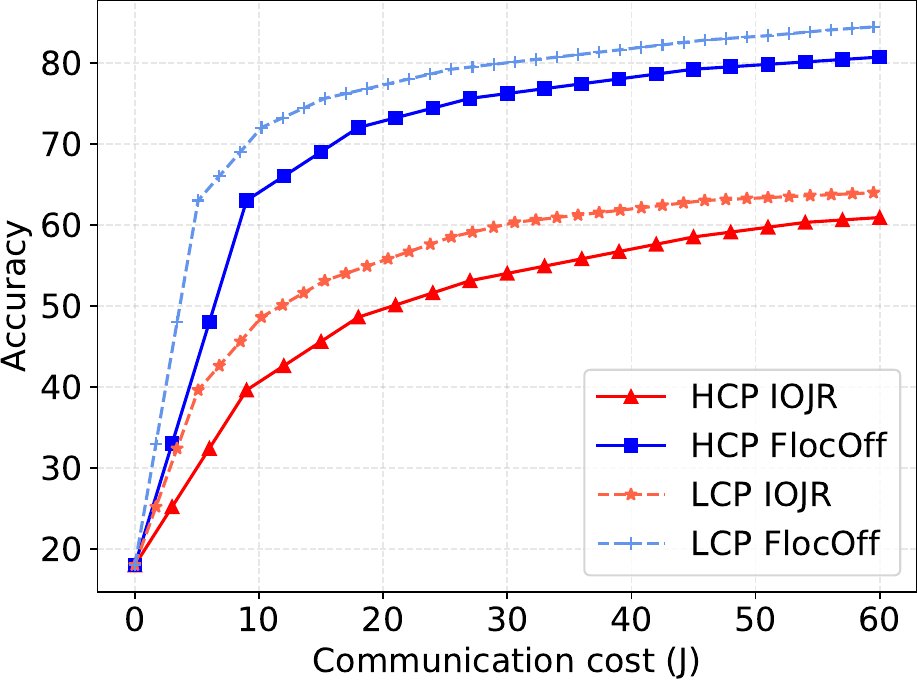}
    \end{minipage}
}
\subfigure[Communication cost performance of MNIST.]
{
 	\begin{minipage}[b]{.98\linewidth}
        \centering
        \includegraphics[scale=0.5]{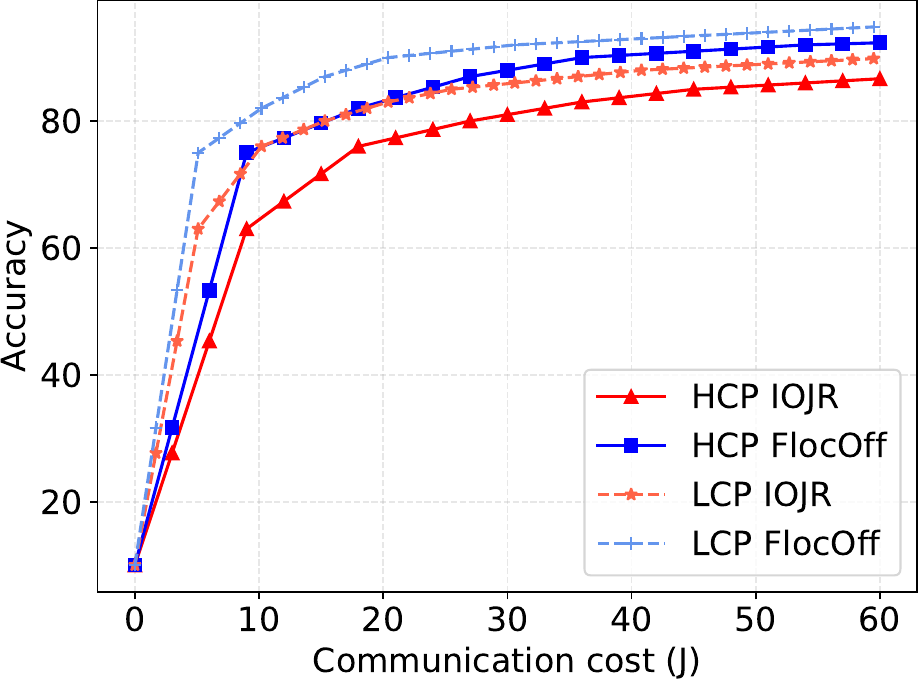}
    \end{minipage}
}
\caption{System communication cost.}
\label{fig17}
\end{figure}

We proceeded to compare the communication cost of the proposed FlocOff framework with that of the IOJR. In edge scenarios, communication between UD and ES is accomplished through OFDMA. The communication cost is evaluated using Eq. (\ref{eq23}). It is worth noting that Eq. (\ref{eq23}) represents the product of the transmission time and the communication power for all tasks. Consequently, the communication cost denotes the energy consumption for transmitting offloaded tasks within the network, measured in Joules. Fig. \ref{fig17} aims to intuitively demonstrate model performance in resource-constrained scenarios. Additionally, we have established two communication scenarios: High Communication Power (HCP) and Low Communication Power (LCP) scenarios. The transmission powers for HCP and LCP are set at 1 watt and 0.5 watts, respectively \cite{ref62, ref63}.

Fig. \ref{fig17} (a) depicts the training performance of FlocOff on the CIFAR-10 dataset. The horizontal axis corresponds to the model accuracy, while the vertical axis represents the communication cost. The blue solid line in the figure represents the performance of FlocOff  in HCP, while the red solid line represents the performance of the IOJR in the Non-IID scenario. Correspondingly, the light blue dashed line represents the performance of FlocOff under the LCP scenario, while the light red dashed line depicts the performance of IOJR in the LCP. In the LCP scenario, both algorithms achieve faster convergence with lower energy consumption. This is shown in the figure as the dotted line is above the solid line, achieving higher accuracy with the same communication cost. Notably, the blue line always lies below the red line, indicating that the FlocOff approach achieves better model training under the same communication cost. Specifically, the communication cost associated with the 60\% accuracy of the IOJR is roughly 51J, whereas the communication cost associated with FlocOff is a mere 9J, with a drop of 82.4\%. This result confirms that the proposed FlocOff framework can significantly alleviate the communication burden.

We carried out a similar experiment on the MNIST dataset, and the outcomes are depicted in Fig. \ref{fig17} (b). Even under the same communication cost, the FlocOff approach still achieves higher accuracy. This can be attributed to two factors: firstly, the proposed offloading strategy is capable of adjusting the data distribution, thereby enhancing the model convergence rate; secondly, the communication power allocation strategy can effectively improve the utilization of edge network resources. We set the benchmark of model accuracy at 80\%, where the IOJR consumes 27J, while FlocOff only needs 14J, with a drop of 48.1\%. This further underscores that the FlocOff framework can substantially reduce communication resources in the system and is thus suitable for edge networks with limited resources.

\section{Conclusion}

The present paper addresses the challenges of federated learning in edge scenarios, namely the heterogeneity of data and the limited communication resources. In particular, we investigate the impact of data distribution on the efficiency of model convergence and highlight the importance of addressing this issue to enhance training efficiency. To this end, we propose two sub-problems: (1) Reshaping the Edge Dataset via Computation Offloading (\textbf{RED-CO}); and (2) Communication Cost Optimization in Edge Environments (\textbf{CCO-EE}). Our proposed FlocOff framework aims to reshape the edge data distribution through offloaded maps. To achieve this, we propose a Minimizes the KL Divergence via Computation Offloading scheduling (\textbf{MKL-CO}) algorithm that adapts the offload decisions based on user device sample distributions. Moreover, we develop an efficient Minimizes the Communication Cost through Resource Allocation (\textbf{MCC-RA}) algorithm to generate power allocation strategies. We demonstrate through extensive experimentation that the FlocOff framework effectively improves the convergence rate and model accuracy of federated learning training while reducing the communication cost of the system.

In summary, our paper makes significant contributions to the field of federated learning in edge scenarios by identifying key challenges and proposing an innovative framework that overcomes these obstacles. The proposed approach leverages advanced techniques, such as offloading algorithms and numerical optimization, to optimize offloading and resource allocation, resulting in a more efficient and effective learning process.

\begin{appendices}

\section{}

We prove the establishment of Assumption 2 based on Assumption 1 and Eq. (\ref{eq5}). Our proof is divided into two steps: (1) \emph{$F(w)$} is a convex function; (2) \emph{$F(w)$} is Lipschitz smooth and calculates Lipschitz continuous gradient.

First, through Assumption 1, for \emph{$F_s$} on any ES, the following inequalities hold:
\begin{equation}
F_s\left(\theta w_1+(1-\theta) w_2\right) \leqslant \theta F_s\left(w_1\right)+(1-\theta) F_s\left(w_2\right),
\end{equation}

\noindent where \emph{$w_1$} and \emph{$w_2$} are the parameters of the model in a certain training cycle. According to Eq. (\ref{eq5}), we can prove the following formula:

\begin{equation}
\begin{aligned}
F\left(\theta w_1+(1-\theta) w_2\right) & =\frac{\sum_{s \in S} D_s F_s\left(\theta w_1+(1-\theta) w_2\right)}{D} \\
& \leqslant \frac{\sum_{s \in S} D_s\left[\theta F_s\left(w_1\right)+(1-\theta) F_s\left(w_2\right)\right]}{D} \\
& =\frac{\theta \sum_{s \in S} D_s F_s\left(w_1\right)}{D} \\
& \ \ \ \ \ +\frac{(1-\theta) \sum_{s \in S} D_s F_s\left(w_2\right)}{D} \\
& =\theta F\left(w_1\right)+(1-\theta) F\left(w_2\right).
\end{aligned}
\end{equation}

Therefore, \emph{$F(w)$} conforms to the definition of a convex function.

The next thing we want to prove is that \emph{$F(w)$} is Lipschitz smooth. For any \emph{$F_s$}, the following inequalities can be proved to hold by Assumption 1:
\begin{equation}
\left\|\nabla F_s(w)-\nabla F_s\left(w^{\prime}\right)\right\| \leq L\left\|w-w^{\prime}\right\|,
\end{equation}

\noindent where \emph{$L_s$} is the Lipschitz continuous gradient of the loss function \emph{$F_s$}. According to Eq. (\ref{eq5}), we can make the following derivation:

\begin{equation}
\begin{aligned}
\left\|\nabla F(w)-\nabla F\left(w^{\prime}\right)\right\| & =\|\frac{\sum_{s \in S} D_s \nabla F_s(w)}{D}- \\
& \ \ \ \ \ \ \ \frac{\sum_{s \in S} D_s \nabla F_s\left(w^{\prime}\right)}{D}\| \\
& =\|\frac{\sum_{s \in S} D_s\left(\nabla F_s(w)-\nabla F_s\left(w^{\prime}\right)\right)}{D}\| \\
& \leqslant \frac{\sum_{s \in S} D_s}{D}\left\|\nabla F_s(w)-\nabla F_s\left(w^{\prime}\right)\right\| \\
& \leqslant \frac{\sum_{s \in S} D_s L_s}{D}\left\|w-w^{\prime}\right\|.
\end{aligned}
\end{equation}

where \emph{$L$} is the Lipschitz continuous gradient of the global loss function \emph{$F(w)$}. Therefore, \emph{$F(w)$} meets the definition of Lipschitz smooth. So far, Assumption 2 has been certified.

~

\section{}

The next thing we want to prove is the establishment of Theorem 1. But before that, referring to the form in the work\cite{ref52}, we give the following definition:

\begin{myDef}

For any training period t of ES, there is an upper bound on the distance between \emph{$w_s^t$} and \emph{$v_s^t$}:
\begin{equation}
\left\|w_s^t-v_s^t\right\| \leqslant \frac{\gamma_s}{L_s}\left(\phi L_s+1\right)^t.
\end{equation}
\end{myDef}

Through Definition 2, we know that the distance between \emph{$w_s^t$} and \emph{$v_s^t$} is related to \emph{$\gamma_s$}, \emph{$L_s$}, learning rate and training cycle t. Next, through Eq. (\ref{eq3}), Eq. (\ref{eq5}), Eq. (\ref{eq6}) and the triangle inequality, we deduce \emph{$\|w-v\|$}:
\begin{equation}
\begin{aligned}
\|w-v\| & =\|w^{\prime}-\phi \frac{\sum_{s \in S} D_s \nabla F_s\left(w_s^{\prime}\right)}{D}-v^{\prime}+\\
& \ \ \ \ \ \ \ \ \phi \frac{\sum_{s \in S} D_s \nabla F_s\left(v_s^{\prime}\right)}{D}\| \\
& =\|\left(w^{\prime}-v^{\prime}\right)-\frac{\phi \sum_{s \in S} D_s\left(\nabla F_s\left(w_s^{\prime}\right)-\nabla F_s\left(v_s^{\prime}\right)\right)}{D}\| \\
& \leqslant\left\|w^{\prime}-v^{\prime}\right\|+\|\frac{\phi \sum_{s \in S} D_s\left(\nabla F_s\left(w_s^{\prime}\right)-\nabla F_s\left(v_s^{\prime}\right)\right)}{D}\|.
\end{aligned}
\end{equation}

Through Assumption 1, Assumption 2 and Definition 2, we deduce the above Eq. (\ref{eq6}):
\begin{equation}
\begin{aligned}
\|w-v\| & \leqslant\left\|w^{\prime}-v^{\prime}\right\|+\frac{\phi \sum_{s \in S} D_s L_s\left(w_s^{\prime}-v_s^{\prime}\right)}{D} \\
& \leqslant\left\|w^{\prime}-v^{\prime}\right\|+\frac{\phi \sum_{s \in S} D_s \gamma_s\left(\phi L_s+1\right)^t}{D}.
\end{aligned}
\end{equation}

Finally, we found that the training speed of the global model is related to \emph{$\gamma_s$}. This means that the dataset distribution directly affects the convergence rate of the model. So far, Theorem 1 has been certified.

~

\section{}

We will prove the relationship between the optimal solution to the problem \textbf{P0} and the optimal solutions to the subproblems \textbf{P1} and \textbf{P2}. We can try to show this relationship through Lagrangian duality. First, construct the Lagrangian function of \textbf{P0}:
\begin{equation}
\begin{split}
L\left(a_{us}, p_u, \lambda, \mu, \nu\right) = \sum_{s \in S} \sum_{u \in U} a_{us} p_u \frac{d_u}{B_u \log_2\left(1+\frac{h_{us}p_u}{\sigma_u^2}\right)}\\
+ \sum_{u \in U} \lambda_u\left(\sum_{s \in S} a_{us} - 1\right) - \sum_{u \in U} \mu_u p_u - \sum_{u \in U} \nu_u (p_u - P_{\text{max}}),
\end{split}
\end{equation}

\noindent where $\lambda$, $\mu$ and $\nu$ are Lagrange multipliers. Now we perform Lagrangian dualization for the subproblems \textbf{P1} and \textbf{P2} respectively. For \textbf{P1}, we treat $p_u$ as a constant and minimize the Lagrangian function with respect to $a_{us}$:
\begin{equation}
\min _{a_{u s}} L\left(a_{u s}, p_u, \lambda, \mu, \nu \right).
\end{equation}

For \textbf{P2}, we treat $a_{us}$ as a constant and minimize the Lagrangian function with respect to $p_u$:
\begin{equation}
\min _{p_{u}} L\left(a_{u s}, p_u, \lambda, \mu, \nu \right).
\end{equation}

Solve the two sub-problems and obtain the optimal $a_{us}^*$ and $p_u^*$. Next, we consider the dual problem. The dual problem is obtained by maximizing the Lagrangian function $L\left(a_{u s}, p_u, \lambda, \mu, \nu\right)$ with respect to the Lagrange multipliers $\lambda$, $\mu$ and $\nu$ of. We assume the optimal solutions to the dual problem $\lambda^*$, $\mu^*$ and $\nu^*$, and the dual optimal value $D^*$. Now we need to prove that $D^*$ is the lower bound of the optimal value of the original problem. That is to say, for a set of Lagrange multipliers $\lambda^*$, $\mu^*$ and $\nu^*$, there is ($a_{us}^*$, $p_u^*$) such that the target value of the original problem Greater than or equal to the dual optimal value $D^*$.

This proof can be done through the following steps:

\begin{itemize}

\item[1)]

Use the Lagrangian duality property to ensure that the dual optimal value $D^*$ is the maximum value of the Lagrangian function $L\left(a_{u s}, p_u, \lambda, \mu, \nu\right)$ , that is, the dual optimal value $D^*=\max _{\lambda, \mu, \nu} L\left(a_{u s}, p_u, \lambda, \mu, \nu\right)$ .
\end{itemize}

\begin{itemize}

\item[2)]

Use the dual optimal value $D^*$ and the optimal solutions of the dual problem $\lambda^*$, $\mu^*$ and $\nu^*$ to prove that there is a set of $\left(a_{u s}^*, p_u^ *\right)$, making the objective function value of the original problem greater than or equal to $D^*$, that is, Primal Optimum $\geq D^*$.
\end{itemize}

\begin{itemize}

\item[3)]

This shows that the optimal solution $\left(a_{u s}^*, p_u^*\right)$ obtained by decoupling can at least reach the dual optimal value $D^*$, which is the optimal value of the original problem will not be less than the optimal value of the dual problem.
\end{itemize}

\begin{itemize}

\item[4)]

Further analyze the conditions of the Lagrangian dual problem and the relationship between the original problem and the dual problem to obtain the optimal solution $\left(a_{u s}^*, p_u^*\right)$ obtained by decoupling It may be close to or reach the optimal solution of problem \textbf{P0}.
\end{itemize}

We know that the Lagrangian function of the problem \textbf{P0} is:
\begin{equation}
\begin{split}
L\left(a_{us}, p_u, \lambda, \mu, \nu\right) = \sum_{s \in S} \sum_{u \in U} a_{us} p_u \frac{d_u}{B_u \log_2\left(1+\frac{h_{us}p_u}{\sigma_u^2}\right)}\\
+ \sum_{u \in U} \lambda_u\left(\sum_{s \in S} a_{us} - 1\right) - \sum_{u \in U} \mu_u p_u - \sum_{u \in U} \nu_u (p_u - P_{\text{max}}).
\end{split}
\end{equation}

We try to maximize the Lagrangian function L with respect to the dual problem of Lagrange multipliers $\lambda$, $\mu$ and $\nu$ as follows:
\begin{equation}
\max _{\lambda, \mu, \nu} \min _{a_{u s}, p_u} L\left(a_{u s}, p_u, \lambda, \mu, \nu\right).
\end{equation}

The dual optimal value is recorded as $D^*$. The dual property shows that $D^*$ is a lower bound on the optimal value of the original problem. Now, we try to find a set of $\left(a_{u s}^*, p_u^*\right)$ such that the target value of the original problem is greater than or equal to the dual optimal value $D^*$, that is, Primal Optimum $ \geq D^*$. Consider minimizing the Lagrangian function $L$ with respect to $a_{u s}$ and $p_u$ . For the minimization sub-problem \textbf{P1}, the optimal solution is $a_{u s}^*$, and for the minimization sub-problem \textbf{P2}, the optimal solution is $p_u^*$ .

We use the optimal solutions $\lambda^*$, $\mu^*$ and $\nu^*$ of the dual problem, as well as the optimal solution $\left(a_{u s}^*, p_u^*\right)$ , to construct the following inequality:
\begin{equation}
L\left(a_{u s}^*, p_u^*, \lambda^*, \mu^*, \nu^*\right) \leq \min _{a_{u s}, p_u} L\left(a_{u s}, p_u, \lambda^*, \mu^*, \nu^*\right).
\end{equation}

This is because for fixed Lagrange multipliers $\lambda^*$, $\mu^*$ and $\nu^*$ , the optimal solution $\left(a_{u s}^*, p_u^*\right)$ is The Langian function $L$ forms the lower bound. Therefore, we have:
\begin{equation}
\begin{split}
L\left(a_{u s}^*, p_u^*, \lambda^*, \mu^*, \nu^*\right) \leq \min _{a_{u s}, p_u} L\left(a_{u s}, p_u, \lambda^*, \mu^*, \nu^*\right) \\
\leq \max _{\lambda, \mu, \nu} \min _{a_{u s}, p_u} L\left(a_{u s}, p_u, \lambda, \mu, \nu\right)=D^* .
\end{split}
\end{equation}

This means that the target value of the original problem at the optimal solution is greater than or equal to the dual optimal value $D^*$ , that is, Primal Optimum $\geq D^*$ .

To sum up, we have proved that the optimal value of the original problem is not less than the optimal value of the dual problem $D^*$. This shows that the optimal solution $\left(a_{u s}^*, p_u^*\right)$ obtained by decoupling can at least reach the dual optimal value $D^*$, that is, they may be close to or reach the problem The optimal solution of \textbf{P0}.

~

\section{}

To establish a more intuitive connection, we can consider the dependency of the local loss function's gradient $\nabla F_s(w_s)$ on its data distribution $Q_s$. For each edge server, the difference between its local data distribution $Q_s$ and the global data distribution $Q_g$ affects the gradient $\nabla F_s(w_s)$. Therefore, we can regard the local gradient as an approximation of the global gradient calculated under the local distribution.

Assume that the local gradient can be written as a perturbation form of the global gradient:
\begin{equation}
\nabla F_s(w_s) = \nabla F(w) + \epsilon_s.
\end{equation}

Here, $\epsilon_s$ represents the perturbation caused by the difference between the local data distribution $Q_s$ and the global data distribution $Q_g$.

Assume that the perturbation $\epsilon_s$ is proportional to the KL divergence between the distributions:
\begin{equation}
\epsilon_s \propto D_{KL}(Q_g \| Q_s),
\end{equation}
this means that if we can reduce $D_{KL}(Q_g \| Q_s)$, we also reduce the perturbation $\epsilon_s$. We then give the proof that the perturbation $\epsilon_s$ is proportional to the KL divergence.

Let $f_c(w)$ be the loss function for category $c$ on a single sample, and let $F_s(w)$ be the loss function on edge server $s$ based on distribution $Q_s$. Then we have:
\begin{equation}
F_s(w) = \sum_{c \in C} Q_s(c) f_c(w),
\end{equation}
with its gradient given by:

\begin{equation}
\nabla F_s(w) = \sum_{c \ in C} Q_s(c) \nabla f_c(w).
\end{equation}

Consider the bias between the global distribution $Q_g$ and the local distribution $Q_s$ on edge server $s$. If the global loss function $F(w)$ were to weigh the sample loss function $f_c(w)$ for category $c$ with the weights from $Q_s$, the perturbation $\epsilon_s$ could be represented as:
\begin{equation}
\epsilon_s = \sum_{c \in C} (Q_s(c) - Q_g(c)) \nabla f_c(w).
\end{equation}

The expression for the KL divergence is:
\begin{equation}
D_{KL}(Q_g \| Q_s) = \sum_{c \in C} Q_g(c) \log \left( \frac{Q_g(c)}{Q_s(c)} \right).
\end{equation}

We observe that, when $Q_s(c)$ is significantly different from $Q_g(c)$, the corresponding $Q_g(c) \log \left( \frac{Q_g(c)}{Q_s(c)} \right)$ is large, which means that category $c$ contributes significantly to $D_{KL}(Q_g \| Q_s)$. With this characteristic of KL divergence, we can establish an intuitive link: the greater the difference between $Q_s(c)$ and $Q_g(c)$, the greater the discrepancy between $Q_s(c) \nabla f_c(w)$ and $Q_g(c) \nabla f_c(w)$.

To express this intuitive assumption mathematically, we arrive at:
\begin{equation}
\epsilon_s = \sum_{c \in C} \beta_c (Q_s(c) - Q_g(c)) \nabla f_c(w),
\end{equation}
where $\beta_c$ is a proportionality coefficient that scales with $Q_g(c) \log \left( \frac{Q_g(c)}{Q_s(c)} \right)$. The global gradient is the weighted average of all local gradients:

\begin{equation}
\nabla F(w) = \frac{\sum_{s \in S} |D_s| \nabla F_s(w_s)}{|D|}.
\end{equation}
By replacing the local gradients with the sum of the global gradient and perturbation, we get:
\begin{equation}
\nabla F(w) = \frac{\sum_{s \in S} |D_s| (\nabla F(w) + \epsilon_s)}{|D|}.
\end{equation}

Since the global gradient $\nabla F(w)$ is constant, we can extract it from the equation above, resulting in:
\begin{equation}
\nabla F(w) = \nabla F(w) + \frac{\sum_{s \in S} |D_s| \epsilon_s}{|D|}.
\end{equation}

Therefore, the gradient difference $\gamma_s$ can be written as:
\begin{equation}
\gamma_s = \left\| \frac{\sum_{s \in S} |D_s| \epsilon_s}{|D|} \right\|.
\end{equation}

This indicates that reducing the perturbation $\epsilon_s$ on each server (i.e., reducing KL divergence) will directly decrease the gradient difference $\gamma_s$. Therefore, reducing the KL divergence $D_{KL}(Q_g \| Q_s)$ will decrease the perturbation $\epsilon_s$ caused by the local distribution, thereby reducing the gradient difference $\gamma_s$. This demonstrates that reducing the KL divergence indeed helps federated learning models to converge better.

~

\section{}

After we replace the variables in the original Eq. (\ref{eq23}), the problem formulation is as follows:
\begin{equation}
\begin{aligned}
\hat{\Lambda}(p_{us}): \min \ & a_{u s} p_u \frac{d_u}{\log _2\left(1+\frac{h_{u s} p_u}{\hat{\sigma_u}^2}\right)}, \\
\text { s.t. } & p_u>0, \quad u \in U, \\
& p_u<P_{\text{max}}, \quad u \in U.
\end{aligned}
\end{equation}

We set \emph{$\epsilon=\frac{a_{u s} d_u}{B_k}$}, \emph{$\kappa=\frac{h_{u s}}{\sigma_u^2}$}. To judge the convexity of the objective function, we calculate its first and second derivatives with respect to \emph{$p_{us}$} respectively. When \emph{$\hat{\Lambda}^{\prime}(p_{us})=0$}, if \emph{$\hat{\Lambda}^{\prime \prime}(p_{us})>0$}, then \emph{$\hat{\Lambda}(p_{us})$} is a quasi-convex function.
\begin{equation}
\hat{\Lambda}^{\prime}(p_{us})=\frac{\epsilon[\ln (1+\kappa p_{us})(1+\kappa p_{us})-\kappa p_{us}]}{\ln 2 \log _2(1+\kappa p_{us})^2(1+\kappa p_{us})},
\end{equation}

\begin{equation}
\hat{\Lambda}^{\prime \prime}(p_{us})=\frac{\epsilon \kappa [\kappa p_{us}-\ln (\kappa p_{us}+1)]}{2 \ln ^2 2 \log _2(\kappa p_{us}+1)^2(\kappa p_{us}+1)^2}. \label{eq41}
\end{equation}

When \emph{$\hat{\Lambda}^{\prime}(p_{us})=0$}, \emph{$\kappa p_{us}=\ln (1+\kappa p_{us})(1+\kappa p_ {us})$} established. Bring this into Eq. (\ref{eq41}). Since \emph{$(1+\kappa p_{us})>0$}, Eq. (\ref{eq41}) is greater than 0. So \emph{$\hat{\Lambda}(p_{us})$} is a quasi-convex function.

\end{appendices}

\begin{IEEEbiographynophoto}{}

\end{IEEEbiographynophoto}

\quad \\

\begin{IEEEbiography}[{\includegraphics[width=1in,height=1.25in,clip,keepaspectratio]{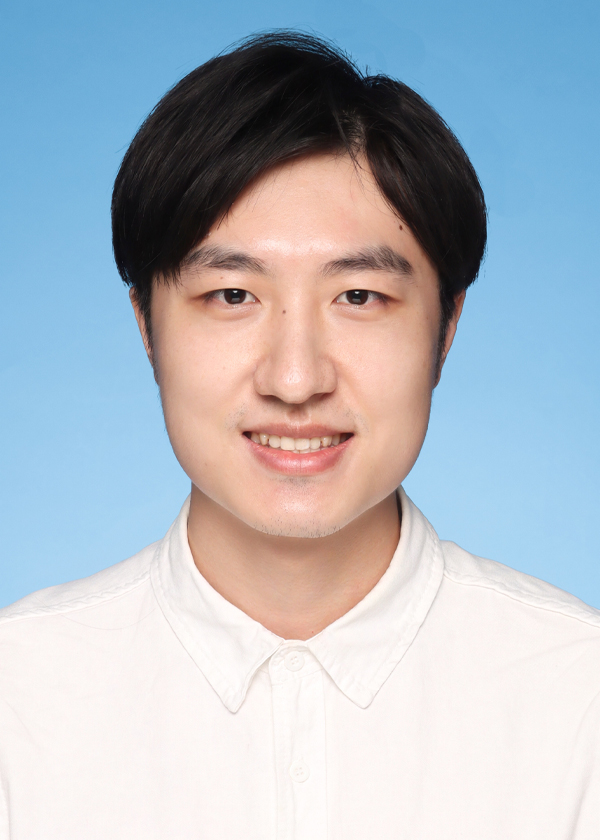}}]{Mulei Ma}
(mma085@connect.hkust-gz.edu.cn) received the B.S. degree from the Southern University of Science and Technology, Shenzhen, China, in 2020, and the master’s degree from ShanghaiTech University, Shanghai, China, in 2023. He is currently pursuing PhD degree in IoT Thrust and Research Center for Digital World with Intelligent Things (DOIT), The Hong Kong University of Science and Technology (Guangzhou), China. His research interests include edge intelligence, federated learning, task scheduling, and IoT.
\end{IEEEbiography}

\begin{IEEEbiography}[{\includegraphics[width=1in,height=1.25in,clip,keepaspectratio]{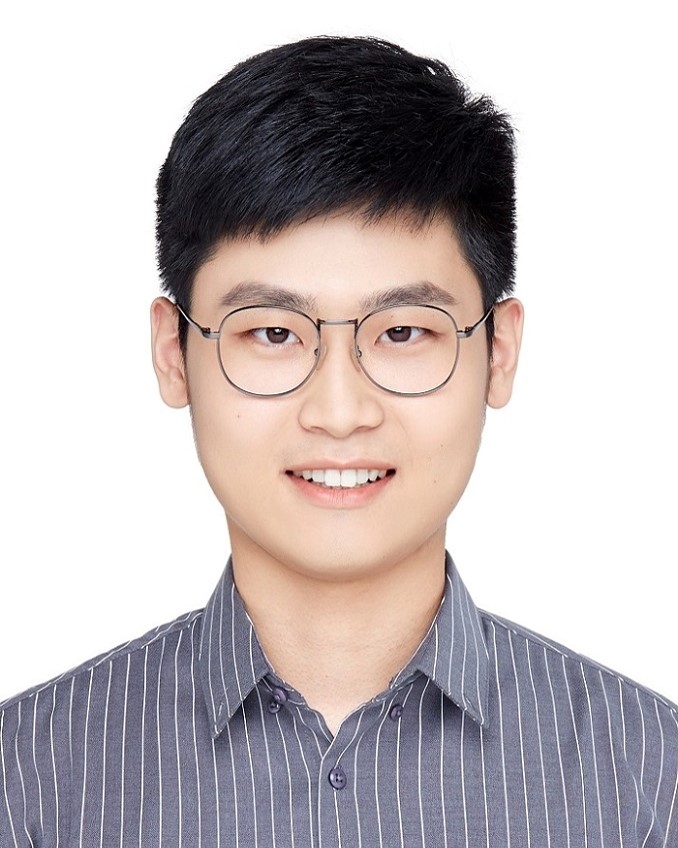}}]{Chenyu Gong}
(cgong040@connect.hkust-gz.edu.cn) received the B.S. degree from Shandong University, Qingdao, China, in 2020, and the master’s degree from ShanghaiTech University, Shanghai, China, in 2023. He is currently pursuing PhD degree in IoT Thrust and Research Center for Digital World with Intelligent Things (DOIT), The Hong Kong University of Science and Technology (Guangzhou), China. His research interests include personalized service, edge intelligence, task scheduling, and IoT.
\end{IEEEbiography}

\begin{IEEEbiography}[{\includegraphics[width=1in,height=1.25in,clip,keepaspectratio]{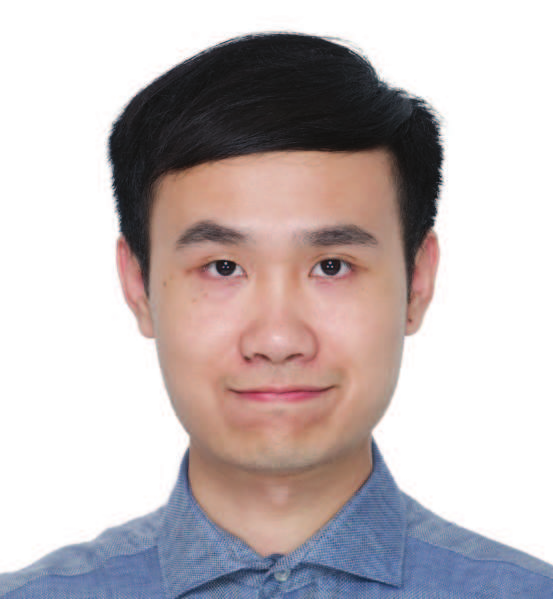}}]{Liekang Zeng}
(liekangzeng@hkust-gz.edu.cn) received the Ph.D. and the B.E. degrees from Sun Yat-sen University, Guangzhou, China. 
His current research interests include edge intelligence, mobile computing, and distributed machine learning systems.
\end{IEEEbiography}

\begin{IEEEbiography}[{\includegraphics[width=1in,height=1.25in,clip,keepaspectratio]{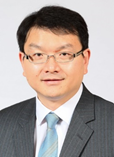}}]{Yang Yang}
[IEEE Fellow] (yyiot@hkust-gz.edu.cn) is a Professor with the IoT Thrust and Director of the Research Center for Digital World with Intelligent Things (DOIT) at the Hong Kong University of Science and Technology (Guangzhou), China. He is also the Chief Scientist of IoT at Terminus Group, an Adjunct Professor with the Peng Cheng Laboratory, and a Senior Consultant for Shenzhen Smart City Technology Development Group, China. His research interests include multi-tier computing networks, 5G/6G systems, and AIoT technologies and applications.
\end{IEEEbiography}

\begin{IEEEbiography}[{\includegraphics[width=1in,height=1.25in,clip,keepaspectratio]{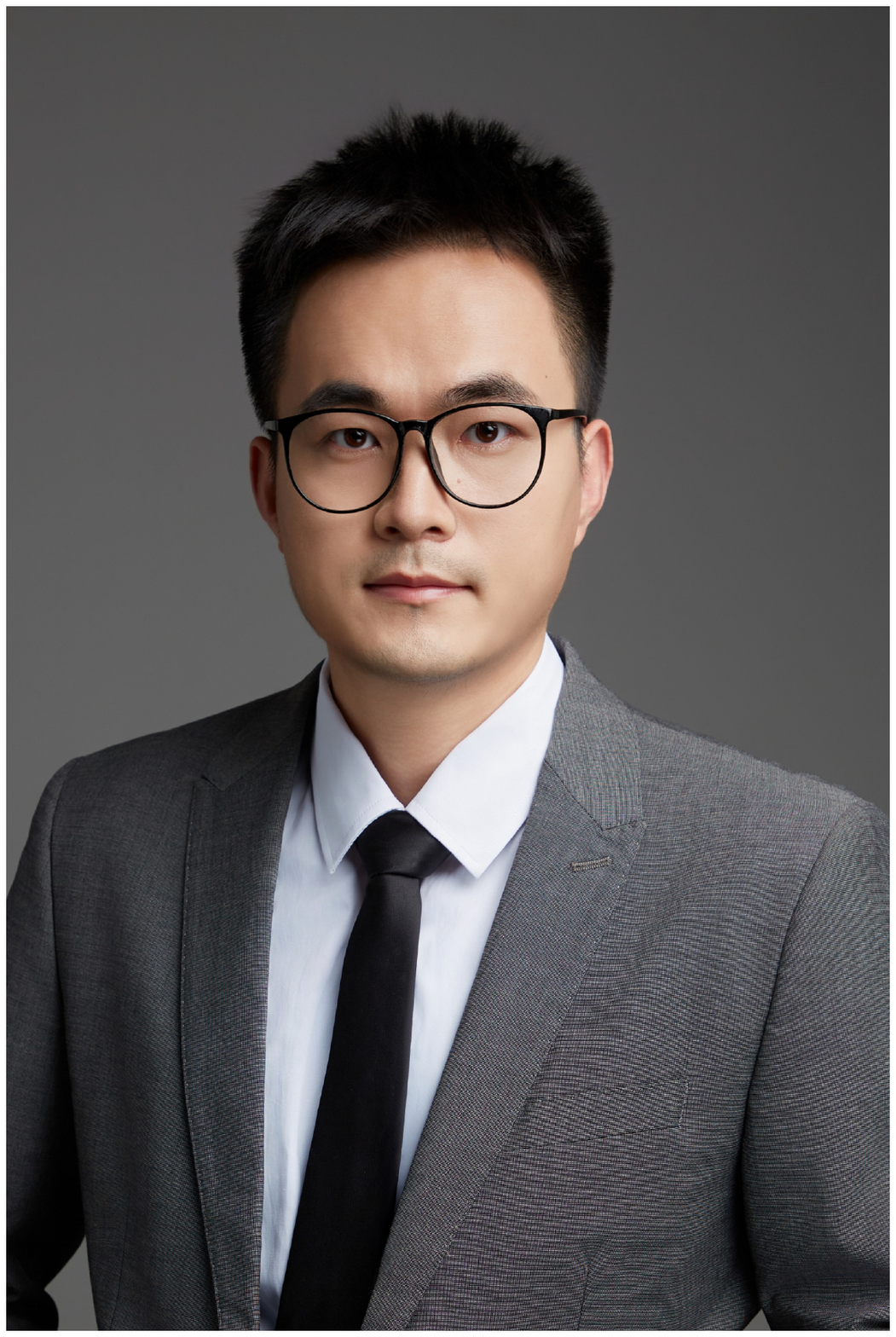}}]{Liantao Wu}
(ltwu@sei.ecnu.edu.cn) received the B.E. degree in automation from Shandong University, China, in 2012, and his Ph.D. degree in Control Science and Engineeringfrom Zhejiang University, China, in 2017.
He is currently an Associate Professor with East China Normal University, China. His research interests include IoT, edge computing, and edge intelligence.
\end{IEEEbiography}

\vfill
\end{document}